\documentclass{article}
\usepackage{comment,amsmath,amsthm,amssymb,array,latexsym,cite,epsfig}
\usepackage{finalT}

\numberwithin{equation}{section}

\input nl.def

\begin{document}


\title{The Newtonian limit for perfect fluids
}
\author{
Todd A. Oliynyk \thanks{Email: todd.oliynyk@monash.edu.au}$\;$\thanks{Present address: School
of Mathematical Sciences, Monash University, VIC 3800, Australia} \\
Max-Planck-Institut f\"{u}r Gravitationsphysik\\
Am M\"{u}hlenberg 1, D-14476 Golm,
Germany}
\date{}
\maketitle


\begin{abstract}
\noindent We prove that there exists a class of non-stationary
solutions to the Einstein-Euler equations which have a Newtonian
limit. The proof of this result is based on a symmetric hyperbolic
formulation of the Einstein-Euler equations which contains a
singular parameter $\ep = v_T/c$ where $v_T$ is a characteristic
velocity scale associated with the fluid and $c$ is the speed of
light. The symmetric hyperbolic formulation allows us to derive
$\ep$ independent energy estimates on weighted Sobolev spaces.
These estimates are the main tool used to analyze the behavior of
solutions in the limit $\ep \searrow 0$.
\end{abstract}

\sect{intro}{Introduction}

The Einstein-Euler equations or, in other words, the Einstein equations
coupled to a simple perfect fluid are given by the following system of equations
\lgath{EEeqn}{
G^{ij} = \frac{8\pi G}{c^4} T^{ij} \label{EEeqn.1} \\
\nabla_{i} T^{ij} = 0 \label{EEeqn.2} } where the stress-energy
tensor for the fluid is given by \lgath{EEdefs}{ T^{ij} = (\rho +
c^{-2} p)v^i v^j + p g^{ij} \,  } with $\rho$ the fluid density,
$p$ the fluid pressure, and $v$ the fluid four-velocity normalized
by $v^iv_i = -c^2$, $c$ the speed of light, and $G$ the
Newtonian gravitational constant. The study of the behavior of solutions to
these equations in the limit that $\ep = v_T/c \searrow 0$ where
$v_T$ is a characteristic velocity scale associated with the
fluid matter is known as the \emph{Newtonian limit}. By suitably rescaling
the gravitational and matter variables (see section \ref{units}), the
Einstein-Euler equations can be written as
\leqn{EEhateqns.intro}{ G^{ij} = 2 \epsilon^4 T^{ij} \quad
\text{and} \quad \nabla_{i} T^{ij} = 0 } where
$v_i v^i$ $=$ $-\ep^{-2}$, and
$t$ $=$ $x^4/v_T$ is a ``Newtonian'' time coordinate.
In the limit $\ep \searrow 0$,  one expects that there exists a class of solutions to Einstein-Euler equations \eqref{EEhateqns.intro} that
approach solutions of the Poisson-Euler equations
\lalign{newtB.intro}{
\del_t \rho + \del_I(\rho w^I) & = 0 \, , && (I,J=1,2,3)\label{newtB.1.intro}\\
\rho(\del_t w^J + w^I\del_I w^J) & =
-(\rho\del^J\Phi + \del^J p) \, , && (\del^I = \delta^{IJ}\del_J )\label{newtB.2.intro} \\
\Delta \Phi &=   \rho \, , && (\Delta = \del_I \del^I) \label{newtB.3.intro} \,
 }
of Newtonian gravity in some sense. As above, $\rho$ and $p$ are the fluid density and pressure,
respectively, while $w^I$ is the fluid (three) velocity.
This problem has
been studied since the discovery of general relativity by many
people and there is a large number of results available in the
literature. The majority of results are based on formal expansions
in the parameter $\ep$ which are used to calculate the
(approximate) values of physical quantities and also to investigate
the behavior of the gravitational and matter fields in the limit
$\ep \searrow 0$. For some classic and
recent
results of this type see \cite{BD86,BFN05,Dau64,EIH38,Chand65,Kun72,Kun76,KD,PW00,Will05} and
reference cited therein.
The main difficulty with the
formal expansions is that they leave completely unanswered the
question of convergence. In the absence of a precise notion of
convergence, it becomes unclear to what extent the formal
expansions actually approximate relativistic solutions.

In this paper, we go beyond formal considerations and supply a
precise notion of convergence for gravitating perfect fluids as
$\ep \searrow 0$. This necessitates introducing suitable
variables that are compatible with the limit $\ep \searrow 0$.
The metric $g_{ij}$, which defines the gravitational field, turns
out to be singular in this limit. To remedy this
problem, we introduce a new  gravitational density $\ufb^{ij}$ which
is related to the metric via the formula
\leqn{metrecA}{
g^{ij} = \frac{\ep}{\sqrt{-\det(Q)}}Q^{ij}
}
where
\leqn{metrecB}{
Q^{ij} = \begin{pmatrix} \delta^{IJ} & 0 \\ 0 & 0 \end{pmatrix}
+  \ep^2 \begin{pmatrix} 4 \ufb^{IJ} & 0 \\ 0 & -1 \end{pmatrix}
+ 4\ep^3 \begin{pmatrix} 0 & \ufb^{I4} \\ \ufb^{J4} & 0 \end{pmatrix}
+ 4 \ep^4\begin{pmatrix} 0 & 0 \\ 0 & \ufb^{44} \end{pmatrix} .
}
From this, it not difficult to see that the density $\ufb^{ij}$ is equivalent to the metric $g_{ij}$
for $\ep > 0$ and is well defined at $\ep =0$.
For the fluid, we also introduce a new velocity variable $w^i$ according
to
\leqn{wdef.intro}{
v^I = w^I \AND v^4=1+\ep w^4\, .
}
For technical reasons,  we only consider isentropic
flow where the pressure is related to the density by an equation
of state of the form $p = f(\rho)$. Moreover, to formulate a symmetric hyperbolic system
for the fluid
variables $\{\rho,v\}$, we need to deal with the well known
problem that the system becomes singular when $\rho + c^{-2} p =
0$. This is a particular problem for fluid balls having compact
support. To get around this problem, we follow Rendall
\cite{Ren92} and use a technique of Makino \cite{Mak} to
regularize the fluid equations so that a class of gravitating fluid
ball solutions can be constructed. Thus as in \cite{Ren92}, we
assume an equation of state of the form \leqn{eos}{ p =
K\rho^{(n+1)/n} } where $K \in \Rbb_{>0}$, $n\in \Nbb$, and we
introduce a new ``density'' variable $\alpha$ via the formula
\leqn{dendef}{ \rho = \frac{1}{\bigl(4Kn(n+1)\bigr)^n}\alpha^{2n}
\, . } As discussed by Rendall, the type of fluid solutions
obtained by this method have freely falling boundaries and hence
do not include static stars of finite radius  and so this
method is far from ideal. However, in trying to understand the
Newtonian limit and post-Newtonian approximations these solutions
are almost certainly general enough to obtain a comprehensive
understanding of the mathematical issues involved in the Newtonian
limit and post-Newtonian approximations. We would also like to
remark that the results contained in this article are largely
independent of the specific structure of the fluid equations. We
therefore expect that the analysis in this paper can be carried
over without much difficulty to other matter models whose
equations can be formulated as a symmetric hyperbolic system and
have a finite propagation speed for the matter density in the
limit $\ep\searrow 0$.

Our approach to analyze the limit $\ep \searrow 0$ is to use the gravitational
and matter variables  $\{\ufb^{ij},w^i,\alpha\}$ along with a harmonic
gauge to put the Einstein-Euler equations into the following form
\leqn{EFsym1.intro}{ b^0(\epsilon V)\partial_t V =
\frac{1}{\epsilon}c^I\partial_I V + b^I(\epsilon,V)\partial_I V +
f(\epsilon,V)V + \frac{1}{\epsilon}g(V)V + h(\ep) \, ,}
where $V$ comprises both the gravitational and matter variables, and
the $c^I$ are  constant matrices.
This system is symmetric hyperbolic and hence by standard
theory there exists local solutions.
However, the difficulty in analyzing the limit $\epsilon \searrow
0$ of such solutions is that the equation contains the singular terms
$\epsilon^{-1}c^I\partial_I V$ and $\epsilon^{-1}g(V)V$. Although, singular limits of
symmetric hyperbolic equations have been previously analyzed in
\cite{BK,KM82,Scho86,Scho88}, these results cannot be directly
applied to the system \eqref{EFsym1.intro}.  There are two main
difficulties in adapting these results to the Einstein-Euler
system. The first is that the Einstein-Euler system \eqref{EFsym1}
must be modified by including an elliptic equation, essentially the
Newtonian Poisson equation, in order
to be of the canonical form required by
\cite{BK,KM82,Scho86,Scho88}. This results in a coupled elliptic-hyperbolic
system of the form
\leqn{EFsym2.intro}{ B^0(\epsilon W)\partial_t W =
\frac{1}{\epsilon}c^I\partial_I W + B^I(\epsilon,W)\partial_I W +
F(\epsilon,W)W + H(\ep) \, ,}
where $W$ is related to $V$ via an elliptic equation and $F$ is a non-local
functional.
The second difficulty is that the
initial data which must include a $1/r$ piece for the metric and
hence it cannot lie in the Sobolev space $H^k$. This $1/r$ type fall-off behavior
is crucial for obtaining the correct limit and is intimately tied to
the elliptic part of our formulation of the Einstein-Euler system.
The standard
procedure in general relativity to deal with this type of fall off, at least for
elliptic systems, is to replace the spaces $H^k$ with the weighted
Sobolev spaces $H^k_\delta$ \cite{Bart86,CBC}. However, the
arguments used in \cite{BK,KM82,Scho86,Scho88} fail for the weighted
spaces as the weight used to define the $H^k_\delta$ spaces
destroys the integration by parts argument which is used
to control the singular term $\epsilon^{-1} c^I\partial_I W$ in \eqref{EFsym2.intro}.
Indeed,  using integration by parts, it follows easily from the definition of
the weighted $L^2_\delta$ inner-product (see \eqref{L2ip} with $\ep=1$) that
\leqn{IP.a}{
\ip{-\ep^{-1}c^I\del_I W}{W}_{L^2_\delta} = -\frac{1}{2\ep}\ip{\del_I(\sigma^{-2\delta-3})c^I W}{W}_{L^2}
}
where $\sigma(x) = \sqrt{1+|x|^2/4}$. In general,
this term will in blow up as $\ep \searrow 0$ unless $\delta = -3/2$ which coincides
with the standard $L^2$ norm. However, to include $1/r$ fall-off, we need to consider $-1<\delta <0$ which introduces a singular
$1/\ep$ term into energy
estimates based on the weighted norm $H^k_\delta$.

To overcome this problem, we
introduce a sequence of weighted spaces $H^k_{\delta,\ep}$
(see appendix \ref{winq} for a definition) by replacing the
weight $\sigma(x)$ with $\sigma_\ep(x)=\sigma(\ep x)$.
Under this replacement, \eqref{IP.a} changes to
\eqn{IP.aa}{
\ip{-\ep^{-1}c^I\del_I W}{W}_{L^2_{\delta,\ep}} \leq C\ip{W}{W}_{L^2_{\delta,\ep}}\, ,
}
which is no longer singular as $\ep \searrow 0$. This allows us
to derive
$\epsilon$ independent energy estimates for solutions to the
Einstein-Euler equations. These estimates can then be used to define a
precise notion of convergence for gravitating perfect fluids
solutions in the limit $\epsilon \searrow 0$ which is essentially
a statement about the validity of the zeroth order expansion in
$\epsilon$. This is formalized in the following theorem; for a
more precise version see propositions \ref{idatA}, \ref{locA} and \ref{convB}, and theorems \ref{convA} and \ref{rocD}.

\begin{thm} \label{introA} \mnote{[introA]}
  Suppose $-1<\delta < -1/2$, $k\geq 3+s$,
  $\beta^j \in \bigcap^{s}_{\ell=0}C^{\ell}([0,T^*],H^{k-\ell}_{\delta-1})$ is a harmonic gauge source
  function, and
  $\alphao,\wo^I\in
H^k_{\delta-1}$, $\zf^{IJ}\in H^{k+1}_{\delta}$, $\zf^{IJ}_4\in
H^k_{\delta-1}$ is the free initial data for the Einstein-Euler equations
where $\text{{\em supp}}$ $\alphao$ $\subset B_R^*$ for some $R^*>0$. Then
for $\ep_0$ small enough, there exists a $T\in (0,T^*]$ independent of $\ep \in (0,\ep_0]$,
and maps
\gath{convB2.intro}{
\ufb^{ij}_\ep(t)-\ufb^{ij}_\ep(0), \; \del_I\ufb^{ij}_\ep(t),
\; \del_t \ufb_\ep(t), \;
\alpha_\ep(t), \;  w^i_\ep(t) \in  \bigcap^{s+1}_{\ell = 0}
C^{\ell}([0,T],H^{k-\ell}_{\delta-1,\ep}) \, \\
\Phi \in  C^{0}([0,T^*],H^{k+2}_{\delta})
\cap C^{1}([0,T^*],H^{k+1}_{\delta})\, , \\
w^I \in C^{0}([0,T^*],H^k_{\delta-1}) \cap C^{1}([0,T^*],H^{k-1}_{\delta-1})\, , \\
\rho \in C^{0}([0,T^*],H^k_{\delta-1}) \cap C^{1}([0,T^*],H^{k-1}_{\delta-2})\, ,
}
such that

\begin{itemize}

\item[(i)]

\alin{idatA1}{ (\ufb^{ij}_\ep(0)) &
=\begin{pmatrix} \ep \zf^{IJ} &
\ep\wf_\ep^I \\
\ep\wf_\ep^J &
\phi_\ep
\end{pmatrix} \, ,\\
(\del_t\ufb^{ij}_\ep(0)) & = \begin{pmatrix} \zf^{IJ}_4 & -
\del_{K}\zf^{KI}+\beta^I(0)
\\ -\del_{K}\zf^{KJ}+\beta^J(0) & -\del_K\wf_\ep^K+\beta^4
\end{pmatrix} \, ,  \\
w^4_\ep(0) &=  -\frac{1}{\ep} +
\frac{-\ep\gb_{4J}\wo^J-\sqrt{\ep^2(\gb_{4J}\wo^J)^2-\gb_{44}
(\ep^2\gb_{IJ}\wo^I\wo^J+1)}}{\ep \gb_{44}}\, , \\
w^I_\ep(0) &= w^I(0) = \wo \, , \\
\alpha_\ep(0) &= \alphao \, , \\
\rho(0) & = \underset{\text{o}}{\rho} = (4Kn(n+1))^{-n}\alphao^{2n}\, ,}
where $\phi_\ep=\phi(\ep,\underset{\text{o}}{\rho},\wo^I,\zf^{IJ}_4,\beta^j(0),\zf^{IJ})$, and $\wf_\ep =
\wf(\ep,\underset{\text{o}}{\rho},\wo^I,\zf^{IJ}_4,\beta^j(0),\zf^{IJ})$ is the initial data determined
by the gravitational constraint equations (see proposition \ref{idatA}), and $\gb_{ij}$ is determined from $\ufb^{ij}_\ep(0)$ by the formulas \eqref{metrecA} and \eqref{bmetricdef},

\item[(ii)]
$\{\ufb_\ep^{ij}(x^I,t),\alpha_\ep(x^I,t),w^i_\ep(x^I,t) \}$
determines, via the formulas \eqref{metrecA}, \eqref{metrecB}, \eqref{wdef.intro},
and \eqref{dendef}, a $1$-parameter family $(0<\ep\leq \ep_0)$ of solutions
to the Einstein-Euler equations \eqref{EEhateqns.intro} in the harmonic gauge
$\ep\del_t\ufb_\ep^{4j}+\del_I\ufb_\ep^{Ij} = \ep \beta^j$
on the common spacetime region $(x^I,t)\in
D=\Rbb^3\times [0,T]$,

\item[(iii)]
$\{\Phi(x^I,t),\rho(x^I,t),w^I(x^I,t)\}$ solves the Euler-Poisson
equations \eqref{newtB.1.intro}-\eqref{newtB.3.intro}
on the spacetime region $D$,
\item[(iv)]
 there exists a constant $R \in (R^*,\infty)$ independent of $\ep \in (0,\ep_0]$ such
 that $\text{{\em supp}}$  $\alpha_{\ep}(t)$, $\text{{\em supp}}$ $\rho(t)$ $\subset$ $B_R$ for all $(t,\ep)\in [0,T]\times (0,\ep_0]$, and
\item[(v)]
there exists a constant $C>0$ independent of $\ep \in (0,\ep_0]$
such that
\gath{rocD1.intro}{
\norm{\ufb_\ep^{ij}(t)-\delta^i_4\delta^i_4\Phi(t)}_{L^6} +
\norm{\del_I\ufb_{\ep}^{ij}(t) - \delta^{i}_4\delta^j_4
\del_I \Phi(t)}_{H^{k-1}}
+\norm{v^I(t)-w^I(t)}_{H^{k-1}} \\
+ \ep^{-1}\norm{v^4(t)-1}_{H^{k-1}} + \norm{\rho_\ep(t) -
\rho(t)}_{H^{k-1}}
+\norm{\del_t\rho_\ep(t)-\del_t\rho(t)}_{H^{k-2}} \leq
C\ep} for all $(t,\ep)\in[0,T^*]\times (0,\ep_0]$.
\end{itemize}
\end{thm}

We remark, that the techniques of this paper can also be
used to derive \emph{convergent} expansions in $\ep$ of the type
considered in theorems 2 and 3 of \cite{KM82} and \cite{Scho88},
respectively. These convergent expansions in general
differ from the formal post-Newtonian expansions.
To get
post-Newtonian expansion to a certain order in $\ep$ requires that
the initial data must be chosen correctly. In the absence of
constraints on the initial data, a general procedure for doing
this is discussed in \cite{BK}. Due to the fact that there are
constraints on the initial data in general relativity, this
becomes a non trivial problem called the
\emph{initialization problem}. See \cite{ILR98} for an extended
discussion. The proof of convergence and a discussion of the initialization
problem will be presented in a separate paper \cite{Oli05d}.

We note that similar results for the
Vlasov-Einstein system have been derived in \cite{Ren94} using a zero shift
maximal slicing gauge. However, unlike
\cite{Ren94}, our approach is able to handle not only higher order
expansions in $\ep$, but also a wide variety of matter models. We also
note that in \cite{FR94,ILR98}, there is another interesting
proposal for analyzing the limit as $\ep \searrow 0$ which is
based on a gauge for which the Einstein equations are again
elliptic-hyperbolic but distinct from \cite{Ren94}. As
in this article, the authors of \cite{FR94,ILR98} also propose to
use the methods of \cite{BK,KM82,Scho86,Scho88}. However, the
required estimates are not proven and it is yet to be verified if
this approach would be successful.

We remark that the results of this and the companion paper \cite{Oli05d}
are local in time
and therefore address the ``near zone'' problem. In the
special case of spherical symmetry, the situation improves and
there are some global results available on the Newtonian limit \cite{Non,RR92}. However,
because spherically symmetric systems
do not generate gravitational radiation, these results do not shed
light on the ``far zone'' problem for post-Newtonian expansions
where radiation plays a crucial role and the $\ep \searrow 0$ limit
must be analyzed  in the region ``close'' to future null infinity.
We plan to investigate the far zone problem in the near future.

Our paper is organized as follows: in section
\ref{units}, we define
dimensionless variables for the Einstein-Euler system. Sections \ref{red} and \ref{eul} are
devoted to introducing variables and a gauge condition that cast
the Einstein-Euler equations into a form suitable for analyzing
the limit $\ep \searrow 0$. Appropriate initial data which is
regular in the limit $\ep \searrow 0$ is constructed in section
\ref{idat} while in section \ref{loc} we prove a local existence
theorem for the Einstein-Euler system on the weighted spaces.
Finally, in section \ref{nlim}, we show that solutions to the
Einstein-Euler system converge as $\ep \searrow 0$ to solutions of
the Poisson-Euler system of Newtonian gravity. A precise
statement of convergence is contained in theorem \ref{rocD}
which is the main result of this paper.

\sect{units}{Units}

Our conventions for units are as follows: \eqn{unitsdef}{ [x^i] =
L\, , \quad [g_{ij}]  = 1\, ,\quad [\rho ] = \frac{M}{L^3} \, ,
\quad [p] = \frac{M}{LT^2} \, , \quad [v^i] = [c]  = \frac{L}{T}
\, ,\quad \text{and} \quad [G] = \frac{L^3}{MT^2} \, . } Note that
with  these choices the stress-energy tensor has units of an
energy density, i.e. $[T^{ij}] = \frac{M}{LT^{2}}$. To introduce
dimensionless variables, we define \eqn{veldef}{ v^i = v_{T} \vh^i
\AND \rho = \rho_T \hat{\rho} } where $v_{T}$ and $\rho_T$  are
``typical'' values for the velocity and the density, respectively.
The Einstein-Euler equations then can be written as
\eqn{EEhateqns}{ \Gh^{ij} = 2 \epsilon^4 \Th^{ij} \quad
\text{and} \quad \hat{\nabla}_{i} \Th^{ij} = 0 } where
\gath{EEhatdefs}{ \epsilon  = \frac{v_T}{c}\, , \quad \kappa  =
\frac{4\pi G\rho_T }{v_T^2} \, , \quad \hat{x}^i = \sqrt{\kappa}\,
x^i \, , \quad \hat{g}_{ij} = g_{ij}\, , \quad
\hat{p}  = \frac{p}{v_T^2\rho_T}\, ,\\
\intertext{and} \hat{T}^{ij}  = (\hat{\rho}+\epsilon^2 \hat{p})\hat{v}^i\hat{v}^j +
\hat{p}\hat{g}^{ij} \, . } The normalization $v_i v^i=-c^2$, implies that
\eqn{vhatnorm}{ \hat{v}_i\hat{v}^i := \hat{g}_{ij}\hat{v}^i\hat{v}^j = -\frac{1}{
\ep^2} \, . } Also, we can introduce a time coordinate $t$ via
\eqn{tcoord}{ t = x^4/v_T \, . } With these choices, we have
\eqn{unitshat}{ [\epsilon] = [\hat{v}^i]  = [\hat{\rho}] = [\hat{p}] = [\hat{g}] =
[\hat{x}^i] = 1 \, , \quad [v_T] = \frac{L}{T} \, , \quad
[t]=[T]\, , \quad \text{and} \quad [\kappa] = \frac{1}{L^2} \, .}
Thus all our dynamical variables and coordinates are dimensionless
and the two constants $v_T$ and $\kappa$ can be used to fix the
length and time scales by using units so that \eqn{setscale}{ v_T
= 1 \AND \kappa = 1 \, .} In this case we can use $t$ and $x^4$
interchangeably as long as we remember that they carry different
units. To simplify notation, we will drop the ``hats'' from the
hatted variables for the remainder of this article.

\sect{red}{Reduced Einstein Equations}

To aid in deriving the appropriate symmetric hyperbolic system for
the gravitational variables, we temporarily introduce a new set of
coordinates related to old ones by the simple rescaling
\eqn{bcoords}{ \xb^J = x^J , \quad \xb^4 = x^4/\epsilon } and let
\eqn{bpartial}{
\partial_i = \frac{\partial\;}{\partial x^i} \, ,
\quad \partialb_i =  \frac{\partial\;}{\partial \xb^i} \, . } In
the new coordinates, the metric $\gb_{ij}$ and its inverse
$\gb^{ij}$ are given by \leqn{bmetricdef}{ (\gb_{ij}) =
\begin{pmatrix}
g_{IJ} & \ep g_{I4} \\
\ep g_{4J} & \ep^2 g_{44}
\end{pmatrix}
\quad \text{and} \quad
(\gb^{ij}) = \begin{pmatrix}
g^{IJ} & \ep^{-1} g^{I4} \\
\ep^{-1} g^{4J} & \ep^{-2} g^{44}
\end{pmatrix}\, .
} Next, consider the metric density \leqn{densdef}{ \gfb^{ij} =
\sqrt{|\gb|}\, \gb^{ij} \quad \text{where} \quad |\gb| =
-\det(\gb_{ij})\, . }
We note that the metric $\gb^{ij}$ is related to the density
$\gfb^{ij}$ by the following formula
\leqn{den2metb}{ \gb^{ij} = \frac{1}{\sqrt{|\gb|}}\gfb^{ij} \quad
\text{where} \quad |\gb| = -\det{\gfb^{ij}}\, , } and hence
\leqn{den2met}{ (g^{ij}) = \frac{1}{\sqrt{|\gb|}}
\begin{pmatrix}
\gfb^{IJ} & \ep \gfb^{I4} \\
\ep \gfb^{4J} & \ep^2 \gfb^{44}
\end{pmatrix} \,.
}
To obtain a gravitational variable that is regular and non-trivial
in the limit $\ep \searrow 0$, we define \leqn{udensdef}{
\ufb^{ij} := \frac{1}{4\epsilon^2}\bigl(\gfb^{ij}-\eta^{ij}\bigr) }
where \eqn{minkowski}{ \eta^{ij} = \begin{pmatrix}
\id_{\!\!3\times 3} & 0 \\
0 & -1
\end{pmatrix}
} is the Minkowski metric density.
As stated in the introduction, for $\ep > 0$,
the metric $g_{ij}$ can be recovered from the density $\ufb^{ij}$ via
the formulas \eqref{metrecA}-\eqref{metrecB}.
As we shall see, even though
the metric $g_{ij}$ is singular in the  limit $\ep \searrow 0 $,
the quantity $\ufb^{ij}$ is well defined at $\ep =0$. We note that
these variables are closely related to the gravitational variables
discovered by J\"{u}rgen Ehlers and
subsequently used in the papers \cite{Heil,Oli05a,Oli05b}
to construct stationary/static solutions
to the Einstein equations coupled to various matter sources.

In the $(\xb^i)$ coordinate system, the Christofell symbols are
given by \leqn{ChristA}{ \Gammab^{k}_{ij} =
\ep^2\bigl(\gfb^{km}(2\gfb_{i\ell}\gfb_{jp} - \gfb_{ij} \gfb_{\ell
p})\partialb_{m}\ufb^{\ell p} + 2( \gfb_{\ell
p}\delta^{k}_{(i}\partialb_{j)}\ufb^{\ell p}-
2\gfb_{\ell(i}\partialb_{j)}\ufb^{k\ell} )\bigr) \, . } We note
that Christofell symbols in the $(x^i)$ coordinate system are
related to the $\Gammab^{k}_{ij}$ as follows \lgath{ChristB}{
\Gamma^A_{44} = \ep^{-2} \Gammab^{A}_{44} \, , \quad
\Gamma^{4}_{44} = \ep^{-1} \Gammab^{4}_{44} \, ,
\quad \Gamma^{4}_{A4} = \Gammab^{4}_{A4}\, ,\label{ChristB.1}\\
\quad \Gamma^{4}_{AB} = \ep \Gamma^{4}_{AB}\, ,\quad \Gamma^A_{B4}
= \ep^{-1}\Gammab^A_{B4} \AND \Gamma^{A}_{BC} = \Gammab^{A}_{BC}
\, .\label{ChristB.2} }

Using \eqref{ChristA}, a straightforward calculation shows that the
Einstein tensor $\Gb^{ij}$ is given in terms of the density
$\ufb^{ij}$ by \leqn{Gb}{ \Gc^{ij}:= \frac{1}{2\ep^2}|\gb|\,
\Gb^{ij} = \gfb^{k\ell}\partialb^2_{k\ell} \ufb^{ij}+
\ep^2\bigl(A^{ij} + B^{ij} + C^{ij}\bigr) + D^{ij} } where
\lalign{Gbdef}{
|\gb| & = -\det(\gfb^{ij})\, , \label{Gbdef.1} \\
A^{ij} & = 2\bigl(\Half \gfb_{k\ell} \gfb_{mn} - \gfb_{km}
\gfb_{\ell n} \bigr) \bigl(\gfb^{ip} \gfb^{jq} - \Half \gfb^{ij}
\gfb^{pq} \bigr)\partialb_{p} \ufb^{k\ell}\partialb_{q}\ufb^{mn}
\label{Gbdef.2}\, ,\\
B^{ij} & = 4\gfb_{k\ell}\bigl(2\gfb^{n(i }
\partialb_{m}\ufb^{j)\ell}\partialb_{n}\ufb^{k m} - \Half
\gfb^{ij}\partialb_{m}\ufb^{k n}\partialb_{n}\ufb^{m \ell}
-\gfb^{mn}\partialb_{m}\ufb^{ik}\partialb_{n}\ufb^{j\ell}\bigr)\label{Gbdef.3} \, , \\
C^{ij} & = 4\bigl(\partialb_k\ufb^{ij}\partialb_{\ell}\ufb^{k\ell}-
\partialb_k\ufb^{i\ell}\partialb_\ell\ufb^{jk})\label{Gbdef.4} \, , \\
D^{ij} & := \gfb^{ij}\partialb^2_{k\ell} \ufb^{k\ell} - 2
\partialb^2_{k\ell}\ufb^{k(i}\gfb^{j)\ell}\, . \label{Gbdef.5} }
To
fix the gauge,  we assume that \leqn{harm}{
\partialb_{i}\ufb^{ij}
= \ep \beta^j } for prescribed spacetime functions
$\beta^j=\beta^j(x^I,x^4)$. For $\ep > 0$, $\partialb_{i}\ufb^{ij}
= \ep \beta^j$ implies that \eqn{harmB}{ \partialb_{i}\gfb^{ij} =
4\ep^3\beta^j \, } or equivalently \eqn{harmBa}{ \del_k \gf^{k4}=
4\ep^3 \beta^4 \AND \del_k \gf^{kA} = 4\ep^2\beta^A }where
$\gf^{ij} = \sqrt{-\det(g_{k\ell})}g^{ij}$ is the metric density
in the $(x^k)$ coordinates. Thus \eqref{harm} is, for $\ep
>0$, a generalized harmonic type gauge and is harmonic if
the functions $\beta^j$ are chosen to be identically zero.
Clearly, if we define \eqn{Edef}{ E^{ij} :=
\gfb^{ij}\delb_k\beta^k-2\delb_k\beta^{(i}\gfb^{j)k}\, , } then
\eqref{harm} implies that \eqn{harmE}{ D^{ij} = \ep E^{ij}. }
Setting \leqn{Gred}{ \Gcr^{ij} :=\Gc^{ij}-D^{ij}+\ep E^{ij} =
\gfb^{k\ell}\partialb^2_{k\ell}\ufb^{ij} + \ep E^{ij} +
\epsilon^2\bigl(A^{ij}+B^{ij}+C^{ij}\bigr)  } and \eqn{streseng}{
\Tc^{ij} := \ep^2 |\gb|\,\Tb^{ij} = |\gb|\begin{pmatrix}
\ep^2T^{IJ} & \ep^{1} T^{I4} \\
\ep^{1}T^{4J} & T^{44}
\end{pmatrix}
} the Einstein equations $G^{ij} = 2\ep^4 T^{ij}$ in the gauge
\eqref{harm} become \leqn{Einred}{ \Gcr^{ij}=  \Tc^{ij} } which we
will refer to as the \emph{reduced Einstein equations}.

To write the reduced Einstein equations in first order form, we
introduce the variables \eqn{fo1}{ \ufb^{ij}_{k} :=
\partialb_{k}\ufb^{ij} = \left\{ \begin{array}{ll}
\partial_{I}\ufb^{ij} & \text{if $k=I$}\\
\ep \partial_{4} \ufb^{ij} & \text{if $k=4$}
\end{array} \right. \, .
} The reduced Einstein equations then become \alin{fo2}{
-\gfb^{44}\partialb_4 \ufb^{ij}_4 & =
2\gfb^{4I}\partialb_I\ufb^{ij}_4 + \gfb^{IJ}\partialb_I\ufb^{ij}_J
+ \ep E^{ij} +\ep^2\bigl(A^{ij}+B^{ij}+C^{ij}) - \Tc^{ij} \, ,\\
\gfb^{IJ}\partialb_4\ufb^{ij}_J & = \gfb^{IJ}\partialb_{J}\ufb^{ij}_4\, , \\
\partialb_4 \ufb^{ij} &= \ufb^{ij}_4 \, , } or equivalently
\alin{fo3}{ -\gfb^{44}\partial_4 \ufb^{ij}_4 & =
\frac{2}{\epsilon}\gfb^{4I}\partial_I\ufb^{ij}_4 +
\frac{1}{\epsilon}\gfb^{IJ}\partial_I\ufb^{ij}_J
+ E^{ij}+\ep\bigl(A^{ij}+B^{ij}+C^{ij}) -\frac{1}{\epsilon} \Tc^{ij} \, ,\\
\gfb^{IJ}\partial_4\ufb^{ij}_J & =
\frac{1}{\ep}\gfb^{IJ}\partial_{J}\ufb^{ij}_4\, , \\ \partial_4
\ufb^{ij} &= \frac{1}{\ep}\ufb^{ij}_4 \, . } Next, define
\leqn{fo4}{ \uf^{ij} := \ep \ufb^{ij} \, \quad
\uf^{ij}_k := \ufb^{ij}_k \,  } and let
\eqn{Vcaldef}{ \Vc = \{ \, (r^{ij}) \in \Mbb_{4\times
4} |\; \det (\eta^{ij} + 4 r^{ij}) > 0 \, \}\, . } Then using
vector notation \eqn{fo5}{ \ufv^{ij} := ( \uf_4^{ij}, \uf_J^{ij},
\uf^{ij})^T \, , } we can write the reduced Einstein equations as
\leqn{fo6}{ A^4 (\ep\ufv)\partial_4\ufv^{ij} =
\frac{1}{\ep}C^{I}\partial_I\ufv^{ij} +
A^I(\ufv)\partial_I\ufv^{ij} + \Fb^{ij}(\ep,\ufv)
-\frac{1}{\ep}(\Tc^{ij},0,0)^T } where
\leqn{fo7}{ A^{4}(\ep\ufv) =
\begin{pmatrix}
1-4\ep\uf^{44} & 0 & 0\\
0 & \delta^{IJ}+4\ep\uf^{IJ} & 0 \\
0 & 0 & 1
\end{pmatrix} \, ,
}
\leqn{fo8}{ C^{I} = \begin{pmatrix}
0 & \delta^{IJ} & 0 \\
\delta^{IJ} & 0 & 0 \\
0 & 0 & 0
\end{pmatrix} \, ,
} \leqn{fo9}{ A^{I}(\ufv) = \begin{pmatrix}
8\uf^{4I} & 4\uf^{IJ} & 0 \\
4\uf^{IJ} & 0 & 0 \\
0 & 0 & 0
\end{pmatrix} \, ,
} and \leqn{fo10}{ \Fb^{ij}(\ep,\ufv) = (
E^{ij} + \ep\fb^{ij}(\ep\uf,\uf_k) ,0,\uf^{ij}_4)^T
\, .
} The functions $\fb^{ij}(\ep\uf,\uf_k)$ are
analytic for $\ep\uf \in \Vc$ and moreover are quadratic in
$\uf_k$. Here we are using the notation \eqn{fo10a}{ \uf =
(\uf^{ij}) \AND \uf_k = (\uf^{ij}_k) \, . } The stress-energy
tensor is given in terms of the $\uf$ variable by \lalign{fo11a}{
(T^{ij}) = \rho (v^i v^j) +\frac{1}{\sqrt{|\gb|}}\begin{pmatrix}
\delta^{IJ} p & 0 \\
0 & 0
\end{pmatrix}
&+ \frac{\ep}{\sqrt{|\gb|}}\begin{pmatrix}
4\uf^{IJ} p & 0 \\
0 & 0
\end{pmatrix} \notag \\
& + \ep^2\left( p(v^i v^j) + \frac{p}{\sqrt{|\gb|}}\begin{pmatrix}
0 & 4\uf^{I4} \\
4\uf^{4J} & -1+4\ep\uf^{44}
\end{pmatrix}
 \right) \, , \label{fo11}
} and hence \leqn{fo12}{ \frac{1}{\ep}(\Tc^{ij}) =
\begin{pmatrix}
0 & 0 \\
0 & \ep^{-1}\rho
\end{pmatrix}
+ \Sc^{ij} } where \lalign{fo13}{ (\Sc^{ij}) & = \rho
\begin{pmatrix}
0 & |\gb|v^{I}v^{4} \\
|\gb|v^{J}v^{4}  & \ep^{-1}\bigl[(|\gb|-1)(v^{4})^2 +((v^4)^2-1)\bigr]
\end{pmatrix} \notag\\
& \quad + \ep |\gb|\begin{pmatrix} (\rho+\ep^2 p) v^{I}v^{J} +
|\gb|^{-1/2}p(\delta^{IJ}+4\ep\uf^{IJ}) &
\ep pv^{I}v^{4} +  4 \ep |\gb|^{-1/2}p\uf^{I4} \\
\ep pv^{J}v^{4} +  4 \ep |\gb|^{-1/2}p\uf^{4J} & p (v^4)^2 +
|\gb|^{-1/2}p(-1+4\ep\uf^{44})
\end{pmatrix} \, .\label{fo13.1}
} We remark that if $v^4-1 = \text{O}(\ep)$, then $S^{ij}$ is
regular in $\ep$ as is easily seen from the above formula and the
expansion \leqn{fo14}{ |\gb| = 1 + 4\ep\eta_{ij}\uf^{ij} +
f(\ep\uf) } where $f(\ep\uf)$ is analytic for
$\ep\uf \in \Vc$ and also satisfies $f(y)=\text{O}(|y|^2)$ as
$y\rightarrow 0$.

\sect{eul}{Regularized Euler equations}

There are various approaches to symmetric hyperbolic formulations
of the relativistic Euler equations \cite{BrKa,Frau,Frie,Ren92,Walt}.
We use the approach of \cite{BrKa} which is based on fluid projection and the
introduction of a Makino variable.

In the coordinates $(\xb^i)$, the Euler equations are given by
\leqn{eul1}{ \nablab_i \Tb^{ij} =0} where $\Tb^{ij} = (\rho + \ep^2
p) \vb^i\vb^j + p \gb^{ij}$ and the fluid velocity $\vb^i$ is
normalized according to \leqn{eul3}{\vb_i\vb^i =
-\frac{1}{\ep^2}\, .} Differentiating \eqref{eul3} yields
\leqn{eul4}{\vb_i \nablab_j \vb^i = 0}  which implies
\leqn{eul5}{\vb^{j}\vb_i \nablab_j \vb^i = 0\, .} Writing out
\eqref{eul1} explicitly, we have \leqn{eul6}{(\delb_i\rho
+\ep^2\delb_i p)\vb^i\vb^j + (\rho+\ep^2 p)(\vb^j\nablab_i\vb^i
+\vb^i\nablab_i\vb^j) + \gb^{ij}\delb_i p = 0\, . } The operator
\eqn{proj}{L^j_i = \delta^j_i + \ep^2 \vb^j\vb_i} projects into
subspace orthogonal to the fluid velocity $\vb^i$, i.e.
$L^j_{i}L^i_k = L^j_k $ and $L^j_i\vb^i=0$.  Using $L^j_k$ to
project the Euler equations \eqref{eul6} into components parallel
and orthogonal to $\vb^i$ yields, after using the relations
\eqref{eul3}-\eqref{eul5}, the following system \lgath{eul9}{
\vb^i\delb_i
\rho +(\rho+\ep^2 p)L^i_j\nablab_i\vb^j = 0 \, ,\label{eul9.1} \\
M_{ij}\vb^k\nablab_k \vb^j + \frac{1}{\rho+\ep^2 p} L^i_j \delb_i
p = 0\, ,\label{eul9.2}} where \eqn{Mdef}{ M_{ij} = \gb_{ij} +
2\ep^2 \vb_i\vb_j\, . }

As discussed in the introduction,
we introduce a new density variable
$\alpha$ via the formula \eqref{dendef}. Multiplying
\eqref{eul9.1} by the square of the function \eqn{hdef}{
h(\ep\alpha) = \left( 1 + \frac{1}{4n(n+1)}(\ep\alpha)^2\right)\,
,} gives \lgath{eul10}{ h^2\vb^i\delb_i
\alpha + h^2(\rho+\ep^2 p)\frac{d\alpha}{d\rho} L^i_j\nablab_i\vb^j = 0 \, ,\label{eul10.1} \\
M_{ij}\vb^k\nablab_k \vb^j + \frac{s^2}{\rho+\ep^2
p}\frac{dp}{d\alpha} L^j_i \delb_j \alpha = 0\, , \label{eul10.2}
} where \eqn{sound}{ s^2 = \frac{dp}{d\rho} =
\frac{1}{4n^2}\alpha^2} is the square of the speed of sound. A
simple calculation shows that \eqn{eul11}{ \frac{s^2}{\rho+\ep^2
p}\frac{dp}{d\alpha} = h^2(\rho+\ep^2 p)\frac{d\alpha}{d\rho} = q
} where \eqn{eul12}{ q = q(\ep,\alpha) = \frac{1}{2n
h(\ep\alpha)}\alpha\, .} This shows that the system
\eqref{eul10.1}-\eqref{eul10.2} is \emph{symmetric}, and moreover
at a point where $\alpha =0$ and hence $p=\rho =0$, it is
\emph{regular} unlike \eqref{eul9.1}-\eqref{eul9.2}. This is the
point of introducing the Makino variable $\alpha$.
 Also note that the pressure is given in terms of the Makino variable by
 \leqn{wdefA.1}{ p=\frac{K}{(4Kn(n+1))^{n+1}} \alpha^{2n+2} .}

Define \eqn{wdef}{ \quad w^I:=\vb^I\, , \AND w^4 :=
\vb^4-\frac{1}{\ep}} so that \leqn{wdefA.2}{ \\
v^I=w^I\, , \AND v^4 = 1+\ep w^4 \, .} Using
vector notation \eqn{wvecdef}{ \wv = (\alpha, w^i)^T, } we can write \eqref{eul10.1} and
\eqref{eul10.2} as \leqn{eul13}{a^4 \del_4 \wv = a^I\del_I\wv + b}
where \lalign{adef}{ a^4 =&
\begin{pmatrix}
h^2(1+\ep w^4) & \ep q L^4_j \\
\ep q L^4_i & M_{ij}(1+\ep w^4)
\end{pmatrix}\, ,\label{adef.1} \\
a^I = & \begin{pmatrix}-h^2w^I & -q L^I_j \\
-q L^I_i & -M_{ij}w^I\end{pmatrix}\, , \label{adef.2}
\intertext{and}
b =& \begin{pmatrix} -q L^i_j \Gammab^j_{i\ell}\vb^\ell\\
-M_{ij}\Gammab^j_{k\ell}\vb^k\vb^\ell \end{pmatrix}\, .
\label{adef.3}} From \eqref{den2metb}, \eqref{udensdef},
\eqref{fo4}, and \eqref{fo14}, we find that \leqn{metexp}{
\gb_{ij} = \eta_{ij} + f_{ij}(\ep\uf) } where the $f_{ij}(y)$ are
analytic and satisfy $f_{ij}(y) = \text{O}(|y|)$ as $y\rightarrow
0$. Also,  \eqref{ChristA} shows that
\leqn{Christexp}{\Gammab^k_{ij}
=\ep\bigl[\eta^{km}\bigl(2\eta_{i\ell}\eta_{jp}-\eta_{ij}\eta_{\ell
p}\bigr)\ep\uf^{lp}_m + 2\bigl(\eta_{\ell p}\delta^k_{(i}\ep\uf^{\ell
p}_{j)}-2\eta_{\ell(i} \ep \uf^{k\ell}_{j)}\bigr)\bigr] + \ep
f^{k}_{ij}(\ep\uf,\ep\uf_m) } for functions $f^k_{ij}(\ep \uf, \ep
\uf_m)$ that are analytic for $\ep\uf \in \Vc$, linear in the $\ep
\uf_m$, and satisfy $f^k_{ij}(0,y)=0$. So then \leqn{Mexp}{ M_{ij}
= \gb_{ij} + 2\ep^2\gb_{ik}\gb_{j\ell}\vb^k\vb^\ell= \delta_{ij} +
m_{ij}(\ep\uf,\ep w^k)} and \leqn{Lexp}{L^j_i = \delta^j_i + \ep^2
\gb_{ik}\vb^k\vb^j = \delta^j_i -\delta_i^4\delta^{j}_{4} +
\ell^j_i(\ep\uf,\ep w^k) } where $\ell^j_i(\ep\uf,\ep w^k)$ and
$m_{ij}(\ep\uf,\ep w^k)$ are analytic for $\ep\uf \in \Vc$ and
$\ell^j_i(0,0)=m_{ij}(0,0) = 0$. Using
\eqref{metexp}-\eqref{Lexp}, the matrices $a^i$ and the vector $b$
can be written as \lalign{aexp}{ a^4 &= \begin{pmatrix} 1 & 0 \\ 0
& \delta_{ij}\end{pmatrix} + \ah^4(\ep\uf,\ep \wv)\, ,
\label{aexp.1}
\\ a^I &= \begin{pmatrix} -w^I & -\frac{\alpha}{2n}\delta^I_j \\
-\frac{\alpha}{2n}\delta^I_i & -\delta_{ij}w^I
\end{pmatrix} +  w^I\ah(\ep\uf,\ep\wv) +
\alpha\ah^I(\ep\uf,\ep\wv)\, , \label{aexp.2}\intertext{and}
b &= \begin{pmatrix} 0 \\
-\eta^{im}\bigl(2\eta_{4\ell}\eta_{4p}+\eta_{\ell
p}\bigr)\uf^{lp}_m - 2\bigl(\eta_{\ell p}\delta^i_4\uf^{\ell
p}_{4}-2\eta_{\ell 4}\uf^{i\ell}_{4}\bigr)
\end{pmatrix}+\begin{pmatrix}\alpha \bh_1(\ep\uf,\ep\wv)\cdot
\ep\uf_k\\
\bh_2(\ep\uf,\ep\wv)\cdot \uf_k\end{pmatrix}\, .} Note that {\bf(i)}
$\ah^4$, $\ah$, $\ah^I$, $\bh_1$, and $\bh_2$ are analytic in all
their variables provided that $\ep\uf \in \Vc$, {\bf(ii)} $\ah^4$,
$\ah$ and $\ah^I$ are symmetric, and {\bf(iii)} $\ah^4(0,0)=0$,
$\ah^I(0,0) = 0$, $\ah(0,0)= 0$, $\bh_1(0,0)= 0$, and
$\bh_2(0,0)=0$. Consequently the system \eqref{eul13} is
\emph{symmetric hyperbolic} on a region where $(\ep\uf,\ep \wv)$
is small enough to ensure that $a^4$ is positive definite. This
can always be arranged by taking $\ep$ small enough and since we
are interested in the limit $\ep \searrow 0$ no generality is lost
in assuming this.

It is important to realize that the derivation above of
\eqref{eul13} required that both the Euler equations \eqref{eul1}
and the fluid velocity normalization \eqref{eul3}  are satisfied.
Alternatively, we can first assume that \eqref{eul13} is satisfied
and then show that \eqref{eul1} and \eqref{eul3} are also
satisfied. To see this, define \leqn{fcon1}{ \Nc := \ep\vb_i\vb^i
+1/\ep = \ep \gb_{44}(1/\ep + w^4)^2 +1/\ep + 2\gb_{4J}(1 + \ep
w^4) w^J +\ep\gb_{IJ}w^Iw^J \, . } Clearly, $\Nc =0$ is equivalent
to $\vb^i\vb_i = -1/\ep^2$ for $\ep > 0$. Furthermore, any
solution of \eqref{eul13} also solves
\eqref{eul9.1}-\eqref{eul9.2} for any $\ep > 0$. So assuming that
$\vb$ is a solution to the system \eqref{eul9.1}-\eqref{eul9.2},
contracting \eqref{eul9.2} with $\vb^i$ yields \eqn{fcon2}{
(1+2\ep^2\vb^i\vb_i)\vb^k\delb_k(\vb^i\vb_i) + \frac{\ep
q\vb^j\delb_j\alpha}{2} \Nc= 0 . } For $(2\ep\Nc
-1)\neq 0$, this implies \leqn{fcon3}{ (1+\ep w^4)
\partial_4\Nc = -w^I\del_I\Nc + \frac{\ep^2 q \vb^j \delb_j\alpha}{4\ep\Nc-1}
\Nc \, . } Clearly, this is a symmetric
hyperbolic equation for $\Nc$ whenever $0<1/C \leq (1+\ep w^4)
\leq C$ for some constant $C$. This can always be arranged at
$x^4=0$ by choosing $\ep$ small enough. Therefore, if initially
$\Nc\bigl|_{x^4=0} = 0$, then $\Nc = 0$ for as long as $(1+\ep
w^4)$ stays absolutely bounded and bounded away from zero.
Consequently, choosing initial data for the system \eqref{eul13}
such that $\Nc\bigl|_{x^4=0} = 0$ will guarantee that the solution
will satisfy the full Euler equations \eqref{eul6} in an open
neighborhood of the hypersurface $x^4=0$. In particular, if
$\{\alpha,w^i\}$ is a solution to \eqref{eul13} with initial data
satisfying $\Nc|_{x^4=0}$, then $\alpha$ is a solution to the
equation \leqn{alphaA}{ \del_4 \alpha + X^I \del_I \alpha +
Y\alpha = 0 } where \leqn{alphaB}{ X^I := \frac{w^I}{1+\ep w^4}\,
, \AND  Y := \frac{\bar{\nabla}_{i}\bar{v}^i}{2n (1+\ep w^4)
h^3(\ep\alpha)} \,. } Observe that \eqn{alphaC}{ Y = \bar{Y}(\ep
w^4,\ep\alpha)(\ep \del_t w^4 + \del_I w^I) +\hat{Y}(\ep\uf,\ep
w^4,\ep \uf_k, \ep w^I,\ep) } where $\bar{Y}(0,0)-1/(2n)$ = 0,
$\hat{Y}(0,\ldots,0)=0$ and $\bar{Y}(\ep w^4,\ep\alpha^4)$,
$\hat{Y}(\ep\uf,\ep w^4, \ep \uf_k,\ep w^I,\ep\alpha^4)$ are
analytic on the region $\ep\uf \in \Vc$ and $1+\ep w^4 > 0$.

\sect{idat}{Newtonian initial data}

Let $S_0\cong \Rbb^3$ be the hypersurface defined by $S_0 :=
\{(x^I,0)\,|\, (x^I)\in \Rbb^3 \}$. The covector $n_i =
\delta_i^4$ is conormal to $S_0$ implying that constraint
equations for the initial data on $S_0$ are given by $n_{i}G^{ij}
= 2\ep^4 n_i T^{ij}$. Defining \eqn{Condef}{ \quad \Cc^J :=
\ep^{-1}(\Gc^{4J}- \Tc^{4J}) \AND \Cc^4 := \Gc^{44} -
\Tc^{44}, } we find that $\Cc^j = 0$ is equivalent to $n_{i}G^{ij}
= 2 \ep^4 n_i T^{ij}$ for $\ep > 0$. Also, by defining
\leqn{Hdef}{ \Hc^j := \delb_{i} \ufb^{ij}-\ep\beta^j\, , } the
generalized harmonic gauge \eqref{harm} can be written as $\Hc^j =
0$.

As will be seen in the proof of the next proposition the equations
$\Cc^j=0$ are regular at $\ep=0$. So to find appropriate initial
data that is well defined at $\ep = 0$, we solve the regularized
constraint equations $\Cc^j=0$. Moreover, we must also ensure that
the harmonic gauge condition $\Hc^j=0$ and the fluid normalization
$\Nc=0$ are satisfied. To solve the constraints $\Cc^j=0$,
$\Hc^j=0$, and $\Nc =0$, we use a implicit function technique
based on the work of Lottermoser \cite{Lott}. We
assume that the fluid velocity can be written as
\eqref{wdefA.1} which is consistent with the expected
behavior of the fluid velocity as $\ep \searrow 0$. We will not
assume that the density and pressure are related by the equation
of state \eqref{eos}. Instead, we will consider them as
independent prescribed fields for the purpose of finding solutions
to the constraint equations. We do this so that
the following proposition remains valid for other equations of state.
\begin{prop}\label{idatA} \mnote{[idatA]}
Suppose $-1<\delta < 0$, $k > 3/2+1$,  $R>0$
and
$(\tilde{\rho},\tilde{p},\tilde{w}^I,\tilde{\zf}_4^{IJ},
\tilde{\beta}^j,\tilde{\zf}^{IJ})
 \in (H^{k-2}_{\delta-2})^2\times H^k_{\delta-1}\times
(H^{k-1}_{\delta-1})^2\times B_R(H^{k}_{\delta}) \,.$
Then there exists an
$\ep_0>0$, an open neighborhood $U$ of
$(\tilde{\rho},\tilde{p},\tilde{w}^I,\tilde{\zf}_4^{IJ},
\tilde{\beta}^j,\tilde{\zf}^{IJ})$,
and analytic maps $(-\ep_0,\ep_0)\times U \rightarrow
H^k_{\delta-1}$ $\; : \;$
$(\ep,\rho,p,w^I,\zf_4^{IJ},\beta^j,\zf^{IJ})$$\mapsto$ $w^4$,
$(-\ep_0,\ep_0)\times U \rightarrow H^k_{\delta}$ $\; : \;$
$(\ep,\rho,p,w^I,\zf_4^{IJ},\beta^j,\zf^{IJ})$$\mapsto$ $\phi$,
$(-\ep_0,\ep_0)\times U \rightarrow H^k_\delta$ $\; : \;$
$(\ep,\rho,p,w^I,\zf^{IJ}_4,\beta^j,\zf^{IJ})$ $\mapsto$ $\wf^I$ such
that for each $(\rho,p,w^I,\zf^{IJ}_4,\beta^j,\zf^{IJ})\in U$,
$(\ep,\rho,p,w^I,w^4,\ufb^{ij}_4,\beta^j,\delb_4\ufb^{ij})$ is a solution to
the three constraints \leqn{idatA2}{ \Cc^j=0\, ,\quad  \Hc^j=0\, ,
\AND \Nc =0\,,} where \lalign{idatA1}{ (\ufb^{ij}) &
=\begin{pmatrix} \ep \zf^{IJ} &
\ep\wf^I \\
\ep\wf^J &
\phi
\end{pmatrix} \label{idatA1.1}\, ,\\
(\del_t\ufb^{ij}) & = \begin{pmatrix} \zf^{IJ}_4 & -
\del_{K}\zf^{KI}+\beta^I
\\ -\del_{K}\zf^{KJ}+\beta^J& -\del_K\wf^K+\beta^4
\end{pmatrix} \qquad (t=x^4)\, , \label{idatA1.2} \intertext{and}
w^4 &=  -\frac{1}{\ep} +
\frac{-\ep\gb_{4J}w^J-\sqrt{\ep^2(\gb_{4J}w^J)^2-\gb_{44}
(\ep^2\gb_{IJ}w^Iw^J+1)}}{\ep \gb_{44}}\, . \label{idatA1.3}}
Moreover, if we let
$\phi_0 = \phi|_{\ep=0}$, $\wf^I_0=\wf^I|_{\ep =0}$,
and $w^4_0 = w^4|_{\ep=0}$,
then $\phi_0$, $\wf^I_0$, and $w^4_0$ satisfy the equations \eqn{idatA3a}{
\Delta\phi_0 = \rho\, , \quad \Delta\wf^I_0 = \del_I\beta^4-\del_L\zf^{LI}_4+ \rho w^I \,
, \AND w^4_0 =  0 \, , } respectively.
\end{prop}
\begin{proof}
Let $\ufb^{44}=\phi$, $\ufb^{IJ} = \ep \zf^{IJ}$, $\ufb^{I4} =
\ep \wf^{I}$, and $\delb_4\ufb^{IJ} = \ep \zf_4^{IJ}$. Solving
 $\Hc^j\bigl|_{S_0}=0$ yields
\leqn{idatA3}{ \delb_4\ufb^{44} = \ep\bigl(-\del_I\wf^{I}+\beta^4)
\AND \delb_4 \ufb^{4J} = \ep(-\del_I \zf^{IJ}+\beta^J) \,  } while
solving $\Nc\bigl|_{S_0} =0$ gives \leqn{idatA4}{ w^4 =
-\frac{1}{\ep} +
\frac{-\ep\gb_{4J}w^J-\sqrt{\ep^2(\gb_{4J}w^J)^2-\gb_{44}
(\ep^2\gb_{IJ}w^Iw^J+1)}}{\ep \gb_{44}}\, . } From
\eqref{den2metb} and \eqref{udensdef}, it is not difficult to
verify that \eqn{idatA5}{ w^4 = \ep^{-1} f(\ep
w^I,\ep^3\zf,\ep^3\wf,\ep^2\phi)} where $f(\yv)$ $(\yv =
(y_1,\ldots,y_4))$ is analytic in a neighborhood of $(0,0,0,0)$
and and moreover $f(\yv)=\text{O}(|\yv|^2)$ as $\yv\rightarrow 0$.

Using the relation \eqref{idatA3} to eliminate $\delb_4\ufb^{44}$
and $\delb_4\ufb^{4J}$ in favour of $\wf^{I}$ and $\zf^{IJ}$, we
find that \alin{idatA6}{ &\gfb^{k\ell}\delb^2_{k\ell}\ufb^{44} +
D^{44} = \Delta \phi -\ep\del^2_{KL}\zf^{KL} + 4\ep^2h^4\, , \\
& \gfb^{k\ell}\delb^2_{k\ell}\ufb^{4J} + D^{4J} =
\ep\bigl(\Delta\wf^J -\del_J\beta^4 +\del_L \zf_4^{LJ} + 4\ep h^J
\bigr) \, , } where \alin{idatA7}{ h^4 &= \ep
\zf^{KL}\del_{KL}\phi + \ep \phi\del^2_{KL}\zf^{KL}
-2\ep^2\wf^L\del^2_{KL}\wf^K \, \\
h^J &= \ep^2\zf^{KL}\del^2_{KL}\wf^J +
\ep^2\wf^J\del^2_{KL}\zf^{KL} -\ep^2\wf^L\del^2_{KL}\zf^{KJ}
-\ep\phi\del_K \zf^{KJ}_4-\ep^2\zf^{JL}\del_L\beta^4\, .} Using this and equation
\eqref{Gb}, \eqref{Gbdef.1}-\eqref{Gbdef.5},
\eqref{fo11}-\eqref{fo13.1}, and \eqref{wdefA.1}-\eqref{wdefA.2},
we see that \lalign{ConI}{ \Cc^I &= \Delta\wf^I + \del_L
\zf_4^{LI} + \ep
h^I \notag \\
&\qquad + \ep f^I(\ep^3\zf,\ep^3\wf,\ep^2\phi,\ep D\zf,\ep D\wf,
D\phi,\ep\zf_4,\ep(-\del_K\wf^K+\beta^4),\ep(-\del_K\zf^{KL}+\beta^L)) - S^{4I}\label{ConI.1} \, , } and
\lalign{Con4}{ \Cc^4 &= \Delta \phi -\rho
-\ep(2w^4+\ep(w^4)^2)\rho
-\ep\del^2_{KL}\zf^{KL} + 4\ep^2h^4+ \notag \\
&\qquad \ep^2f^4(\ep^3\zf,\ep^3\wf,\ep^2\phi,\ep D\zf,\ep D\wf,
D\phi,\ep\zf_4,\ep(-\del_K\wf^K+\beta^4),\ep(-\del_K\zf^{KL}+\beta^L)) -\ep S^{44}\, , \label{Con4.1} } where the
functions $f^j(\yv)$ ($\yv = (y_1,\ldots,y_9)$) are analytic in a
neighborhood of $\{(0,0,0)\}\times U$ where $U$ is any open set
and are quadratic in $(y_4,\ldots,y_9)$. Note that \eqn{idatA8}{
\Sc^{44} = \rho S_1^{4}(\ep,w^I,\ep^2\zf,\ep^2\wf,\ep\phi) + p
S_1^{4}(\ep,w^I,\ep^2\zf,\ep^2\wf,\ep\phi) } and \eqn{idatA9}{
\Sc^{4I} = \rho w^I + \ep\rho
S^{I}_1(\ep,w^I,\ep^2\zf,\ep^2\wf,\ep\phi)+ \ep
pS_2^{I}(\ep,w^I,\ep^2\zf,\ep^2\wf,\ep\phi) } where the functions
$S^j(\yv)$ $(\yv = (y_1,\ldots,y_7))$ are analytic in a
neighborhood of $U\times\{(0,0,0)\}$ for any open set $U$.

Using lemma \ref{ciG} and
proposition 3.6 of \cite{Heil}, we see from the above considerations that
for any $R>0$ there exists an $\ep_0>0$ such that the maps
\eqn{idatA10}{ (-\ep_0,\ep_0) \times B_R(H^k_{\delta-1})\times B_R(H^k_\delta)^3
\longrightarrow H^k_\delta \; : \; (\ep,w^I,\zf,\wf,\phi)
\longmapsto w^4 } and \eqn{idatA11}{ (-\ep_0,\ep_0)\times
 (H^{k-2}_{\delta-2})^2\times B_R(H^k_{\delta-1})\times (H^{k-1}_{\delta-1})^2\times
 B_R(H^k_{\delta})^3
   \longrightarrow H^{k-2}_{\delta-2} \; : \; (\ep,\rho,p,w^I,\zf_4,\beta,\zf,\wf,\phi)\longmapsto \Cc^j}are
analytic. Since \leqn{idatA15}{\Cc^I|_{\ep=0} = \Delta \wf^I
-\del_I\beta^4+\del_L\zf_4^{LI}- \kappa \rho w^I,\qquad
\Cc^4|_{\ep=0} = \Delta\phi - \kappa \rho} and for $-1<\delta < 0$
the Laplacian $\Delta : H^{k}_\delta \rightarrow
H^{k-2}_{\delta-2}$ is an isomorphism (see \cite{Bart86},
proposition 2.2), we can use the analytic version of the implicit
function theorem (see \cite{Deim} theorem 15.3) to conclude,
shrinking $\ep_0$ if necessary, that there exists an open
neighborhood $U$ of any point in $(H^{k-2}_{\delta-2})^2\times
B_R(H^k_{\delta-1})\times (H^{k-1}_{\delta-1})^2\times
B_R(H^k_{\delta})$ and analytic maps
\eqn{idatA12}{(-\ep_0,\ep_0)\times U \longrightarrow H^k_\delta \;
:\; (\ep,\rho,p,w^I,\zf_4,\beta,\zf) \longmapsto \phi } and
\eqn{idatA13}{(-\ep_0,\ep_0)\times U \longrightarrow H^k_\delta \;
:\; (\ep,\rho,p,w^I,\zf_4,\beta,\zf) \longmapsto \zf } such the
constraints are satisfied, i.e. \eqn{idatA14}{
C^j(\ep,\rho,p,w^I,\zf_4,\beta,\zf,\wf(\ep,\rho,p,w^I,\zf_4,\zf),\phi(\ep,\rho,p,w^I,\zf_4,\zf))=0
} for all $(\ep,\rho,p,w^I,\zf_4,\beta,\zf)\in
(-\ep_0,\ep_0)\times U$.
\end{proof}

\sect{loc}{Local existence for the Einstein-Euler system}

The combined systems \eqref{fo6} and \eqref{eul13} can be written
as \leqn{EFsym1}{ b^0(\ep U,\ep V)\del_t V =
\frac{1}{\ep}c^I\del_I V + b^I(\ep,U,V)\del_I V + f(\ep,U,V)V +
\frac{1}{\ep}g(V)V + h_\ep  \qquad (t=x^4) } where
\lalign{EFsym2}{ U & := ( 0 , 0 ,\ufo^{ij}, 0, 0 )^T\, , \qquad
\qquad \ufo^{ij} := \uf^{ij}|_{t=0} = \ep
\ufb^{ij}|_{t=0} \, , \label{EFsym2.1} \\
V & := ( \uf^{ij}_{4} , \uf^{ij}_{J} , \delta\uf^{ij} , \alpha ,
w^i )^T\, ,  \qquad \qquad \delta\uf^{ij}:=
\uf^{ij}-\ufo^{ij}\, , \label{EFsym2.2} \\
b^0(\ep U,\ep V) & := \begin{pmatrix} A^4(\ep\uf) & 0 \\
0 & a^4(\ep \uf,\ep w^i,\ep \alpha)  \end{pmatrix} \, , \label{EFsym2.3} \\
c^I & := \begin{pmatrix}
C^I & 0 \\
0 & 0
\end{pmatrix}
\label{EYsym2.4} \, ,\\
b^I(\ep,U,V) & := \begin{pmatrix}  A^I(\uf) & 0 \\
0 & a^I(\ep,\ep \uf, w^i, \alpha) \end{pmatrix}  \, , \label{EFsym2.5}  \\
f(\ep,U,V)V & := \begin{pmatrix}
\ep \fb^{ij}(\ep \uf,\uf_k) - S^{ij} +4\ep \delta\uf^{ij}\delb_k\beta^k-8\ep\delb_k\beta^{(i}\delta\uf^{j)k} \\
0 \\
\uf^{ij}_4 \\
b(\ep,\ep\uf,\uf_k,w^i,\alpha)
\end{pmatrix}   \, , \label{EFsym2.6} \\
g(V)V & := ( -\delta^{i}_4\delta^{j}_4 \rho(\alpha) ,
0,\ldots,0)^T \, , \label{EFsym2.7} \intertext{and} h_\ep & :=
\begin{pmatrix}
4\ep \ufo^{ij}\delb_k\beta^k-8\ep\delb_k\beta^{(i}\ufo^{j)k} + \eta^{ij} \delb_k \beta^k - 2\delb_k \beta^{(i}\eta^{j)k} \\
0
\end{pmatrix} \, . \label{EFsym2.8}
} For initial data, we will use the following notation: given a
function $z$ that depends on time $t$, we define \eqn{t0}{
\underset{o}{z} := z|_{t=0}\, .}

To fix a region on which the system \eqref{EFsym1} is well
defined, we note from \eqref{fo7}, \eqref{aexp.1}, and the
invertibility of the Lorentz metric $(\eta^{ij})$ that there
exists a constant $K_0>0$ such that \leqn{contA1}{ -\det(\eta^{ij}
+ 4\ep\uf^{ij}) > 1/16\, , \quad 1+\ep w^4
> 1/16\, ,} \leqn{contA2}{ \quad A^4(\ep \uf) \geq \frac{1}{16} \id\, ,
\quad a^4(\ep\uf,\ep w,\ep \alpha) \geq \frac{1}{16}\id } and
\leqn{contA3}{ |A^4(\ep \uf)|\leq 16 \, , \quad |a^4(\ep\uf,\ep
w,\ep \alpha)| \leq 16} for all $|\ep\uf|\leq 2K_0$, $|\ep
w^i|\leq 2K_0$, $|\ep \alpha| \leq 2K_0$. The choice of the bounds
$1/16$ and $16$ is somewhat arbitrary and they can be replaced by
any number of the form $1/M$ and $M$ for any $M > 1$ without
changing any of the arguments presented in the following sections.
However, since we are interested in the limit $\ep \searrow 0$, we
lose nothing by assuming $M=16$.

\begin{prop} \label{locA} \mnote{locA}
Suppose $-1<\delta < 0$, $k \geq 3+s$, $\alphao,\wo^I\in
H^k_{\delta-1}$, $\zf^{IJ}\in H^{k+1}_{\delta}$, $\zf^{IJ}_4\in
H^k_{\delta-1}$, $\beta^j \in C^{1}([-T,T],H^k_{\delta-1})$. Let
$\ufbo^{ij}_{\ep}$, $\del_t\ufbo^{ij}_{\ep}$ and $\wo_\ep^4$ be
the initial data constructed in proposition \ref{idatA} which, by
choosing $\ep_0\leq 1$ small enough, satisfies \eqn{locA1}{ |\ep
\wo^i|\, , |\ep \alphao|\, , |\ep\ufbo^{ij}_\ep| \leq K_0 \quad
\text{for all $\ep \in (0,\ep_0]$}.} Then
\begin{itemize} \item[(i)] for each $\ep \in (0,\ep_0]$, there
exists $T_{1}(\ep), T_{2}(\ep)>0$ and a unique solution
\eqn{locA2}{ \Ve \in
\bigcap_{\ell=0}^{s+1}C^\ell((-T_{1}(\ep),T_{2}(\ep)),H^{k-\ell}_{\delta-1})}
to the system \eqref{EFsym1} with initial data \eqn{locA3}{
\underset{o}{\Ve} = ( \ep\del_t\ufbo_\ep^{ij} , \del_J
\ufbo_\ep^{ij} , 0 , \alphao , \wo^i )\, . } \item[(ii)] The
identities \eqn{locA4}{ \del_t\ufb^{ij}_\ep =
\frac{\uf^{ij}_{4,\ep}}{\ep}\, , \AND \uf^{ij}_{J,\ep} =
\del_J\ufb^{ij}_{\ep}} hold where by definition $\ufb^{ij}_{\ep} =
\ep^{-1}\uf^{ij}_{\ep}$, $\uf^{ij}_\ep = \ufo_\ep + \delta
\uf^{ij}_\ep $, and $\ufo_\ep = \ep \ufbo_\ep$. \item[(iii)] The
triple $\{\ufb^{ij}_\ep, w_\ep^i, \alpha_\ep\}$ determines, via
the formulas \eqref{dendef}, \eqref{den2met}, \eqref{udensdef},
and \eqref{wdefA.2},  a solution to the full Einstein-Euler system
\eqref{EEeqn.1}-\eqref{EEeqn.2} that satisfies the 
constraints \eqn{locA5}{ \ep\del_t
\ufb^{4j}_\ep + \del_I\ufb^{Ij}_\ep = \ep \beta^j \AND v^i v_i =
-\frac{1}{\ep^2}\, . } \item[(iv)] For some constant $C>0$
independent of $\ep$, the initial data $\Vo_\ep$ satisfies the
estimate \eqn{locA5a}{ \norm{\Vo_\ep - \Vo_0}_{H^k_{\delta-1,\ep}}
\leq C\norm{\Vo_\ep-\Vo_0}_{H^k_{\delta-1}} \leq C\ep } while
\eqn{locA5b}{ \norm{\del_t\Ve(0)}_{H^{k-1}_{\delta-1,\ep}} \leq
\norm{\del_t\Ve(0)}_{H^{k-1}_{\delta-1}} \leq C \, } for all $\ep
\in (0,\ep_0]$. \item[(v)] If $\: \sup_{0\leq
t<T_{2}(\ep)}\norm{\Ve(t)}_{W^{1,\infty}} < \infty$ and for all
$(x,t) \in \Rbb^3\times [0,T_{2}(\ep))$ $|\ep\delta \uf_\ep(x,t)|
< K_0$, $|\ep w^i(x,t)| < 2K_0$, and $|\ep\alpha_\ep(x,t)| <
2K_0$, then there exists a $T_*
> T_{2}(\ep)$ such that the solution $\Ve$ can be continued to the
interval $(-T_{1}(\ep),T_*)$.
\end{itemize}
\end{prop}
\begin{proof}
\noindent\textbf{(i)} Follows directly from
theorem \ref{leA}, proposition \ref{leB}, and corollary \ref{leC},  
where we use the initial data from proposition
\ref{idatA}.
\smallskip

\noindent \textbf{(ii)} This follows from standard arguments on
reductions of $2^{\text{nd}}$ order hyperbolic equations to
$1^{\text{st}}$ order symmetric hyperbolic systems. See
\cite{TayIII}, section 16.3 for details.
\smallskip

\noindent \textbf{(iii)} By part (ii), the triplet
$\{\ufb^{ij}_\ep,w_\ep^i,\alpha_\ep\}$ satisfies the reduced
Einstein equations \eqref{Einred} and the fluid equations
\eqref{eul13}. By construction,  $\{\ufb^{ij}_\ep|_{t=0},w_\ep^i
|_{t=0},\alpha_\ep |_{t=0}\}$  satisfies the constraints
$\Nc|_{t=0} =0$, $\Hc^j|_{t=0} =0$, and $(\Gc^{4i} -
\Tc^{4i})|_{t=0}=0$. The reduced Einstein equations \eqref{Einred}
can be written in terms of the Einstein density $\Gc^{ij}$ as
\leqn{locA6}{ \Gc^{ij} - \gfb^{ij}\delb_k \Hc^k+2\delb_k\Hc^{(i}
\gfb^{j)k} = \Tc^{ij}\, . } Using $(\Gc^{4i} -
\Tc^{4i})|_{t=0}=0$, we see that \eqn{locA7a}{ \bigl(-
\gfb^{4j}\delb_k \Hc^k+2\delb_k\Hc^{(4} \gfb^{j)k}\bigr)_{t=0} = 0
\,. } A straightforward calculation then shows that this implies
that $\del_t \Hc^j|_{t=0} = 0$. As discussed in sections \ref{eul}
(see \eqref{fcon3}), $\Nc$ satisfies a linear symmetric hyperbolic
system and hence by uniqueness, it follows that $\Nc = 0$ for all
$(x^I,t) \in \Rbb^3\times (-T_1(\ep),T_2(\ep)$. Thus
$\{w_\ep^i,\alpha_\ep\}$ determine a solution, via the formulas
\eqref{wdefA.2}, to the Euler equation which are equivalent to
$\nablab_i \Tc^{ij}=0$. So taking the divergence of \eqref{locA6}
while using $\nablab_i\Tc^{ij}=\nablab_i\Gc^{ij} = 0$ shows that
$\Hc^j$ satisfies an equation of the form \eqn{locA7}{
\gfb^{ik}\delb^j_{ik}\Hc^j + Q^{jp}_q(\gfb,\delb_k\gfb)\delb_p
\Hc^q = 0 } where the $Q^{jp}_{q}$ are analytic in $\gfb$ and
$\delb_k\gfb$. Clearly, this is a linear, $2^{\text{nd}}$ order
hyperbolic equation for $\Hc^j$. Since $\Hc^j|_{t=0} =
\del_t\Hc^j|_{t=0} = 0$, we must have $\Hc^j = 0$ for all $(x^I,t)
\in \Rbb^3\times (-T_1(\ep),T_2(\ep))$.
\smallskip

\noindent \textbf{(iv)} We know from proposition \ref{idatA} that
the map $(0,\ep_0]$ $\ni$ $\ep$ $\rightarrow$ $\Vo_\ep$ $\in$
$H^k_{\delta-1,\ep}$ is analytic which implies the estimate
$\norm{\Vo_\ep - \Vo_0}_{H^k_{\delta-1}} \leq C\ep$ for some fixed
constant $C>0$. So then \eqn{locA8}{ \norm{\Vo_\ep -
\Vo_0}_{H^k_{\delta-1,\ep}} \leq \norm{\Vo_\ep -
\Vo_0}_{H^k_{\delta-1}} \leq C\ep } by lemma \ref{ciF}. Since
$\{\ufb_\ep,w^i,\alpha_\ep\}$ solves the reduced Einstein
equations \eqref{Einred}, we have that  \eqn{locA9}{ \ep
\gfb_\ep^{44}\del_t \delb_4 \ufb_\ep^{Ij} + 8\ep^2\del_{L}\delb_4
\ufb_\ep^{Ij} + \gfb_\ep^{KL}\del^2_{KL}\ufb^{Ij}_{\ep} + \ep^2
f^{Ij}(\ep^2 \ufb_\ep,\delb_4\ufb_\ep,\del_L \ufb_\ep) = \ep^2
S^{Ij}(\ep^2\ufb_\ep,\alpha_\ep,w^i_\ep) } where the $f^{IJ}$ are
analytic and quadratic in $\del_4\ufb_\ep$ and $\del_k\ufb_\ep$
while $S^{IJ}$ are also analytic and linear in $\alpha_\ep$ and
$w^i_\ep$. Evaluating this equation at $t=0$, and using the
following facts from proposition \ref{idatA} \leqn{locA10}{
\ep^{-1}\norm{\ufbo_\ep^{Ij}}_{H^{k+1}_{\delta}} +
\norm{\ufbo^{44}_\ep}_{H^{k+1}_{\delta}}+
\norm{\del_t\ufbo^{ij}}_{H^{k}_{\delta-1}} +
\norm{\alphao_\ep}_{H^k_{\delta-1}} +
\norm{\wo^i}_{H^k_{\delta-1}} \leq C\, , } we find upon solving
for $\del_t\delb_4\ufb^{Ij}$ that \leqn{locA11}{
\norm{\del_t\delb_4\ufb^{Ij}(0)}_{H^{k-1}_{\delta-1}} \leq C \quad
\forall \; \ep \in (0,\ep_0] } by the calculus inequalities of
appendix \ref{winq}. But from part (iii), we get that $\delb_4
\ufb^{44}_\ep + \del_I \ufb^{I4}_{\ep} = 0$ and hence
differentiating this with respect to $t$ and evaluating at $t=0$
yields \leqn{locA12}{
\norm{\del_t\delb_4\ufb^{44}_\ep(0)}_{H^{k-1}_{\delta-1}}
=\norm{\del_I \del_t\ufbo^{I4}_{\ep}}_{H^{k-1}_{\delta-1}} \leq C
\quad \forall \; \ep \in (0,\ep_0] \, .
 }
From the estimates \eqref{locA10},  the fluid equations
\eqref{eul13} and similar arguments as above show that
\leqn{locA13}{ \norm{\del_t \alpha_\ep(0)}_{H^{k-1}_{\delta-1}}
\leq C + \norm{\del_t w^i_\ep(0)}_{H^{k-1}_{\delta-1}} \leq C
\quad \forall\; \ep \in (0,\ep_0]\, .} Estimates
\eqref{locA10}-\eqref{locA13} and lemma \ref{ciF} then imply that
$\norm{\del_t V_\ep(0)}_{H^{k-1}_{\delta-1,\ep}}$ $\leq$
$\norm{\del_t V_\ep(0)}_{H^{k-1}_{\delta-1}}\leq C $ for all $\ep$
$\in$ $(0,\ep_0]$.
\smallskip

\noindent \textbf{(v)} This is just a statement of the continuation
principle of theorem \ref{leB}.
\end{proof}

\sect{nlim}{The Newtonian limit}

Let $\{\Ve, 0<\ep\leq \ep_0\}$ be the sequence of solutions from
theorem \ref{locA} where we will always assume that \eqn{nlim0}{
 -1 < \delta < -1/2
\AND \text{$\text{supp}\, \alphao \subset B_R$ for some $R
> 0$.} } If we let $T_m(\ep)$ denote the maximal time of existence for
the solution $V_\ep$, then \leqn{nlim0e}{ \Ve \in
\bigcap_{\ell=0}^{s+1}C^\ell([0,T_m(\ep)),H^{k-\ell}_{\delta-1})
\subset
\bigcap_{\ell=0}^{s+1}C^\ell([0,T_m(\ep)),H^{k-\ell}_{\delta-1,\ep})\,
. }
 So $\alpha_\ep \in
\bigcap_{\ell=0}^{s+1}C^\ell([0,T_m(\ep)),H^{k-\ell}_{\delta-1})$
and hence proposition 3.6 of \cite{Heil} and lemma \ref{ciG} imply
that \eqn{nlim2}{ \rho_\ep=\rho(\alpha_\ep) \in
\bigcap_{\ell=0}^{s+1}C^\ell([0,T_m(\ep)),H^{k-\ell}_{\delta-2}).
} Using proposition 2.2 of \cite{Bart86}, we can solve the
equation \leqn{nlim2a}{ \Delta \Phi_\ep = \rho_\ep } to find
\eqn{nlim3}{ \Phi_\ep \in
\bigcap_{\ell=0}^{s+1}C^\ell([0,T_m(\ep)),H^{k+2-\ell}_{\delta}).
} To obtain the Newtonian limit, we use $\Phi_\ep$ to take care of
the singular term $\ep^{-1}g(\Ve)\Ve$ in \eqref{EFsym1} by
introducing the new variable \leqn{nlim4}{ \We := (
\uf^{ij}_{4,\ep} , u^{ij}_{J,\ep} , \delta\uf^{ij}_\ep ,
\alpha_\ep , w_\ep^i ) \qquad \qquad u^{ij}_{J,\ep} :=
\uf^{ij}_{J,\ep} - \delta^i_4\delta^j_4\del_J \Phi_\ep \, . }
Observe that \eqn{nlim5}{ \Ve = \We + d\Phi_\ep } where
\eqn{nlim6}{ d\Phi_\ep := ( 0 , \delta^{i}_4\delta^{j}_4
\del_J\Phi_\ep , 0 , 0 , 0)\, . } Noting that \leqn{nlim7}{
b^0(\ep\Ue,\ep\Ve) = b^0(\ep\Ue,\ep \We) \AND b^I(\ep,\Ue,\Ve) =
b^I(\ep,\Ue,\We), } $\We$ satisfies the equation \leqn{EFreg}{
b^0(\ep \Ue,\ep \We)\del_t \We = \frac{1}{\ep}c^I\del_I \We +
b^I(\ep,\Ue,\We)\del_I \We + f(\ep,\Ue,\We+d\Phi_\ep)\We + \He }
where \eqn{Hdefa}{ \He := h_\ep -b^0(\ep \Ue,\ep \We)\del_t
d\Phi_\ep + b^I(\ep,\Ue,\We)\del_I d\Phi_\ep +
f(\ep,\Ue,\We+d\Phi_\ep)d\Phi_\ep. }

By construction the initial data $\underset{o}{\Ve}$ is bounded in
$H^k_{\delta-1}$ as $\ep \searrow 0$. Therefore by lemma
\ref{ciF}, there exists a constant $K_1$ such that \leqn{Wb1}{
\text{$\norm{W_\ep|_{t=0}}_{H^k_{\delta-1,\ep}}\leq K_1$ for all
$\ep \in (0,\ep_0]$.}} Also by definition of $W_\ep$ and lemma
\ref{ciD}, \leqn{Wb2}{
\max\{\norm{\delta\uf_\ep}_{L^\infty},\norm{\alpha_\ep}_{L^\infty},\norm{w_\ep^i}_{L^\infty}\}
\leq \norm{\We}_{C^1_b} \leq
\Csob\norm{W_\ep}_{H^k_{\delta-1,\ep}} } where $\Csob$ is
the constant from lemma \ref{ciD} that is $\ep$ independent.
Shrinking $\ep_0$ if necessary, we can always assume that
\leqn{nlim0b}{ 2 \ep_0 \Csob  K_1 < K_0\, .} Define \leqn{taudef}{
\tau_\ep := \min\bigl\{\sup\bigl\{ \tau
> 0 |\; \sup_{0\leq t \leq \tau}
\norm{\We(t)}_{H^k_{\delta-1,\ep}} \leq 2K_1 \AND \sup_{0\leq t
\leq \tau}\norm{\Ve}_{H^k_{\delta-1,\ep}} < \infty \bigr\}\, , \;
1 \; \bigr\}.} From the continuation principle in theorem
\ref{locA}, it is clear that $\tau_\ep$ satisfies \eqn{taudef1}{
0<\tau_\ep \leq T_{m}(\ep).}

\subsect{es}{Energy estimates}

We will now use energy estimates on the $H^k_{\delta-1,\ep}$
spaces to show that $\tau_\ep$ is bounded below by a constant
independent of $\ep$. The strategy we use is that of
\cite{BK,KM82} adapted to the $H^k_{\delta,\ep}$ spaces. All of
the results below will be derived under the assumption that the
1-parameter family $\Ve$ of solutions has the additional regularity
\eqn{nlim0ee}{ \Ve \in
\bigcap_{\ell=0}^{s+1}C^\ell([0,\tau_\ep],H^{k+1-\ell}_{\delta-1})
\, . } It is then not difficult to use solution of this type to
approximate solutions of the regularity type \eqref{nlim0e} and
thereby show that all of the following results also hold for
solutions with the regularity \eqref{nlim0e}. Since these sort of
approximation arguments are standard, we will leave the details to
the interested reader.

The next lemma contains the basic energy estimate which is the key
to deriving estimates independent of $\ep$. We note that this type
of estimate has been derived previously for the standard Sobolev
spaces in \cite{BK,KM82}. It also makes clear why we need to
introduce the variables $\We$ and $\Phie$ to put the
Einstein-Euler equations into the form \eqref{EFreg}.
\begin{lem} \label{nlimA} \mnote{[nlimA]}
Suppose $\ep_0 \geq  0$, $a^0 \in C^1([0,\tau],W^{1,\infty})$,
$a^I\in C^0([0,\tau],W^{1,\infty})$, $g\in
C^0([0,\tau],L^2_{\lambda,\ep})$, and that $w\in
C^1([0,\tau],H^1_{\lambda,\ep})$ is a solution to the
linear equation \eqn{nlimA.1}{ a^0\del_t w = a^I\del_I w + g \, .
} Then there exists a constant $C>0$ independent of $\ep\in [0,\ep_0]$ such
that \eqn{nlimA.2}{ \frac{d\,}{dt} \ip{w}{a^0
w}_{L^2_{\lambda,\ep}} \leq C\bigl[\bigl(\norm{\emph{\Div}
a}_{L^\infty} + \ep \norm{\vec{a}}_{L^\infty}\bigr)
\norm{w}_{L^2_{\lambda,\ep}}^2 +
\norm{g}_{L^2_{\lambda,\ep}}\norm{w}_{L^2_{\lambda,\ep}}\bigr] }
where $\emph{\Div} a = \del_t a^0 + \del_I a^I$ and $\vec{a} =
(a^1,a^2,a^3)$.
\end{lem}
\begin{proof} Let $\bar{\sigma} = \sigma_\ep^{-2\lambda -3}$. Then $\norm{\bar{\sigma}^{-1}\del_j
\bar{\sigma}}_{L^\infty}
\leq \ep C$ for some constant $C>0$ that is independent of $\ep$. Using this,
the proof follows by
a standard integration by parts argument as in the proof of lemma \ref{engA}.
\end{proof}

To continue, we estimate, in terms of $K_1$, how much the support
of $\alpha_\ep$ can change as $\ep \searrow 0$.
\begin{lem} \label{nlimB} \mnote{[nlimB]}
$\;$
 \eqn{nlimB0}{
 \emph{ \text{supp}}\,\alpha_\ep(t) \subset B_{R+32 K_1} } for
all $(t,\ep)\in [0,\tau_\ep]\times (0,\ep_0]$.
\end{lem}
\begin{proof}
Letting $X^I$, $\bar{Y}$ and $\hat{Y}$ be as in section \ref{eul}
(see \eqref{alphaB}), we define \eqn{nlimB1}{ \Xe^I(t) := X^I(\ep
\we^4(t),\we^J(t)) } and \eqn{nlimB2}{ \Ye^I(t) :=
\bar{Y}(\ep\we^4(t),\ep\alphae(t))\bigl(\ep\del_t \we^4(t) +\del_I
w^I_\ep(t)\bigr) + \hat{Y}(\ep(\ufo_\ep+\delta\uf(t)),\ep
w^4(t),\ep \uf_k(t),\ep w^J(t),\ep\alpha(t)). } Using
\eqref{contA1}, \eqref{Wb2}, \eqref{nlim0b}, and \eqref{taudef},
we obtain the bound \leqn{nlimB3}{ \norm{\Xe^I(t)}_{L^\infty} \leq
32 K_1\, \quad \forall \;(t,\ep) \in [0,\tau_\ep]\times
(0,\ep_0].} From lemmas \eqref{ciD} and \eqref{Lip}, and
\eqref{nlim0e}, it follows that $\Xe^I \in
C^0([0,\tau_\ep],C^1_b)$ and $\Ye^I \in C^0([0,\tau_\ep],C^0_b)$.
Therefore the vector field $\Xe^I$ can be integrated to get a
$C^1$ flow $\psi^I_\ep(t,x)$ that is well defined for all $(t,x)
\in [0,\tau_\ep] \times \Rbb^3$. For each $x\in \Rbb^3$, define
$\alpha^x_\ep(t)$ $:=$ $\alpha_\ep(t,\psi_\ep(t,x))$. Then
$\del_t\psi^I_\ep(t,x) = \Xe^I(t,\psi_\ep(t,x))$ together with the
evolution equation \eqref{alphaA} implies that \eqn{nlimB14}{
\frac{d\;}{dt}\alpha^x_{\ep}(t) +
Y(t,\psi_\ep(t,x))\alpha^x_\ep(t) = 0 \, . } By assumption
$\text{supp}\,\alpha_0 \subset B_R$ and hence $\alpha^x_\ep(0) =
\alphao(x) = 0$ for $x\in E_R:= \Rbb^3\setminus B_R$. Therefore
\leqn{nlimB15}{ \alpha_\ep(t,\psi_\ep(t,x)) = 0\quad \text{all
$x\in E_R$}} by the uniqueness of solutions to ODEs. But
\eqn{nlimB16}{ |\psi_\ep(t,x)-x| \leq
\int_{0}^{\tau_\ep}|\del_t\psi_\ep(t,x)| =
\int_{0}^{\tau_\ep}|\Xe(t,\psi_\ep(t,x))| \leq 32 K_1\tau_\ep \leq
32 K_1 } by \eqref{nlimB3} and $0<\tau_\ep \leq 1$. From this,
\eqref{nlimB15}, and the fact that for each $t$ the map $\Rbb^3
\ni x\mapsto \psi_\ep(t,x)\in \Rbb^3$ defines a $C^1$
diffeomorphism, it follows that $\text{supp}\, \alpha_\ep(t)
\subset B_{R+32 K_1}$ for all $(t,\ep) \in [0,\tau_\ep]\times
(0,\ep_0]$.
\end{proof}

Next,  we estimate $\norm{\Phie}_{H^{k+2}_{\delta}}$ in
terms of $\norm{\We}_{H^k_{\delta-1,\ep}}$.

\begin{lem} \label{nlimC} \mnote{[nlimC]}
Let $\bar{R} = R+32 K_1$ and \eqn{nlimC0b}{ C_1 =
(1+\bar{R})^{-(\delta-2)-3/2}\sqrt{1+(1+\bar{R})^{2k}}. } Then
there exists a constant $C>0$ such that
 \eqn{nlimC0}{
\norm{\Phie(t)}_{H^{k+2}_\delta} \leq C C_1
\norm{\We(t)}_{H^k_{\delta-1,\ep}}^{n} } for all $(t,\ep) \in
[0,\tau_\ep]\times (0,\ep_0]$.
\end{lem}
\begin{proof}
By lemma \ref{nlimB}, the $\text{supp}\,\alphae(t) \subset B_{R+32
K_1}$ for  all $(t,\ep)\in [0,\tau_\ep]\times (0,\ep_0]$. Letting
$\bar{R} = R+32 K_1$, it follows directly from the definition of
the weighted norms that
 \eqn{nlimC0a}{
\norm{u}_{L^2} \leq \norm{u}_{L^2_{\eta,\ep}} \leq
(1+\bar{R})^{-\eta-3/2}\norm{u}_{L^2} }for all functions $u$ whose
support is contained in $B_{\bar{R}}$ and for any $\ep \in (0,1]$
and $-\eta -3/2 \geq 0$. Therefore \eqn{nlimC1}{
\norm{\rho_\ep}_{H^{k}_{\delta-2}} \leq C C_1 \norm{
\rho_\ep}_{H^{k}_{\delta-1,\ep}} } where $C>0$ is a constant
independent of $\ep$ and \eqn{nlimC1a}{ C_1 =
(1+\bar{R})^{-(\delta-2)-3/2}\sqrt{1+(1+\bar{R})^{2k}}. } Since
$\Delta : H^{k+2}_{\delta} \rightarrow H^{k}_{\delta-2}$ is an
isomorphism and $\Delta \Phie = \rho_\ep$, we have
$\norm{\Phie}_{H^{k+2}_{\delta}} \leq
C\norm{\rho_\ep}_{H^{k}_{\delta-2}}$ and hence, by lemma \ref{ciG}
(see also \eqref{dendef} and \eqref{nlim4}) and the above estimate
that \eqn{nlimC2}{ \norm{\Phie}_{H^{k+2}_\delta} \leq C
C_1\norm{\rho_\ep}_{H^{k}_{\delta-2,\ep}} \leq C
C_1\norm{\alpha_\ep}_{H^{k}_{\delta-1,\ep}}^{n}  \leq C
C_1\norm{\We}_{H^k_{\delta-1,\ep}}^{n} \, . }
\end{proof}

We note that for the remainder of this section, all of the
constants appearing in the estimates may depend on the fixed
constant $K_1$. We will often use $C$ to denote constants that
depend on $K_1$ and that may change from line to line.

Let $\We^\alpha = D^\alpha\We$ $(|\alpha|\geq 0)$, $\bep^0 =
b^0(\ep\Ue,\ep\We)$, $\bep^I = b^I(\ep,\Ue,\Ve)$ and $\fep =
f(\ep,\Ue,\We+d\Phie)\We$. The evolution equation \eqref{EFreg}
implies that \leqn{nlim8}{ \del_t \We =
(\bep^0)^{-1}\left(\frac{1}{\ep} c^I+\bep^I\right)\del_I\We +
(\bep^0)^{-1}\fep + (\bep^0)^{-1}\He\, .} Differentiating this
equation yields \leqn{nlim9}{ \bep^0\partial_t\We^{\alpha} =
\frac{1}{\ep}c^I\del_I\We^\alpha +\bep^I\del_I \We^\alpha +
q^\alpha \quad |\alpha|\geq 0 } where  \leqn{nlim10}{q^\alpha =
\bep^0[\Der^\alpha,(\bep^0)^{-1}(\ep^{-1}c^I+\bep^I)]\del_I \We +
\bep^0\Der^\alpha\bigl((\bep^0)^{-1} \fep\bigr) + \bep^0
\Der^\alpha\bigl( (\bep^0)^{-1} \He \bigr) \, .}

From lemma \ref{ciF}, we know, since $-1<\delta<-1/2$, that
$\norm{\ep\ufbo_\ep}_{H^{k+1}_{\delta,\ep}} \leq
\ep^{|\delta+1/2|}\norm{\ufbo_\ep}_{H^{k+1}_\delta}$. Since
$\norm{\ufbo_\ep}_{H^{k+1}_{\delta}}$ is uniformly bounded in
$\ep$, we get, by lemmas \ref{ciD} and \ref{ciF}, that
\leqn{nlim11}{ \norm{\Ue}_{C_b^{1,\infty}} \leq \Csob
\norm{\Ue}_{H^{k+1}_{\delta,\ep}} \leq C\ep^{|\delta+1/2|}} for
some constant $C>0$ independent of $\ep$. So \leqn{nlim12}{
\norm{b^i_\ep(t)}_{W^{1,\infty}} \leq C\quad \forall\: (t,\ep)\in
[0,\tau_\ep]\times (0,\ep_0]} by \eqref{nlim7}, \eqref{Wb2},
\eqref{taudef} and \eqref{nlim11} . Also, note that \eqn{nlim8a}{
\norm{d\Phie}_{L^\infty} + \norm{Dd\Phie}_{L^\infty} \leq
C\norm{\Phie}_{H^{k+2}_{\delta}} \leq C  \AND \norm{\del_t
d\Phie}_{L^\infty} \leq C\norm{\Phie}_{H^{k+1}_\delta} } by
\eqref{diff}, \eqref{inclA} and lemmas \ref{nlimC} and \ref{ciD}.
The evolution equation \eqref{nlim8} then implies that
\leqn{nlim13b}{ \norm{\del_t \bep^0}_{L^\infty} =
\norm{\ep Db^0(\ep\Ue,\ep \We)\cdot \del_t\We }_{L^\infty} \leq
C(1+\norm{\del_t d\Phie}_{H^{k+1}_{\delta}} )\, . } Together
\eqref{nlim12} and \eqref{nlim13b} establish the
existence of a  constant $C >0$ such that \leqn{nlim13}{ \norm{\Div
b_\ep(t)}_{L^\infty} \leq C(1 + \norm{\del_t
\Phie(t)}_{H^{k+1}_{\delta}} ) \quad \forall\: (t,\ep) \in
[0,\tau_\ep]\times (0,\ep_0].  }

Differentiating $(\bep^0)^{-1}$ yields \eqn{nlim14}{
\del_J(\bep^0)^{-1} = -\ep (\bep^0)^{-1}\bigl(Db^0(\ep \Ue,\ep
\We)\cdot (\del_J \Ue,\del_J\We)\bigr)(\bep^0)^{-1}\, .} This
along with \eqref{nlim11}, \eqref{nlim12}, \eqref{diff},
\eqref{inclA}, and lemmas \ref{ciD} and \ref{ciE} can be used to
control the singular term in \eqref{nlim10} and results in the
following estimate (see also appendix \ref{energy})\leqn{nlim15}{
\norm{q^\alpha(t)}_{L^2_{\delta-1-|\alpha|,\ep}} \leq
P_\alpha(\norm{\We(t)}_{H^k_{\delta-1,\ep}},\norm{\Phie(t)}_{H^{k+2}_{\delta}},
\norm{\del_t \Phie(t)}_{H^{k+1}_{\delta}})\quad \forall\: t\in
[0,\tau_\ep] } where $P_{\alpha}(y_1,y_2,y_3)$ is a polynomial
that is  independent of $\ep$ and satisfies $P(0)=0$. Note that in
deriving this results, we have used the estimate\leqn{nlim15a}{
\norm{d\Phie}_{H^{k}_{\delta-1,\ep} }+\norm{D
d\Phie}_{H^{k}_{\delta-2,\ep} } \leq
C\norm{\Phie}_{H^{k+2}_{\delta}} \AND \norm{\del_t
d\Phie}_{H^k_{\delta-1,\ep}} \leq C
\norm{\del_t\Phie}_{H^{k+1}_{\delta}} } for some $C$ independent
of $\ep$ which follows from \eqref{diff}, \eqref{inclA}, and lemma
\ref{ciF}.

Define \eqn{nlim16}{ \nnorm{\We}^2_{k,\delta-1,\ep} :=
\sum_{|\alpha|\leq k} \ip{\del^\alpha \We}{\bep^0 \del^\alpha \We
}_{L^2_{\delta-|\alpha|,\ep}}\, . } Then \leqn{nlim17}{
\frac{1}{4}\norm{\We(t)}_{H^k_{\delta-1,\ep}} \leq
\nnorm{\We(t)}_{k,\delta-1,\ep} \leq 4
\norm{\We(t)}_{H^k_{\delta-1,\ep}} \quad \forall \: t\in
[0,\tau_\ep] \,  } by \eqref{contA2} and \eqref{contA3}. Lemma
\ref{nlimA} combined with the estimates \eqref{nlim12},
\eqref{nlim13}, and \eqref{nlim15} implies that \eqn{nlim18}{
\frac{d\;}{dt}\nnorm{\We}^2_{k,\delta-1,\ep} \leq
P(\nnorm{\We}_{k,\delta-1,\ep}, \norm{\Phie}_{H^{k+2}_{\delta}},
\norm{\del_t\Phie}_{H^{k+1}_{\delta}}
 ) \nnorm{\We}_{k,\delta-1,\ep}  }
or equivalently \leqn{nlim19}{
\frac{d\;}{dt}\nnorm{\We(t)}_{k,\delta-1,\ep} \leq
P(\nnorm{\We(t)}_{k,\delta-1,\ep},\norm{\Phie(t)}_{H^{k+2}_{\delta}},
\norm{\del_t\Phie(t)}_{H^{k+1}_{\delta}} ) \quad \forall \: t\in
[0,\tau_\ep] }
 for a $\ep$ independent  polynomial $P(y_1,y_2,y_3)$ satisfying $P(0)=0$.
By lemma \ref{nlimC}, $\norm{\Phie}_{H^{k+2}_{\delta}}$ can be
bounded by a polynomial of $\norm{\We}_{H^k_{\delta-1,\ep}}$ that
is independent of $\ep$ and vanishes for
$\norm{\We}_{H^k_{\delta-1,\ep}}=0$. The differential inequality
\ref{nlim19} shows that if we can do the same for
$\norm{\del_t\Phie}_{H^{k+1}_{\delta}}$ then we get an estimate
for $\nnorm{\We(t)}_{k,\delta-1,\ep}$ independent of $\ep$.
\begin{lem} \label{nlimCa} \mnote{[nlimCa]}
There exists a polynomial $P(y)$ with coefficients independent of
$\ep$  such that $P(0) = 0$ and \eqn{nlimCa1}{
\norm{\del_t\Phie(t)}_{H^{k+1}_\delta} \leq
P(\norm{\We(t)}_{H^k_{\delta-1,\ep}}) } for all $(t,\ep)\in
[0,\tau_\ep]\times (0,\ep_0]$.
\end{lem}
\begin{proof}
By \eqref{eul13},  $\wv_\ep := (\alpha_\ep, w^i_\ep)^T$ satisfies
an equation of the form  \eqn{nlimCa4}{ a^4(\ep \Ue,\ep \We)\del_t
\wv_\ep = a^I(\ep \Ue, \ep \We)\del_t \wv_\ep + b_1(\ep \Ue,\ep
\We)\We + b_2(\ep \Ue,\ep \We)d\Phie } and so \eqn{nlimC5}{ \del_t
\wv_\ep = (a^4)^{-1} a^I \del_I\wv_\ep + (a^4)^{-1}b_{1}\We +
(a^4)^{-1}b_{2}d\Phie \, . } Thus \eqn{nlimCa6}{
\norm{\del_t\wv_\ep}_{H^{k-1}_{\delta-1,\ep}} \leq \norm{
(a^4)^{-1} a^I}_{H^{k-1}_{1,\ep}}
\norm{D\We}_{H^{k-1}_{\delta-2,\ep}}
+\norm{(a^4)^{-1}b_{1}}_{H^{k-1}_{0,\ep}}\norm{\We}_{H^{k-1}_{\delta-1,\ep}}
+
\norm{(a^4)^{-1}b_{2}}_{H^{k-1}_{0,\ep}}\norm{d\Phie}_{H^{k-1}_{\delta-1,\ep}}
} by lemma \ref{ciG}. Also by \eqref{nlim11}, \eqref{diff},
\eqref{inclA}, and lemmas \ref{ciD} and \ref{ciE}, we have that
\gath{nlimCa7}{
 \norm{ (a^4)^{-1} a^I}_{H^{k-1}_{1,\ep}} \leq P(\norm{\We}_{H^k_{\delta-1,\ep}}) \, ,
\quad \norm{D\We}_{H^{k-1}_{\delta-2,\ep}} \leq \norm{\We}_{H^k_{\delta-1,\ep}} \, , \\
\norm{(a^4)^{-1}b_{1}}_{H^{k-1}_{0,\ep}}  \leq
P(\norm{\We}_{H^k_{\delta-1,\ep}}) \, , \AND
\norm{(a^4)^{-1}b_{2}}_{H^{k-1}_{0,\ep}} \leq
P(\norm{\We}_{H^k_{\delta-1,\ep}}) } for some polynomial $P(y)$
that is independent of $\ep$. The above two inequalities along
with \eqref{nlim15a} and lemma \ref{nlimC} show that
\eqn{nlimCa8}{ \norm{\del_t \alpha_\ep}_{H^{k-1}_{\delta-1,\ep}}
\leq \norm{\del_t\wv_\ep}_{H^{k-1}_{\delta-1,\ep}} \leq
P(\norm{\We}_{H^k_{\delta-1,\ep}}) } for a polynomial $P(y)$
independent of $\ep$ and satisfying $P(0)=0$. Using lemma
\ref{ciG}, the above estimate implies that \eqn{nlimCa5}{
\norm{\del_t\rho_\ep}_{H^{k-1}_{\delta-2,\ep}} \leq
P(\norm{\We}_{H^{k}_{\delta-1,\ep}} ) } where as above $P(y)$ is a
polynomial that is independent of $\ep$. Since $\Delta \del_t\Phie
= \del_t\rho_\ep$, the same arguments used in the proof of lemma
\ref{nlimC} can be used to conclude \eqn{nlimCa6a}{
\norm{\del_t\Phie}_{H^{k+1}_\delta} \leq
C\norm{\del_t\rho_\ep}_{H^{k-1}_{\delta-1,\ep}} \leq
P(\norm{\We}_{H^{k}_{\delta-1,\ep}} ) \, . }
\end{proof}
Lemmas \ref{nlimC} and \ref{nlimCa} combined with the estimate
\eqref{nlim19} yield \leqn{nlim20}{
\frac{d\;}{dt}\nnorm{\We(t)}_{k,\delta-1,\ep} \leq
P(\nnorm{\We(t)}_{k,\delta-1,\ep})\nnorm{\We(t)}_{k,\delta-1,\ep}
\quad \forall \: t\in [0,\tau_\ep] } for a polynomial $P(y)$ that
is independent of $\ep$ and whose coefficients depend only on
$K_1$. By Gronwall's inequality there exists a time $T^*\in
(0,1)$, independent of $\ep$, such that if $y(t)\geq 0$ is  $C^1$
and satisfies $dy/dt \leq P(y)y$, then $y(t) \leq e^{K_3 t}y(0)$
where $K_3$ is a constant that depends on $K_1$. Therefore
\leqn{nlim20a}{ \nnorm{\We(t)}_{k,\delta-1,\ep} \leq e^{K_3
t}\nnorm{\We(0)}_{k,\delta-1,\ep} \quad \text{for all $(t,\ep) \in
[0,\min\{T^*,\tau_\ep\}]\times (0,\ep_0]$.} } Shrinking $T^*$ if
necessary, we conclude that
\leqn{nlim21}{\nnorm{\We(t)}_{k,\delta-1,\ep} \leq \frac{3}{2} K_1
\quad \text{ for all $(t,\ep)\in [0,\min{T^*,\tau_\ep}]\times
(0,\ep_0]$.} } Note also that
\leqn{nlim24}{\norm{\Ve(t)}_{{H^k}_{\delta-1,\ep}} \leq C \quad
\text{for all $(t,\ep) \in [0,\min\{T^*,\tau_\ep\}]\times
(0,\ep_0]$} } by \ref{nlim15a}, \ref{nlim17} and lemma
\ref{nlimC}. Therefore by the definition of $\tau_\ep$, we must
have $0<T^* < \tau_\ep $ for all $0 < \ep \leq \ep_0$.

Differentiating \eqref{nlim8} with respect to $t$, shows that
$\Wde := \del_t\We$ and $d\Phide := \del_t d\Phie$ satisfy the
equation \alin{nlim26}{ b^0(\ep\Ue,\ep \We)\del_t \Wde =
\frac{1}{\ep}c^I\del_I \Wde + &b^I(\ep,\Ue,\We)\del_I \Wde +
\bar{f}_1(\ep,\Ue,\We,D\We, d\Phie,D d\Phie,d\Phide)\Wde \\ & +
\bar{f}_2(\ep,\Ue,\We,d\Phie,D d\Phi,d\Phide,D d\Phide,\del_t
d\Phide) + \del_t h \,  } for analytic functions $f_1$, $f_2$ with
$f_2$ linear in the last 3 variables. This equation has the same
structure \eqref{EFreg} and is not difficult to show that the
arguments used to derive \eqref{nlim20a} can also be used to obtain
the estimate \leqn{nlim27}{
\norm{\Wde(t)}_{H^{k-1}_{\delta-1,\ep}}
 \leq C \quad \forall \; (\ep,t) \in (0,\ep_0]\times [0,T^*] }
under the assumption that
$\norm{\Wde(0)}_{H^{k-1}_{\delta-1,\ep}}$ is bounded as $\ep
\searrow 0$. But this is clear from proposition \ref{locA} and
lemma \ref{nlimC} and so the estimate holds. We have proved the
following proposition.
\begin{prop} \label{nlimD} \mnote{[nlimD]}
For $\ep_0>0$ small enough, there exists a $T^*>0$ independent of
$\ep \in (0,\ep_0]$ such that the one parameter family of
solutions $\Ve$ exist, for all $\ep \in (0,\ep_0]$, on a common
time interval $[0,T^*]$. Moreover,  there exists constants
$C>0$, $\bar{R}>0$ such that  \gath{nlimD1}{
\max\{\norm{\delta\uf_\ep}_{L^\infty},\norm{\alpha_\ep}_{L^\infty},\norm{w_\ep^i}_{L^\infty}\}
\leq \frac{K_0}{\ep_0} \quad \norm{\Ve(t)}_{H^k_{\delta-1,\ep}}
\leq C\, , \quad
\norm{\del_t\Ve(t)}_{H^{k-1}_{\delta-1,\ep}} \leq C \, , \\
\norm{\Phie(t)}_{H^{k+2}_\delta} \leq C \, , \quad
\norm{\del_t\Phie(t)}_{H^{k+1}_\delta}\leq C\, , } and $
\emph{\text{supp}}\, \alpha_\ep(t) \subset B_{\bar{R}}$ for all
$(\ep,t) \in (0,\ep_0]\times [0,T^*]$.
\end{prop}

\subsect{cog}{Properties of the limit equations}

To fully understand the limit equations of section \ref{conv}, we first
need to consider the following system \lalign{newtA}{
\del_t\alphah &= -\wh^I\del_I
\alphah -\frac{\alphah}{2n}\del_I\wh^I \label{newtA.1} \\
\del_t \wh^J &= -\frac{\alphah}{2n}\del^J\alphah - \wh^I\del_I
\wh^J -\del^J\Phih \label{newtA.2} \\
\Delta \Phih &= \rhoh \label{newtA.3} } with initial data
\leqn{newtA.4}{ \alphah(0) = \alphao \AND \wh^I(0) = \wo^I} where
$\alphao$ and $\wo^I$ are as defined in proposition \ref{locA}.
This system is precisely the Poisson-Euler equation written using
the Makino variable $\rhoh =
\frac{1}{(4Kn(n+1))^{-n}}\alphah^{2n}$. Indeed, a straightforward
calculation shows that $(\rhoh,\wh^I)$ satisfy the Poisson-Euler
equations of Newtonian gravity \lalign{newtB}{
\del_t \rhoh + \del_I(\rhoh\wh^I) & = 0 \, , \label{newtB.1}\\
\rhoh(\del_t \wh^J + \wh^I\del_I \wh^J) & =
-(\rhoh\del^J\Phih + \del^J\ph) \, , \label{newtB.2} \\
\Delta \Phih &=   \rhoh \, , \label{newtB.3}
 }
where $\ph = K\rhoh^{(n+1)/n}$.

\begin{prop} \label{cogA} \mnote{[cogA]}
There exists a $T>0$ and a solution \gath{cogA1}{ \alphah, \wh^I
\in C^0([0,T],H^k_{\delta-1})\cap
C^{1}([0,T],H^{k-1}_{\delta-1})\, , \\
\Phih \in C^0([0,T],H^{k+2}_{\delta})\cap
C^{1}([0,T],H^{k+1}_{\delta})\, , \quad \del_t\Phih \in
C^{0}([0,T],H^{k+1}_{\delta-1}) } to the initial value problem
\eqref{newtA.1}-\eqref{newtA.4} where $\alphah(t)$ has compact
support for all $t \in [0,T]$. Moreover \begin{itemize} \item[(i)]
this solution is unique in the class \eqn{cogA2}{ \alphat, \wt \in
C^0([0,T],H^k)\cap C^{1}(\Rbb^n\times [0,T]) \quad \Phit \in
C^0([0,T],H^{k+2}_{\delta})\cap C^{1}(\Rbb^n\times [0,T]) } where
$\alphat(t)$ has compact support for all $t\in [0,T]$, and
\item[(ii)] the solution also satisfies \gath{cogA2a}{\alphah,
\wh^I \in \cap_{\ell=0}^{s+1}
C^{\ell}([0,T],H^{k-\ell}_{\delta-1}) \, , \\
\Phih \in \cap_{\ell=0}^{s+1}
C^{\ell}([0,T],H^{k+2-\ell}_{\delta}) \, , \quad \del_t \Phih \in
\cap_{\ell=0}^{s}C^{\ell}([0,T],H^{k+1-\ell}_{\delta-1})\, . }
\end{itemize}

\end{prop}
\begin{proof}
Writing the system \eqref{newtA.1}-\eqref{newtA.3} as
\eqn{cogA3}{ \del_t
\begin{pmatrix}\alphah \\ \wh^J \end{pmatrix}
 = \begin{pmatrix} -\wh^I  & -\frac{\alphah}{2n}\delta^{I}_J \\
- \frac{\alphah}{2n}\delta^{IJ} & - \wh^I \end{pmatrix}
\del_I\begin{pmatrix} \alphah  \\ \wh^J \end{pmatrix}
 -\frac{1}{(4Kn(n+1))^n}\begin{pmatrix} 0 \\
 \del^J(\Delta^{-1}\alpha^{2n}) \end{pmatrix}\, , } we
see that this system is symmetric hyperbolic  with a non-local
source term. Since $\Delta : H^{k+2}_\delta \rightarrow
H^{k}_{\delta-2}$ is an isomorphism, it is not difficult to adapt
the approximation scheme and energy estimates of appendices
\ref{galerk} and \ref{energy} to this system. Then as in appendix
\ref{le},  this is enough to produce an existence theorem.
Consequently, there exists a $T>0$ and a solution \leqn{cogA6}{
\alphah, \wh^I \in C^0([0,T],H^k_{\delta-1})\cap
C^1([0,T],H^{k-1}_{\delta-1}). } Therefore \leqn{cogA7}{\rhoh \in
C^0([0,T],H^k_{\delta-2})\cap C^1([0,T],H^{k-1}_{\delta-2})} and
hence $\Phit$ $=$ $\Delta^{-1}\rhoh$ $\in$
$C^0([0,T],H^{k+2}_{\delta})$ $\cap$
$C^1([0,T],H^{k+1}_{\delta})$.

Differentiating \eqref{newtB.3} with respect to $t$ and using
\eqref{newtB.1} yields \leqn{cogA4}{ \Delta \del_t\Phih =-
\del_I(\rhoh\wh^I)\, . } But, \eqref{cogA6} implies that
$\rhoh\wh^I \in C^{0}([0,T],H^k_{\delta-2})$ and hence
$\Delta^{-1}(\rhoh\wh^I)$ $\in$ $C^{0}([0,T],H^{k+2}_\delta)$.
Taking the divergence then gives $\del_I(\Delta^{-1}\rhoh\wh^I )$
$\in$ $C^{0}([0,T],H^{k+1}_{\delta-1})$. However, \eqref{cogA4}
implies that $\del_t\Phih$ $=$ $- \Delta^{-1} \del_I(\rhoh \wh^I)$
$=$ $- \del_I( \Delta^{-1} ( \rhoh \wh^I))$ and so $\del_t \Phih $
$\in$ $C^{0}([0,T],H^{k+1}_{\delta-1})$.

The statement about compact support follows from the symmetric
hyperbolic equation satisfied by $\alphah$ and the property of
finite propagation speed. Uniqueness follows from a slight
modification of standard arguments, see \cite{TayIII} proposition
1.3, section 16.1.
\end{proof}


\subsect{conv}{Convergence as $\ep\searrow 0$}

In this section, we identify the limit of the relativistic
solutions as $\ep \searrow 0$. To accomplish this, we adapt the
arguments of \cite{Scho86},  section III. Define \alin{conv1}{ \Vt
& := (\uft_4^{ij}, \uft^{ij}_{J}, \delta \uft^{ij}, \alphat,
\wt^i)^T
\, , \\
\at^I &:= \begin{pmatrix}
-\wt^I & \frac{\alphat}{2n}\delta^I_j \\
-\frac{\alphat}{2n}\delta^I_i & -\delta_{ij}w^I
\end{pmatrix} \, , \\
\bt^I & :=
\begin{pmatrix} A^I(\delta \uft^{ij}) & 0 \\
0 & \at^I
\end{pmatrix}\, , \\
\Sct^{ij} &:= \rho \begin{pmatrix} 0 & \wt^I \\
\wt^J & 4\eta_{ij}\delta\uft^{ij}
\end{pmatrix} \, ,  \\
\bt &:= \begin{pmatrix} 0\\
-\eta^{im}\bigl(2\eta_{4\ell}\eta_{4p}+ \eta_{\ell
p}\bigr)\uft^{\ell p}_{m} - 2\bigl( \eta_{\ell
p}\delta^i_4\ufb^{\ell p}_4 -2 \eta_{\ell 4}\ufb^{i\ell}_4 \bigr)
\end{pmatrix}\, , \\
\ft(\Vt)\Vt &:= ( - \Sct^{ij}, 0, \uft^{ij}_4, \bt)^T
\, , \\
\intertext{and} \tilde{h} &:= (
\eta^{ij}\del_I\beta^I -
2\del_I\beta^{(i}\eta^{j)I}, 0,\ldots,0)^T \, .
}

\begin{thm} \label{convA} \mnote{[convA]}
For any $r > 0$, $\Phie$ and $\Ve$ converge in
$C^0([0,T^*],H^{k+1-r}_{\emph{\text{loc}}})$ and
$C^0([0,T^*],H^{k-r}_{\emph{\text{loc}}})$ as $\ep \searrow 0$ to
$\Phit \in C^1(\Rbb^3\times [0,T^*])\cap
C^0([0,T^*],H^{k+2}_\delta)$ and the unique solution $\Vt \in
C^1(\Rbb^3\times [0,T^*])\cap C^0([0,T^*],H^k)$ of the system
\gath{convA1}{ \Pbb\bigl( \del_t\Vt -\bt^I\del_I\Vt - \ft(\Vt)\Vt
- \tilde{h} \bigr) =
0 \, ,\\
c^I\del_I(\Vt-d\Phit) =0\, , \\
\Vt(0) = \underset{o}{V_0}(0)\, ,\\
\Delta \Phit = \rhot\, ,} where $\Pbb$ is the projection onto the
$L^2$ orthogonal complement of $\{\, c^I\del_I W =0 \, | \: W\in
H^1\}$. Moreover,
\begin{itemize} \item[(i)] there exists a $\bar{R}>0$ such that
$\emph{\text{supp}}\,\alphat(t) \subset B_{\bar{R}}$ for all $t\in
[0,T^*]$,
\item[(ii)] there exists a $\omega \in C^0([0,T^*],H^k_{\emph{\text{loc}}})$ such
that $\del_I\omega \in C^0([0,T^*],H^{k-1})$ and
\leqn{convA2}{ \del_t\Vt -\bt^I\del_I\Vt - \ft(\Vt)\Vt -
\tilde{h} - c^I\del_I \omega = 0\, , }
\item[(iii)] and for $\delta_1\geq -1/2$, there exists
a $\ufbt \in C^{0}([0,T],L^6_{\delta_1})$ such that
\eqn{convA2a}{
\uft^{ij}_J = \del_J \ufbt^{ij} \, .
}
\end{itemize}
\end{thm}
\begin{proof}
By assumption $-1<\delta < -1/2$, and so it follows directly from
the definition of the weighted norms that for every $\ell \geq 0$,
\leqn{convA3}{\norm{u}_{H^\ell}\leq
\norm{u}_{H^\ell_{\delta-1,\ep}} \quad \text{for all $u\in
H^\ell_{\delta-1,\ep}$}. } So by proposition \ref{nlimD},
\eqn{Vbound}{\Ve
\in C^0([0,T^*],H^k)\cap C^{1}([0,T^*],H^{k-1})
\subset C^0([0,T^*],H^k_{\delta-1,\ep})\cap
C^{1}([0,T^*],H^{k-1}_{\delta-1,\ep})
} and $\Phie \in
C^0([0,T^*],H^{k+2}_\delta)\cap C^{1}([0,T^*],H^{k+1}_{\delta})$ are
uniformly bounded for $\ep \in (0,\ep_0]$. Therefore by the
Banach-Alaoglu theorem there exists subsequences of $\Phie$ and
$\Ve$, which we still denote by $\Phie$ and $\Ve$, and $\Phit \in
L^{1,\infty}([0,T^*],H^{k+1}_\delta )\cap
\text{Lip}([0,T^*],H^k_\delta)$, $\Vt \in
L^{1,\infty}([0,T^*],H^k )\cap \text{Lip}([0,T^*],H^{k-1})$ such
that $\Phie$ and $\Ve$ converge weakly to $\Phit$ and $\Vt$,
respectively, as $\ep \searrow 0$.

By proposition \ref{nlimD},  the support of $\alpha_\ep$ is
uniformly bounded in $\ep$ and hence the support of the weak limit
$\alphat$ must also be bounded. From proposition \ref{locA}, we
have that $\uf^{ij}_{J,\ep} = \del_J \ufb^{ij}_\ep$. So by lemmas
\ref{ciD} and  \ref{ciF}, and \eqref{nlim11}, we find that for
$\delta_1 \geq -1/2\geq \delta$ \eqn{convA4}{
\norm{\ufb^{ij}_\ep}_{L^6_{\delta_1}}\leq
C\norm{\ufb^{ij}_\ep}_{L^6_{\delta}}  \leq
C\norm{\ufb^{ij}_\ep}_{L^6_{\delta,\ep}} \leq
C\bigl(\norm{\uf^{ij}_{J,\ep}}_{L^2_{\delta-1,\ep}} +
\norm{\uf^{ij}_\ep}_{L^2_{\delta,\ep}}\bigr) \leq C(1 +
\norm{\Ve}_{H^k_{\delta-1,\ep}}) } for a constant $C$ independent
of $\ep$. It follows that $\ufb_\ep^I$ converges weakly to a
$\ufbt^{ij} \in L^{1,\infty}([0,T^*],L^6_{\delta_1})$ for which
$\partial_J \ufbt^{ij} = \uft_J$.

Now, $\Ve$ satisfies \leqn{convA5}{ b^0(\ep \Ue,\ep \Ve)\del_t \Ve
- \frac{1}{\ep}c^I\del_I (\Ve - d\Phie) + b^I(\ep,\Ue,\Ve)\del_I
\Ve - f(\ep,\Ue,\Ve)\Ve  - h(\ep \Ue) =0 \, ,} and hence it
follows from the boundedness of $\Phie$ and $\Ve$ that
\eqn{convA5a}{\norm{c^I\del_I (\Ve - d\Phie)}_{H^{k-1}} \leq
\norm{c^I\del_I(\Ve-d\Phie)}_{H^{k-1}_{\delta-1,\ep}}\leq C\ep \,
.} Letting $\ep \searrow 0$ yields \eqn{convA5aa}{ c^I\del_I (\Vt -
d\Phit) = 0\, . } Next, applying the projection $\Pbb$ (note that
$\Ve-d\Phie \in H^1$) to \eqref{convA5} gives \eqn{convA6}{
\Pbb\bigl( b^0(\ep \Ue,\ep \Ve)\del_t\Ve  - b^I(\ep,\Ue,\Ve)\del_I
\Ve - f(\ep,\Ue,\Ve)\Ve - h(\ep \Ue) \bigr ) =0 } or equivalently
\eqn{convA7}{ \Pbb b^0_\ep \Pbb \del_t \Ve + \Pbb b^0_\ep (\id
-\Pbb)\del_t\Ve  -\Pbb\bigl(b^I_\ep\del_I \Ve - f_\ep - h_\ep\bigr
) = 0 } where we set $b^0_\ep =  b^0(\ep \Ue,\ep \Ve)$, $b^I_\ep =
b^I(\ep,\Ue,\Ve)\del_I \Ve$, $f_\ep = f(\ep,\Ue,\Ve)\Ve$, and
$h_\ep =  h(\ep \Ue)$. Suppose $\psi \in C^\infty_0$ and let
$\ip{u}{v} = \int_{\Rbb^3} uv \,d^3 x$ be the standard $L^2$ norm.
Then \leqn{convA8}{ \ip{\psi}{\Pbb b^0_\ep (\id-\Pbb)\del_t \Ve}
 = \ip{(\id-\Pbb) b_\ep^0\Pbb \psi}{\del_t \Ve} \,  }
 as $\Pbb$ is a self-adjoint projection operator. Since the imbedding
 $H^k(B_R)\rightarrow H^{k-r}(B_R)$ $(r>0)$ is compact for any
 ball $B_R$, $V_\ep$ and $\Phi_\ep$ converge in $C^{0}([0,T^*],H^{k-r}_{\text{loc}})$
 and $C^{0}([0,T^*],H^{k+2-r}_{\text{loc}})$
 to $\Vt$ and $\Phit$, respectively, as $\ep \searrow 0$. Using
 this strong convergence and \eqref{nlim11}, we find that $
(\id-\Pbb) b_\ep^0\Pbb \psi$ $\rightarrow$ $(\id-\Pbb)\Pbb \psi
 $ $=$ $0$ in $L^2$ as $\ep\searrow 0$
and hence $\ip{\psi}{\Pbb b^0_\ep (\id-\Pbb)\del_t \Ve}$
 $\rightarrow$ $0$
by \eqref{convA8} and the fact that $\norm{\del_t\Ve}_{L^2}$ is
uniformly bounded in $\ep$. Therefore, we have established that
\eqn{convA9}{ \Pbb b^0_\ep (\id-\Pbb)\del_t \Ve \longrightarrow 0
\quad \text{weakly in $L^2$ as $\ep\searrow 0$.} } The remainder
of the proof follows from a straightforward adaptation of the
proof of Theorem 2 in \cite{Scho86}.
\end{proof}
From the block diagonal form of the matrix $c^I$, it is clear that
$\omega$ can be written as \eqn{omdef}{ \omega =
(\omega^{ij}_4,\omega^{ij}_I,0,\ldots,0)^T \, . } Using this, we
can write the system \eqref{convA2} as \lalign{Nsys}{
\del_t\alphat & = -\wt^I\del_I \alphat
-\frac{\alphat}{2n}\del_I \wt^I \label{Nsys.1}\\
\del_t \wt^J &= -\frac{\alphat}{2n}
\del^J \alphat - \wt^I\del_I \wt^J -\left[
\delta^{IJ}\bigl(\uft^{44}_I + \delta_{KL}\uft^{KL}_I\bigr)+4\uft^{J4}_4\right]
\label{Nsys.2} \\
\del_t \wt^4 &= -\wt^I\del_I \wt^4 -
\bigl(\uft^{44}_4 + \delta_{IJ}\uft^{IJ}_4\bigr) \label{Nsys.3} \\
\del_t \uft^{ij}_4 &= 4\delta\uft^{4I}\del_I \uft_4^{ij}
+ 4\delta\uft^{IJ}\del_I\uft^{ij}_J
+ \eta^{ij}\del_I\beta^I-
2\del_I \beta^{(i}\eta^{j)I}
-\Sct^{ij} + \del^I \omega_I^{ij} \label{Nsys.4} \\
\del_t\uft^{ij}_I &
= 4\delta\uft^{IJ}\del_J \uft^{ij}_4 + \del_I \omega^{ij}_4 \label{Nsys.5}\\
\del_t\delta \uft^{ij}
& = \uft^{ij}_4 \label{Nsys.6} \\
\del_J \uft^{ij}_4 & =  0 \label{Nsys.7} \\
\del^J\uft_J^{ij} & = \delta^i_4\delta^j_4 \Delta \Phit\label{Nsys.8}\\
\Delta \Phit & =  \rhot \label{Nsys.9} } with initial conditions
\lgath{Nidat}{ \uft^{ij}_4(0) = 0\, , \quad \uft_I^{iJ}(0) = 0 \,
, \quad \uft_I^{44} = \del_I\phi  \quad (\phi :=
\Delta^{-1}\rhot(0))\, ,
\label{Nidat.1} \\
\alphat(0) = \alphao \, , \quad \wt^I = \wo^I \, , \quad \wt^4 = 0
\, . } Equation \eqref{Nsys.7} immediately implies that
\leqn{conv2}{ \uft^{ij}_4 = 0 \, , } and hence, by uniqueness and
the fact that $\delta\uf^{ij}(0) = 0$, it follows from
\eqref{Nsys.6} that \leqn{conv3}{ \delta\uft^{ij} = 0 \, . } Since
$\uft^{ij}_J = \del_J\ufbt^{ij}$, we get from \eqref{Nsys.8} that
$\Delta\ufbt^{ij} = \delta^i_4\delta^j_4 \Delta \Phit$. But
$\ufbt^{ij} \in L^6_{\delta_1}$ and $\Delta \Phit \in
L^2_{\delta-2}$ and so by theorem 1.2 and proposition 1.6 of
\cite{Bart86}, we find that $\ufbt^{ij} \in H^k_{\delta_2}$ for $0
> \delta_2 > \delta_1\geq -1/2 >\delta > -1$. Since the Laplacian
$\Delta : H^k_{\delta_2} \rightarrow H^{k-2}_{\delta_2-1}$ is
injective for $\delta_2 <0$ (see \cite{Bart86}, proposition 2.2),
we must have $\ufbt^{ij} = \delta^{i}_4\delta^{j}_4\Phit$ and
hence \leqn{conv4}{ \uft^{ij}_J = \delta^{i}_4\delta^{j}_4 \del_J
\Phit \, . } Substituting \eqref{conv2}-\eqref{conv4} into
\eqref{Nsys.1}-\eqref{Nsys.8} yields \lalign{NsysA}{ \del_t\alphat
& = -\wt^I\del_I \alphat
-\frac{\alphat}{2n}\del_I\wt^I\label{NsysA.1}
\, ,\\
\del_t \wt^J &= -\frac{\alphat}{2n}\del^J \alphat - \wt^I\del_I \wt^J -
\del^J\Phit \, , \label{NsysA.2} \\
\Delta \Phit & =  \rhot \label{NsysA.3} \, , \intertext{and}
\del_t \wt^4 &= -\wt^I\del_I \wt^4 \label{NsysA.4} \, , \\
 \del^I \omega_I^{ij} &=  \eta^{ij}\del_I\beta^I-
2\del_I \beta^{(i}\eta^{j)I}
+ \Sct^{ij} \label{NsysA.5} \, , \\
\del_I \omega^{Jk}_4 &= 0 \label{NsysA.6} \, , \\
\del_t \del_I\Phit & = \del_I \omega^{44}_4 \label{NsysA.7} \, . }
Since $\wt^4(0) = 0$, uniqueness of solutions to hyperbolic
equations implies that \leqn{conv5}{ \wt^4 = 0 \, . } Proposition
\ref{cogA} and \eqref{NsysA.1}-\eqref{NsysA.3} imply that
$\{\Phit,\wt^I,\alphat\}$ must satisfy \leqn{conv6}{ \alphat,
\wt^I \in C^0([0,T^*],H^k_{\delta-1})\cap
C^{1}([0,T^*],H^{k-1}_{\delta-1})} and \lgath{conv6a}{  \Phit \in
C^0([0,T^*],H^{k+2}_{\delta})\cap
C^{1}([0,T^*],H^{k+1}_{\delta})\, , \label{conv6a.1} \\
\del_t\Phit \in C^{0}([0,T^*],H^{k+1}_{\delta-1})\cap
C^{1}([0,T^*],H^{k}_{\delta-1}) \, . \label{cov6a.2}} We then get
from \eqref{NsysA.6} and \eqref{NsysA.7} that \leqn{conv7}{
\omega^{44}_4 =  \del_t \Phit \in C^{1}([0,T^*],H^{k}_{\delta-1})
} and \leqn{conv8}{ \omega^{4J}_4 = 0 \, . } Equations
\eqref{conv3} and \eqref{conv5} imply that $\Sct^{ij}$ can be
written as $\Sct^{ij} = 2\delta^{(i}_I\delta^{j)}_4 \rhot \wt^I$. We
then find from \eqref{NsysA.5} that \leqn{conv9}{ \omega_{I}^{ij}
= \del_I\Omega^{ij} } where \leqn{conv10}{ \Omega^{ij} =
\Delta^{-1}\bigl(
 \eta^{ij}\del_I\beta^I-
2\del_I \beta^{(i}\eta^{j)I} +2 \delta^{(i}_I\delta^{j)}_4 \rhot \wt^I
\bigr)\, . } Note that \eqn{conv11}{\Omega^{ij} \in
C^{1}([0,T^*],H^{k+1}_\delta)} since $\del_I\beta^j \in
C^{1}([0,T^*],H^{k-1}_{\delta-2})$ and $\Sct^{ij} \in
C^{1}([0,T^*],H^{k-1}_{\delta-2})$ by \eqref{conv6}. Therefore
\leqn{conv12}{ \omega^{ij}_I = \del_I\Omega^{ij} \in
C^{1}([0,T^*],H^k_{\delta-1}) \, . } We collect the above results
in the following proposition.
\begin{prop} \label{convB} \mnote{[convB]}
The limit solution $\{\Vt,\Phit\}$
from theorem \ref{convA} satisfies
\gath{convB2}{
\delta \ut^{ij} = \ut_4^{ij} = \wt^4 = 0 \, , \\
\Phit \in  C^{0}([0,T^*],H^{k+2}_{\delta})
\cap C^{1}([0,T^*],H^{k+1}_{\delta})\, ,
\quad \del_t\Phi \in C^{0}([0,T^*],H^{k+1}_{\delta-1})\cap C^{1}([0,T^*],H^k_{\delta-1}) \, ,\\
\ut_J^{ij} = \delta^i_4\delta^j_4\del_J\Phit \in
C^{1}([0,T^*],H^{k+1}_{\delta-1})\cap C^{1}([0,T^*],
H^k_{\delta-1}) \, ,\\
\alphat, \wt^I \in C^{0}([0,T^*],H^k_{\delta-1})
\cap C^{1}([0,T^*],H^{k-1}_{\delta-1})\, ,
}
while  $\{\Phit,\alphat,\wt^I\}$ solves the equations
\eqref{NsysA.1}-\eqref{NsysA.3}.
Moreover, the $\omega$ from theorem \ref{convA}
is given by
\eqn{convB3}{
\omega = (\omega^{ij}_4,\omega^{ij}_I,0,\ldots,0)^T
}
where
\gath{convB4}{
\omega^{ij}_4 = \delta_4^i\delta_4^j\del_t \Phit
\in C^{1}([0,T^*],H^{k}_{\delta-1}) \, , \\
\omega^{ij}_I = \del_I\Delta^{-1}\bigl(
 \eta^{ij}\del_J\beta^IJ-
2\del_J \beta^{(i}\eta^{j)J} +2 \delta^{(i}_J\delta^{j)}_4 \rhot \wt^J
\bigr) \in C^1([0,T^*],H^k_{\delta-1})\, . }
\end{prop}

\subsect{roc}{Error Estimate}

To get an error estimate which measures the difference between the
relativistic and Newtonian solutions, we adapt the arguments
of \cite{Scho86}, section IV. Define \eqn{rocA1}{ \Ze := \Ve-\Vt +
d\Phie - d\Phit - \ep \omega \AND \gammae := \alphae-\alphat \, .
} A simple but useful observation is that \leqn{rocA2}{
\norm{\gammae}_{H^{k-1}_{\delta-1,\ep}} =
\norm{\alphae-\alphat}_{H^k_{\delta-1,\ep}} \leq
\norm{\Ze}_{H^{k-1}_{\delta-1,\ep}} \AND
\norm{\we^i-\wt^i}_{H^{k-1}_{\delta-1,\ep}} \leq
\norm{\Ze}_{H^{k-1}_{\delta-1,\ep}}\, . }

\begin{lem} \label{rocA} \mnote{[rocA]}
There exists an $\ep$ independent constant $C>0$ such that
\gath{rocA0}{ \norm{d\Phie(t)-d\Phi(t)}_{H^{k-1}_{\delta-1,\ep}}
+\norm{Dd\Phie(t)-Dd\Phi(t)}_{H^{k-1}_{\delta-2,\ep}} \leq
\norm{\Phie(t)-\Phit(t)}_{H^{k+1}_{\delta}} \leq
C\norm{\Ze(t)}_{H^{k-1}_{\delta-1,\ep}} \\
 \norm{\del_t d\Phie(t)
-\del_t d\Phit(t)}_{H^{k-1}_{\delta-1,\ep}} \leq
\norm{\del_t\Phie(t) -\del_t\Phit(t)}_{H^{k}_{\delta}} \leq
C\norm{\Ze(t)}_{H^{k-1}_{\delta-1,\ep}} + C\ep \intertext{and}
\norm{\del_t\gamma_\ep}_{H^{k-2}_{\delta-1,\ep}} \leq
C\norm{\Ze(t)}_{H^{k-1}_{\delta-1,\ep}} + C\ep } for all
$(t,\ep)\in [0,T^*]\times (0,\ep_0]$.
\end{lem}
\begin{proof}
Since the support of $\alphae(t)$ and $\alphat(t)$ are both
bounded for all $(t,\ep)\in [0,T^*]\times (0,\ep_0]$, there exists
a $\ep$ independent constant $C>0$ such that \eqn{rocA10a}{
C^{-1}\norm{\rhoe-\rhot}_{H^{k-1}_{\delta-2}} \leq
\norm{\rhoe-\rhot}_{H^{k-2}_{\delta-1,\ep}} \leq C
 \norm{\rhoe-\rhot}_{H^{k-1}_{\delta-2}}\, . }
Also, $\Delta\Phie = \rhoe$, $\Delta\Phit = \rhot$, and $\Delta :
H^{k+1}_{\delta}\rightarrow H^{k-1}_{\delta-1}$ is an isomorphism,
and therefore
 \leqn{rocA10}{ \norm{\Phi_\ep - \Phit}_{H^{k+1}_\delta} \leq
\norm{\rhoe-\rhot}_{H^{k-1}_{\delta-2}} \leq
C\norm{\rhoe-\rhot}_{H^{k-1}_{\delta-1,\ep}} \leq
C\norm{\gammae}_{H^{k-1}_{\delta-1,\ep}} \leq
C\norm{\Ze}_{H^{k-1}_{\delta-1,\ep}}\,  } by proposition
\ref{nlimD} and lemma \ref{Lip}. From \eqref{alphaA} and
\eqref{NsysA.1},  it follows that $\gammae$ satisfies
\leqn{rocA3}{ \del_t\gammae = -X^I\del_I\gammae - Y\gammae +
 (X^I-\wt^I)\del_I\alphat + \bigl(Y-\frac{\del_I\wt^I}{2n}\bigr)\alphat \, ,
} where $X^I$ and $Y$ are given by \eqref{alphaB}. But
$X^I=X^I(\ep\we^4,\we^I)$ and $\wt^I=X^I(0,\wt^I)$ and hence
\leqn{rocA4}{ \norm{X^I-\wt^I}_{H^{k-2}_{\delta-1,\ep}} \leq
C\norm{\we^I-\wt^I}_{H^{k-2}_{\delta-1,\ep}} \leq
C\norm{\Ze}_{H^{k-1}_{\delta-1,\ep}}} by \eqref{rocA2},
\eqref{inclA}, lemma \ref{Lip} and proposition \ref{nlimD}. Next,
\eqn{rocA5}{ Y-\frac{\del_I\wt^I}{2n} =
(\bar{Y}(\ep\Ve)-\frac{1}{2n})(\ep\del_t\we^4+\del_I\we^I) +
\frac{1}{2n}\ep\del_t\we^4 +\frac{1}{2n}( \del_I\we^I-\del_I\wt^I)
+ \hat{Y}(\ep\Ue,\ep\Ve) } where $\hat{Y}(0)=0$ and
$\bar{Y}(0)-1/(2n) = 0$. Using \eqref{nlim11}, \eqref{diff},
\eqref{inclA}, proposition \ref{nlimD}, and lemmas \ref{ciD}
-\ref{Lip},  we can estimate each of the above terms as follows
\alin{rocA6}{ \norm{
(\bar{Y}(\ep\Ve)-\frac{1}{2n})&(\ep\del_t\we^4+\del_I\we^I)}_{H^{k-2}_{\delta,\ep}}
 \leq \norm{ (\bar{Y}(\ep\Ve)-\frac{1}{2n})}_{H^{k-2}_{\delta-1,\ep}}
(\ep\norm{\del_t\we^4}_{H^{k-2}_{\delta-1,\ep}} + \norm{\we^I}_{H^{k-2}_{\delta-1,\ep}}) \\
& \leq C\ep\norm{\Ve}_{H^k_{\delta-1,\ep}}(\ep\norm{\del_t\Ve}_{H^{k-1}_{\delta-1,\ep}} +
\norm{\Ve}_{H^k_{\delta-1,\ep}})
\leq C\ep \, , \\
&\norm{ \frac{1}{2n}\ep\del_t\we^4 }_{H^{k-2}_{\delta,\ep}}
\leq C\ep\norm{\del_t\Ve}_{H^{k-1}_{\delta-1,\ep}} \leq  C\ep \,  \\
&\norm{\frac{1}{2n}(
\del_I\we^I-\del_I\wt^I)}_{H^{k-2}_{\delta,\ep}} \leq
C\norm{\Ze}_{H^{k-1}_{\delta-1,\ep}} \, , \intertext{and}
&\norm{\hat{Y}(\ep\Ue,\ep\Ve)}_{H^{k-2}_{\delta,\ep}} \leq
C\ep(\norm{\Ue}_{H^{k}_{\delta,\ep}}+\norm{\Ve}_{H^k_{\delta-1,\ep}})\leq
C\ep\, . } Therefore \leqn{rocA7}{
\norm{Y-\frac{\del_I\wt^I}{2n}}_{H^{k-2}_{\delta,\ep}} \leq
C\norm{\Ze}_{H^{k-1}_{\delta-1,\ep}}+ C\ep \, . } We can also
estimate $X^I$ and $Y$ as follows \lalign{rocA8}{
&\norm{X^I}_{H^{k-2}_{\delta-1,\ep}}\leq C\norm{\Ve}_{H^k_{\delta-1,\ep}}\leq C\, ,\label{rocA8.1} \\
&\norm{Y}_{H^{k-2}_{\delta,\ep}}\leq C(\norm{\Ue}_{H^k_{\delta,\ep}}
+ \norm{\Ve}_{H^k_{\delta-1,\ep}}+\norm{\del_t\Ve}_{H^{k-1}_{\delta-1,\ep}})\leq C \, .
\label{rocA8.2}}
The estimates \eqref{rocA2}, \eqref{rocA4}, \eqref{rocA7}, \eqref{rocA8.1}, \eqref{rocA8.2} along with
lemma \ref{ciG} imply via the equation \eqref{rocA3}  that
\leqn{rocA9}{
\norm{\del_t\gammae}_{H^{k-2}_{\delta-1,\ep}} \leq C\norm{\Ze}_{H^{k-1}_{\delta-1,\ep}} + C\ep \, .
}
Since $\Delta\del_t\Phie = \delta\rhoe$ and $\Delta \del_t \Phit = \del_t\rhot$, the same arguments used
to establish the estimate \eqref{rocA10} can be used in conjunction with \eqref{rocA9} to show
\leqn{rocA11}{
\norm{\del_t\Phie -\del_t\Phit}_{H^{k}_{\delta}} \leq C\norm{\Ze}_{H^{k-1}_{\delta-1,\ep}} + C\ep \, .
}
Finally from \eqref{rocA10}, \eqref{rocA11}, and lemma \ref{ciF}, we get the desired estimates
\gath{rocA12}{
\norm{d\Phie-d\Phi}_{H^{k-1}_{\delta-1,\ep}}
+\norm{Dd\Phie-Dd\Phi}_{H^{k-1}_{\delta-2,\ep}}
\leq \norm{\Phie-\Phit}_{H^{k+1}_{\delta}} \leq C\norm{\Ze}_{H^{k-1}_{\delta-1,\ep}}\intertext{and}
\norm{\del_t d\Phie -\del_t d\Phit}_{H^{k-1}_{\delta-1,\ep}} \leq \norm{\del_t\Phie -\del_t\Phit}_{H^{k}_{\delta}}
\leq C\norm{\Ze}_{H^{k-1}_{\delta-1,\ep}} + C\ep
}
for some constant $C$ independent of $\ep$.
\end{proof}

\begin{lem} \label{rocB} \mnote{[rocB]}
There exists a constant $C>0$ such that
\eqn{rocB1}{\norm{\del_t\alpha_\ep-\del_t\alphat}_{H^{k-2}_{\delta-1,\ep}}
+ \norm{\Ve(t)-\Vt(t)}_{H^{k-1}_{\delta-1,\ep}} \leq C\ep \quad
\text{for all $(t,\ep)\in [0,T^*]\times(0,\ep_0]$. } }
\end{lem}
\begin{proof}
From the evolution equation \eqref{EFsym1}, we find that $\Ze$
satisfies the equation \leqn{rocB4}{ b^0_\ep\del_t\Ze =
\frac{1}{\ep}c^I\del_I\Ze + b^I_\ep \del_I\Ze + F_\ep} where
$b^0_\ep = b^0(\ep\Ue,\ep \Ve)$, $b^I_\ep = b(\ep,\Ue,\Ve)$ and
\lalign{rocB5}{ F_\ep = -b^0_\ep\del_t (d\Phit&-d\Phie) - \ep
b^0_\ep \del_t \omega + b^I_\ep(\del_I d\Phit-\del_I d\Phie)+ \ep
b^I_\ep \del_I\omega \notag \\ & -(b^0_\ep-\id)\del_t\Vt +
(\bt^I-b^I_\ep)\del_I\Vt + f(\ep,\Ue,\Ve)\Ve - \ft(\Vt)\Vt + h_\ep
-\tilde{h} \, . \label{rocB5.1}}

Using \eqref{nlim11}, \eqref{diff}, \eqref{inclA}, lemmas
\ref{rocA}, \ref{ciD}-\ref{ciE}, and propositions \ref{nlimD} and
\ref{convB}, we get the following estimates \lgath{rocB7}{
\norm{b^0_\ep-\id}_{H^k_{\delta,\ep}} \leq C\ep(
\norm{\Ue}_{H^k_{\delta,\ep}}+\norm{\Ve}_{H^k_{\delta-1,\ep}})
\leq C\ep
\label{rocB7.1} \, ,\\
\norm{b^0_\ep\del_t(d\Phit-d\Phie)}_{H^{k-1}_{\delta-1,\ep}} \leq
\norm{ (b^0_\ep-
\id)\del_t(d\Phit-d\Phie)}_{H^{k-1}_{\delta-1,\ep}} +
\norm{\del_t(d\Phit-d\Phie)}_{H^{k-1}_{\delta-1,\ep}} \notag \\
\leq C(\norm{b^0_\ep-
\id}_{H^{k-1}_{\delta,\ep}}+1)\norm{\del_t(d\Phit-d\Phie)}_{H^{k-1}_{\delta-1,\ep}}
\leq C\norm{\Ze}_{H^{k-1}_{\delta-1,\ep}} + C\ep \label{rocB7.2}
\, ,
\\
\norm{\ep b^0_\ep \del_t \omega }_{H^{k-1}_{\delta-1,\ep}} \leq
\ep
C(\norm{b^0_\ep-1}_{H^{k-1}_{\delta,\ep}}+1)\norm{\del_t\omega}_{H^{k-1}_{\delta-1,\ep}}
\leq C\ep \label{rocB7.3} \, , \\
\norm{\ep b^I_\ep \del_I\omega}_{H^{k-1}_{\delta-1,\ep}} \leq C\ep
\norm{b^I_\ep}_{H^{k-1}_{\delta,\ep}}
\norm{\del_I\omega}_{H^{k-1}_{\delta-2,\ep}} \leq C\ep
\norm{b^I_\ep}_{H^{k-1}_{\delta,\ep}}\norm{\omega}_{H^{k}_{\delta-1,\ep}}
\leq C\ep \label{rocB7.4} \, , \\
\norm{h_\ep-\tilde{h}}_{H^{k-1}_{\delta-1,\ep} }\leq C\ep \,
.\label{rocB7.5}}

To estimate the term $b^I_\ep -\bt$, we first note that
\eqn{rocB8}{ b^I_\ep - \bt =
\begin{pmatrix}A^I(\ufo_\ep^{iJ})+A^I(\delta \uf_\ep^{iJ}) & 0 \\
0 & a^I(\ep\uf_\ep,\ep \we^i,\ep
\alphae,\we^I,\alphae)-a^I(0,0,0,\wt^i,\alphat) \end{pmatrix} \, }
where the map $a^I$ is analytic. Next, the estimate \eqref{nlim11}
implies that \leqn{rocB9}{
\norm{\uf^{ij}_\ep}_{H^{k-1}_\delta,\ep} \leq
\norm{\ufo_\ep^{ij}}_{H^{k-1}_{\delta,\ep}}+
C\norm{\delta\uf_\ep^{ij}}_{H^{k-1}_{\delta-1,\ep}} \leq C +
C\norm{\Ze}_{H^{k-1}_{\delta-1,\ep}}\, . } From proposition
\ref{idatA2} and lemma \ref{ciF}, we see that the $\ufo_\ep^{iJ}$
can be estimated by
\leqn{rocB10}{\norm{\ufo_\ep^{iJ}}_{H^{k+1}_{\delta,\ep}} =
\norm{\ep\ufbo_\ep^{iJ}}_{H^{k+1}_{\delta,\ep}} \leq
\ep^{|\delta+1/2|}\norm{\ufbo_\ep^{iJ}}_{H^{k+1}_\delta} \leq
C\ep^{|\delta+1/2|+1} \, . } Also, from proposition \ref{convB}
and lemma \ref{rocA}, we obtain
\leqn{rocB10a}{\norm{\del_t\alpha_\ep-\del_t\alphat}_{H^{k-2}_{\delta-1,\ep}}
+ \norm{\Ve-\Vt}_{H^{k-1}_{\delta-1,\ep}} \leq
\norm{\Ze}_{H^{k-1}_{\delta-1,\ep}}+
\norm{d\Phie-d\Phit}_{H^{k-1}_{\delta-1,\ep}} +
\ep\norm{\omega}_{H^{k-1}_{\delta-1,\ep}} \leq
C\norm{\Ze}_{H^{k-1}_{\delta-1,\ep}} + C\ep \, . }

The three estimates \eqref{rocB9}-\eqref{rocB10a} along with
lemmas \ref{ciE} and \ref{Lip}, and proposition \ref{nlimD} and
\ref{convB}, show that
\eqn{rocB11}{\norm{A^I(\ufo_\ep^{iJ})+A^I(\delta\uf_\ep^{iJ})
}_{H^{k-1}_{\delta,\ep}} \leq
C\norm{\ufo_\ep^{iJ}}_{H^k_{\delta,\ep}} +
C\norm{\delta\uf^{iJ}_\ep}_{H^{k-1}_{\delta-1,\ep} }\leq C\ep +
C\norm{\Ze}_{H^{k-1}_{\delta-1,\ep}} \, ,  } and
\alin{rocB12}{\norm{a^I(\ep\uf_\ep,\ep \we^i,\ep
\alphae,\we^I,\alphae)-a^I(0,0,0,\wt^i,\alphat)}_{H^{k-1}_{\delta,\ep}}
&\leq C(\ep \norm{\uf_\ep^{ij}}_{H^{k-1}_{\delta,\ep}} +
\norm{\alphae-\alphat}_{H^{k-1}_{\delta-1,\ep}} \\
& +\norm{\we^i-\wt^i}_{H^{k-1}_{\delta-1,\ep}} \leq C\ep +
C\norm{\Ze}_{H^{k-1}_{\delta-1,\ep}}) \, . } Therefore
\eqn{rocB13}{ \norm{b^I_\ep-\bt}_{H^{k-1}_{\delta-1,\ep}} \leq
C\ep + C\norm{\Ze}_{H^{k-1}_{\delta-1,\ep}} } and hence
\leqn{rocB14}{
\norm{(\bt-b^I_\ep)\del_I\Vt}_{H^{k-1}_{\delta-1,\ep}} \leq
C\norm{\bt-b^I_\ep}_{H^{k-1}_{\delta,\ep}}\norm{D\Vt}_{H^{k-1}_{\delta-2,\ep}}
\leq C\ep + \norm{\Ze}_{H^{k-1}_{\delta-1,\ep}}\, . }

Next, we notice that \eqn{rocB15}{f(\ep,\Ue,\Ve)\Ve-\ft(\Vt)\Vt =
- \rho_\ep \mathcal{F}_\ep + \hat{f}(\Ve)\Ve-\hat{f}(\Vt)\Ve + \ep
\bar{f}(\ep,\Ue,\Ve)\Ve  \, } where \eqn{rocB16}{\mathcal{F}_\ep
:= -4  \rho_\ep (
\ep\delta^i_4\delta^j_4\eta_{pq}\ufbo_\ep^{pq},0,\ldots,0)^T } and
$\hat{f}$ and $\bar{f}$ are analytic. We obtain \leqn{rocB17}{
\norm{\hat{f}(\Ve)\Ve-\hat{f}(\Vt)\Ve + \ep
\bar{f}(\ep,\Ue,\Ve)\Ve}_{H^{k-1}_{\delta-1,\ep}}\leq
C\norm{\Ze}_{H^{k-1}_{\delta-1,\ep}} + C\ep  } by the arguments
used above. Also, the boundedness of the support of $\alphae(t)$
implies that
\leqn{rocB18}{\norm{\mathcal{F}_\ep}_{H^{k-1}_{\delta-1,\ep}} \leq
C\ep\norm{\rhoe\eta_{ij}\ufbo^{ij}_\ep}_{H^{k-1}_\delta}\leq
C\ep\norm{\rhoe}_{H^{k-1}_\delta} \norm{\ufbo}_{H^{k-1}_\delta}
\leq C\ep
\norm{\rhoe}_{H^{k-1}_{\delta-1,\ep}}\norm{\ufbo}_{H^{k-1}_\delta}
 \leq C\ep \, . }
So then \leqn{rocB19}{
\norm{f(\ep,\Ue,\Ve)\Ve-\ft(\Vt)\Vt}_{H^{k-1}_{\delta-1,\ep}} \leq
C\norm{\Ze}_{H^{k-1}_{\delta-1,\ep}} + C\ep } by \eqref{rocB17}
and \eqref{rocB18}. Combining the estimates
\eqref{rocB7.1}-\eqref{rocB7.5}, \eqref{rocB14}, \eqref{rocB17},
and \eqref{rocB19} yields \leqn{rocB20}{
\norm{F_\ep}_{H^{k-1}_{\delta-1,\ep}} \leq
C\norm{\Ze}_{H^{k-1}_{\delta-1,\ep}} + C\ep \, .}

Letting $\Ze^\alpha = D^\alpha\Ze$ and differentiating the
equation \eqref{rocB4} yields \eqn{rocB21}{ b^0_\ep
\del_t\Ze^\alpha  = \frac{1}{\ep}c^I\del_I \Ze^\alpha + b^I_\ep
\del_I \Ze^\alpha + q^\alpha \qquad 0\leq |\alpha|\leq k-1 \, , }
where \eqn{rocB21a}{ q^\alpha = -
[D^\alpha,b^0_\ep]\del_t\Ze^\alpha +
[D^\alpha,b^I_\ep]\del_I\Ze^\alpha + D^\alpha F_\ep \, . } Using
the estimates above along with  propositions \ref{nlimD} and
\ref{convB} and the calculus inequalities from appendix
\ref{winq}, we find \gath{rocB22}{
\norm{\del_t\Ze}_{H^{k-1}_{\delta-1,\ep}} \leq
C\norm{\Ze}_{H^{k-1}_{\delta-1,\ep}}
+ C\ep \, , \\
\norm{[D^\alpha,b_\ep^0]\del_t\Ze^{\alpha}}_{L^2_{\delta-1-|\alpha|,\ep}}
\leq
C \norm{b^0_\ep-\id}_{H^{k-1}_{\delta,\ep}}\norm{\del_t\Ze}_{H^{k-1}_{\delta-1,\ep}}
\leq C \norm{\Ze}_{H^{k-1}_{\delta-1,\ep}} + C \ep \, , \\
\norm{[D^\alpha,b^I_\ep]\del_I\Ze^\alpha}_{L^2_{\delta-1-|\alpha|,\ep}}
\leq
C\norm{b^I_\ep}_{H^{k-1}_{\delta,\ep}}\norm{D\Ze}_{H^{k-2}_{\delta-2,\ep}}
\leq C\norm{\Ze}_{H^{k-1}_{\delta-1,\ep}} \, , } and hence
\eqn{rocB22}{ \norm{q^\alpha}_{H^{k-1}_{\delta-1,\ep}} \leq
C\norm{\Ze}_{H^{k-1}_{\delta-1,\ep}} + C\ep \, . } Combining this
estimate with the estimates \eqn{rocB23}{ \norm{\del_t b^0_\ep +
\del_I b^I_\ep}_{L^\infty} \leq C \, , \quad
\norm{b^I_\ep}_{L^\infty} \leq C \, , } and lemma \ref{nlimA},
shows that \eqn{rocB24}{ \frac{d\,}{dt} \ip{\Ze^\alpha}{b^0_\ep
\Ze^\alpha}_{L^2_{\delta-1-|\alpha|,\ep}} \leq
C(\norm{\Ze}_{H^{k-1}_{\delta-1,\ep}} +\ep )
\norm{\Ze}_{H^{k-1}_{\delta-1,\ep}} \quad 0\leq |\alpha|\leq k-1
\, . } Summing over $\alpha$ and using Gronwall's inequality, we
get \eqn{rocB25a}{ \norm{\Ze(t)}_{H^{k-1}_{\delta-1,\ep}} \leq
C\norm{\Ze(0)}_{H^{k-1}_{\delta-1,\ep}} + C\ep \quad \text{for all
$(t,\ep) \in [0,T^*]\times (0,\ep_0]$.} } This estimate and
\eqref{rocB10a} then prove the proposition since
$\norm{\Ze(0)}_{H^{k-1}_{\delta-1,\ep}}$ $\leq$ $C \ep$ by
proposition \ref{locA}.
\end{proof}
We are now ready to prove a precise error estimate for the
difference between the relativistic and Newtonian solutions.
\begin{prop} \label{rocC} \mnote{[rocC]}
Suppose $-1< \delta < -1/2$ and $k\geq 3$. Then there exists a
constant $C> 0$ such that \gath{rocC1}{
\norm{\ufb_\ep^{ij}(t)-\delta^i_4\delta^i_4\Phit(t)}_{L^6_{\delta,\ep}}
+ \norm{\del_I\ufb_{\ep}^{ij}(t) - \delta^{i}_4\delta^j_4
d\Phit(t)}_{H^{k-1}_{\delta-1,\ep}}
+\norm{v^I(t)-\wt^I(t)}_{H^{k-1}_{\delta-1,\ep}} \\
+ \ep^{-1}\norm{v^4(t)-1}_{H^{k-1}_{\delta-1,\ep}} +
\norm{\rho_\ep(t) - \rhot(t)}_{H^{k-1}_{\delta-1,\ep}}
+\norm{\del_t\rho_\ep(t)-\del_t\rhot(t)}_{H^{k-2}_{\delta-1,\ep}}
\leq C\ep} for all $(t,\ep)\in[0,T^*]\times (0,\ep_0]$.
\end{prop}
\begin{proof}
From the evolution equations and proposition \ref{convB}, we have
\eqn{rocCx.1}{
\ep\partial_t\bigl(\ufb_\ep^{ij} -\delta^i_4\delta^j_4\Phit)
= \uf^{ij}_{4} - \ep \omega^{ij}_4
}
and hence integrating yields
\leqn{rocCx.2}{
\ep \norm{\ufb^{ij}_\ep(t) -\delta^i_4\delta^j_{4}\Phit(t)}_{L^2_{\delta,\ep}}
 \leq \ep\norm{\ufbo^{ij}_{\ep} -\delta^i_4\delta^j_4\phi}_{L^2_{\delta,\ep}}
+ \int_{0}^{t}\norm{\uf^{ij}_{4}(s) - \ep \omega^{ij}_4(s)}_{L^2_{\delta,\ep}} ds \,.
}
But
\leqn{rocCx.3}{
 \int_{0}^{t}\norm{\uf^{ij}_{4}(s) - \ep \omega^{ij}_4(s)}_{L^2_{\delta,\ep}} ds
\leq \int_{0}^{t}\norm{V_\ep(s)-\Vt(s)}_{H^{k-1}_{\delta-1,\ep}}
+ \ep\norm{\omega^{ij}(s)}_{H^{k-1}_{\delta-1,\ep}} ds
}
and
\leqn{rocCx.4}{
\ep\norm{\ufbo^{ij}_{\ep} -\delta^i_4\delta^j_4\phi}_{L^2_{\delta,\ep}}
\leq C\ep^{3/2}
}
by the calculus inequalities of appendix \ref{winq} and
proposition \ref{idatA}.
Also, by lemma \ref{ciB} and $\uf_{I,\ep} = \del_I\ufb_\ep$,
we have
\leqn{rocC2}{ \norm{\ufb^{ij}_\ep -
\delta^i_4\delta^j_4\Phit}_{L^6_{\delta,\ep}} \leq
C\norm{\uf_{I,\ep}^{ij}-\delta^i_4\delta^j_4
d\Phit}_{L^2_{\delta-1,\ep}} + \ep
\norm{\ufb^{ij}_\ep(t) -\delta^i_4\delta^j_{4}\Phit(t)}_{L^2_{\delta,\ep}} +
\ep \norm{\ufb^{ij}_\ep(t) -\delta^i_4\delta^j_{4}\Phit(t)}_{L^2_{\delta,\ep}} \, .
} Recall that
$\rho_\ep$ $=$ $(4Kn(n+1))^{-n}\alpha_\ep^{2n}$ and $\rhot$ $=$
$(4Kn(n+1))^{-n}\alphat^{2n}$. Since
$\norm{\alpha_\ep}_{H^k_{\delta-1,\ep}}$ is bounded as
$\ep\searrow 0$, we obtain \leqn{rocC3}{ \norm{\rho_\ep -
\rhot}_{H^{k-1}_{\delta-1,\ep}} \leq
C\norm{\alpha_\ep-\alphat}_{H^{k-1}_{\delta-1,\ep}} \leq C
\norm{\Ve - \Vt}_{H^{k-1}_{\delta-1,\ep}} \,  } by lemma
\ref{Lip}. We also have that
$\norm{\del_t\alpha_\ep}_{H^{k-2}_{\delta-1,\ep}}$ is bounded as
$\ep \searrow 0$, so the formulas \eqn{rocC4}{ \del_t\rho_\ep =
\frac{2n}{(4Kn(n+1))^{n}}\alpha_\ep^{2n-1}\del_t\alpha_\ep \, ,
\quad \del_t \rhot =
\frac{2n}{(4Kn(n+1))^{n}}\alphat^{2n-1}\del_t\alphat \, , } and
the calculus inequalities of appendix \ref{winq} imply that
\lalign{roc5}{
\norm{\del_t\rho_\ep-\del_t\rhot}_{H^{k-2}_{\delta-1,\ep}} & \leq
C(\norm{\alpha_\ep-\alphat}_{H^{k-2}_{\delta-1,\ep}} +
\norm{\del_t\alpha_\ep - \del_t\alphat}_{H^{k-2}_{\delta-1,\ep}} )
\notag \\  &\leq C(\norm{\Ve-\Vt}_{H^{k-1}_{\delta-1,\ep}} +
\norm{\del_t\alpha_\ep - \del_t\alphat}_{H^{k-2}_{\delta-1,\ep}}
)\, .\label{rocC5.1} } Finally, from the definition of $\Ve$ and
$\Vt$, we have \leqn{roc6}{ \norm{\del_I\ufb_{\ep}^{ij}(t) -
\delta^{i}_4\delta^j_4 d\Phit(t)}_{H^{k-1}_{\delta-1,\ep}}
+\norm{v^I(t)-\wt^I(t)}_{H^{k-1}_{\delta-1,\ep}}  +
\ep^{-1}\norm{v^4(t)-1}_{H^{k-1}_{\delta-1,\ep}} \leq
C\norm{\Ve-\Vt}_{H^{k-1}_{\delta-1,\ep}} \, . } The proof now
follows as a direct consequence of lemma \ref{rocB} and
\eqref{rocCx.2}-\eqref{roc6}.
\end{proof}
In the above error estimate, the norm itself depends on $\ep$. We
now show how to choose norms independent of $\ep$ which are
compatible with the error estimate above. First, for any $\eta \in
\Rbb$ define a norm by \eqn{rocD.2}{\norm{u}_{\ell,p,\eta} :=
\sum_{|\alpha|\leq \ell} \norm{D^\alpha u}_{L^p_\eta}  \, .}
Recalling that $-1<\delta<-1/2$,  fix $\eta$ $\in$
$[\delta,-1/2]$. Then from \eqref{inclA}and lemma \ref{ciF}, we
get that \leqn{rocD.3}{ \norm{u}_{\ell,2,\eta-1} \leq
C\ep^{\eta+1/2}\norm{u}_{H^\ell_{\delta-1,\ep}} \AND
\norm{u}_{0,6,\eta} \leq
C\ep^{\eta+1/2}\norm{u}_{L^6_{\delta,\ep}}
 } for some constant $C>0$ independent of $\ep$.
Combining \eqref{rocD.3} with corollary \ref{rocC}
yields the following theorem which is our main result.
\begin{thm} \label{rocD} \mnote{[rocD]}
Suppose $-1<\delta < -1/2$, $-\delta\leq \eta \leq -1/2$ and $k
\geq 3$. Then there exists a constant $C>0$ such that
\gath{rocD1}{
\norm{\ufb_\ep^{ij}(t)-\delta^i_4\delta^i_4\Phit(t)}_{0,6,\eta} +
\norm{\del_I\ufb_{\ep}^{ij}(t) - \delta^{i}_4\delta^j_4
\del_I\Phit(t)}_{k-1,2,\eta-1}
+\norm{v^I(t)-\wt^I(t)}_{k-1,2,\eta-1} \\
+ \ep^{-1}\norm{v^4(t)-1}_{k-1,2,\eta-1} + \norm{\rho_\ep(t) -
\rhot(t)}_{k-1,2,\eta-1}
+\norm{\del_t\rho_\ep(t)-\del_t\rhot(t)}_{k-2,2,\eta-1} \leq
C\ep^{\eta+3/2} } for all $(t,\ep)\in[0,T^*]\times (0,\ep_0]$.
\end{thm}
Note that for $\eta = -1/2$, we have
\eqn{rocD.4}{\norm{u}_{0,6,-1/2} = \norm{u}_{L^6}\AND
\norm{u}_{\ell,2,-3/2} = \norm{u}_{H^\ell} } where
$\norm{u}_{H^\ell}$ is the standard Sobolev norm. So the above
theorem shows that the difference between the relativistic and
Newtonian solutions is of order $\ep$ with respect to the norms
$\norm{\cdot}_{L^6}$ and $\norm{\cdot}_{H^{k-1}}$.

\bigskip


\appendix
\sect{winq}{Weighted calculus inequalities}

In this and the following sections $C$ will denote a
constant that may change value from line to line but whose exact
value is not needed.

Let $V$ be a finite dimensional vector space with inner product
$\ipe{\cdot}{\cdot}$ and corresponding norm $\enorm{\cdot}$. For
$u\in L^p_{\text{loc}}(\Rbb^n,V)$, $1\leq p \leq \infty$, $\delta
\in \Rbb$, and $\ep \in \Rbb_{\geq 0}$,  the \emph{weighted
$L^{p}$ norm} of $u$ is defined by \leqn{wLpdef}{
\norm{u}_{L^{p}_{\delta, \ep}} := \left\{\begin{array}{ll}
\norm{\sigma_\ep^{-\delta-n/p}\,u}_{L^p} &
\text{if $ 1\leq p < \infty$}\\
\\
\norm{\sigma_\ep^{-\delta}\,u}_{L^{\infty}} & \text{if $p=\infty$}
\end{array} \right. }where $\displaystyle{\sigma_{\ep}(x) := \sqrt{1+\frac{1}{4}|\ep x|^2}}$. The
\emph{weighted Sobolev norms} are then defined by \leqn{wSobdef}{
\norm{u}_{W^{k,p}_{\delta,\ep}} := \left\{
\begin{array}{ll} \displaystyle{\Bigl(\sum_{|\Ic|\leq k}
\norm{\Der^{\Ic}u}^{p}_{L^{p}_{\delta-|\Ic|,\ep} } \Bigr)^{1/p}} &
\text{if $1\leq p < \infty$} \\
\\ \displaystyle{\sum_{|\Ic|\leq k}
\norm{\Der^{\Ic}u}_{L^{\infty}_{\delta-|\Ic|,\ep} }} & \text{if
$p=\infty$}
\end{array}\right. } where $k\in \Nbb_0$, $\Ic = (\Ic_{1},\ldots, \Ic_{n}) \in
\Nbb_{0}^{n}$ is a multi-index and $\Der^{\Ic} =
\partial_{1}^{\Ic_{1}}\ldots\partial_{n}^{\Ic_{n}}$. Here
\eqn{pardef}{\partial_i = \frac{\partial\;}{\partial x^i}} where
$(x^1,\ldots,x^n)$ are the standard Cartesian coordinates on
$\Rbb^n$.

The weighted Sobolev spaces are then defined as \eqn{wsobdef}{
W^{k,p}_{\delta,\ep} = \{\, u \in W^{k,p}_{\text{loc}}(\Rbb^n,V)
\, | \, \norm{u}_{W^{k,p}_{\delta,\ep}} < \infty \,  \}\, .  }
Directly from this definition, we observe the simple but useful
inequality \leqn{diff}{ \norm{\del_j u}_{W^k_{\delta,\ep}} \leq
\norm{u}_{W^{k+1}_{\delta+1,\ep}}\, . }
We note
that $W^{k,p}_{\delta,0}$ are the standard Sobolev spaces and for
$\ep > 0$, the $W^{k,p}_{\delta,\ep}$ are equivalent to the
radially weighted Sobolev spaces \cite{Bart86,CBC}.
For $p=2$, we use the alternate notation $H^{k}_{\delta,\ep} :=
W^{k,2}_{\delta,\ep}$. The spaces $L^{2}_{\delta,\ep}$ and
$H^{k}_{\delta,\ep}$ are Hilbert spaces with inner products
\leqn{L2ip}{ \ip{u}{v}_{L^{2}_{\delta,\ep}} := \int_{\Rbb^{n}}
\ipe{u}{v} \sigma_\ep^{-2\delta-n}d^{n}x\, ,} and \leqn{Hkip}{
\ip{u}{v}_{H^{k}_{\delta,\ep}} := \sum_{|\Ic|\leq k}
\ip{\Der^{\Ic}u}{\Der^{\Ic}v}_{L^{2}_{\delta-|\Ic|,\ep} }, }
respectively. When $\ep =1 $, we will also use the notation
$W^{k,p}_{\delta} = W^{k,p}_{\delta,1}$ and $H^k_\delta = H^k_{\delta,1}$.

Let $B_R$ be the open ball of radius $R$ and  $a_R$ and $A_R$ denote the annuli
$B_{2R}\setminus B_{R}$ and $B_{4R}\setminus B_{R}$, respectively.
Let $\{\phi_j\}_{j=0}^{\infty}$ be a smooth partition of
unity satisfying \eqn{unityA}{ \text{supp}\, \phi_0 \subset B_2\,
, \quad \text{supp}\, \phi_j \subset A_{2^{j-1}} \quad (j\geq 1)
\, ,\quad \text{and} \quad \phi_{j}(x) := \phi_1(2^{1-j}x) (j\geq
1) \, . } Scaling gives a one parmeter family
of smooth partitions of unity \eqn{unityB}{ \phi^\ep_j(x) :=
\phi_{j}(\ep x) \quad (j\geq 0) } which satisfy \leqn{unityC}{ \text{supp}\, \phi^\ep_0
\subset B_{2/\ep}\, , \quad \text{supp}\, \phi^\ep_j \subset
A_{2^{j-1}/\ep} \quad (j\geq 1) \, ,\quad \text{and} \quad
S_j\phi_{j}^\ep(x) := \phi_1^\ep(2^{1-j}x) (j\geq 1) \, . }

Define a scaling operator by \leqn{scalopA}{ S_{j}u(x) := u(2^{j-1}x)\, . }
This operator satisfies the following simple, but useful
identities \lgath{scalopB}{ S_1 = \id \, ,\quad S_j\circ S_k = S_k
\circ S_j = S_{k+j-1}\, , \label{scalopB.1} \\
S_{j}\phi_j^\ep = \phi_1^\ep \quad (j\geq 1)\, , \label{scalopB.2} \\
\norm{S_{j}u}_{L^p} = 2^{\frac{n(1-j)}{p}}\norm{u}_{L^p} \, ,\label{scalopB.3}\\
\intertext{and} S_{j}\circ \Der^{\Ic} = 2^{(1-j)|\Ic|} \Der^{\Ic}
\circ S_{j} \, . \label{scalopB.4} }

\begin{lem} \label{equivA} \mnote{[equivA]}
For $1\leq p < \infty$, there exists a constant $C>0$ independent
of $\ep \geq 0$ such that \eqn{equivA.1}{\frac{1}{C}
\norm{u}^{p}_{L^p_{\delta,\ep} } \leq \norm{\phi^\ep_0 u}^p_{L^p}
+ \sum_{j=0}^{\infty} \norm{S_j(\phi_j^\ep u)}^p_{L^p} \leq C
\norm{u}^{p}_{L^p_{\delta,\ep}}\, . }
\end{lem}
\begin{proof}
 From the identity \eqn{equivA.2}{ \norm{u}^{p}_{L^p} =
\int_{B_{4/\ep} }|u|^p\, d^n x + \sum_{j=1}^{\infty}
\int_{a_{2^{j+1}/\ep}}|u|^p \, d^n x } and a simple change of
variables, it follows that
\leqn{equivA.3}{\norm{u}_{L^p_{\delta,\ep}}^p =
\norm{\sigma_\ep^{-\delta-n/p}u}^p_{L^p(B_{4/\ep}) } +
\sum_{j=1}^{\infty} 2^{n(j-1)}
\norm{S_j(\sigma_\ep^{-\delta-n/p}u)}^p_{L^p(a_{4/\ep})}\, . }
This identity and \alin{equivA.4}{\max_{x\in \overline{B}_{4/\ep}
} \sigma_\ep(x)^{-\delta p -n}  &= \left\{\begin{array}{ll}
2^{\frac{-\delta p -n}{2}} &
\text{if $-\delta p -n \geq 0$} \\
1 & \text{if $-\delta p -n < 0$}
\end{array} \right. \, , \\
\min_{x\in \overline{B}_{4/\ep} } \sigma_\ep(x)^{-\delta p-n} & =
\left\{\begin{array}{ll} 1 &
\text{if $-\delta p -n \geq 0$} \\
2^{\frac{-\delta p -n}{2}} & \text{if $-\delta p -n < 0$}
\end{array} \right. \, , \\
\max_{x\in \overline{a}_{4/\ep} } (S_j\sigma_\ep)(x)^{-\delta p-n}
& = \left\{\begin{array}{ll} (1+2^{2j})^{\frac{-\delta p -n}{2}} &
\text{if $-\delta p -n \geq 0$} \\
(1+2^{2(j-1)})^{\frac{-\delta p-n}{2}} & \text{if $-\delta p -n <
0$}
\end{array} \right. \, , \\
\min_{x\in \overline{a}_{4/\ep} } (S_j \sigma_\ep)(x)^{-\delta p
-n}  &= \left\{\begin{array}{ll} (1+2^{2(j-1)})^{\frac{-\delta
p-n}{2}} &
\text{if $-\delta p -n \geq 0$} \\
(1+2^{2j})^{\frac{-\delta p -n}{2}} & \text{if $-\delta p -n \leq
0$}
\end{array} \right. \, ,
 }
show that \leqn{equivA.5}{\frac{1}{C}\norm{u}^p_{L^p(B_{4/\ep})}
\leq \norm{\sigma_\ep^{-\delta-n/p}u}^p_{L^p(B_{4/\ep})} \leq C
\norm{u}^p_{L^p(B_{4/\ep})} } and \leqn{equivA.6}{ \frac{1}{C}
2^{-p\delta(j-1)}\norm{S_j u}^p_{L^p(a_{4/\ep})} \leq
2^{n(j-1)}\norm{S_j(\sigma_\ep^{-\delta-n/p}u) }_{L^p(a_{4/\ep})}
\leq C 2^{-p\delta(j-1)}\norm{S_j u}^p_{L^p(a_{4/\ep})}}for some
constant $C>0$ which is independent of $\ep \geq 0$. Using a change
of variable, the inequality \eqref{equivA.6} can be written
as
\leqn{equivA.7}{
\frac{1}{C}
2^{-p\delta(j-1)}2^{(1-j)n}\norm{u}^p_{L^p(a_{2^{j+1}/\ep})} \leq
2^{n(j-1)}\norm{S_j(\sigma_\ep^{-\delta-n/p}u) }_{L^p(a_{4/\ep})}
\leq C 2^{-p\delta(j-1)}2^{(1-j)n}\norm{u}^p_{L^p(a_{2^{j+1}/\ep})} \, .
}
From
\leqn{equivA.8}{
\sum_{k=0}^{2}\phi_k^{\ep}\bigl|_{B_{4/\ep}} = \id_{\!B_{4/\ep}}
\quad \text{and} \quad
\sum_{k=0}^{2}\phi^\ep_{j+k}\bigl|_{a_{2^{j+1}/\ep}} =
\id_{\!a_{2^{j+1}/\ep}}
}
and \eqref{scalopB.3}, we obtain
\leqn{equivA.9}{
\norm{u}^{p}_{L^p(B_{4/\ep})}
\leq C\bigl( \norm{\phi_0^\ep u}^{p}_{L^p} + \sum_{k=1}^{3}
\norm{S_k(\phi^{\ep}_ku)}_{L^p}^p \bigr) \, ,
}
and
\leqn{equivA.10}{
\norm{u}^{p}_{L^p(a_{2^{j+1}/\ep})} \leq C \sum_{k=0}^2 2^{n(j+k)}
\norm{S_{j+k}(\phi^{\ep}_{j+k}u)}_{L^p}^p \, .
}
Combining \eqref{equivA.3} with the inequalities
\eqref{equivA.5}, \eqref{equivA.7}, \eqref{equivA.9} and \eqref{equivA.10}
yields
\leqn{equivA.11}{
\norm{u}^p_{L^p_{\delta,\ep}}
\leq C \bigl(\norm{\phi^\ep_0 u}^p_{L^p}
+ \sum_{j=1}^\infty 2^{-p\delta(j-1)}\norm{S_j(\phi^\ep_j u)}^p_{L^p}\bigr) \,
}
for some constant $C>0$ independent of $\ep \geq 0$.

Since $\text{supp}\, \phi^\ep_0 \subset B_{4/\ep}$ and
$\norm{\phi^{\ep}_0}_{L^\infty} = \norm{\phi_{0}}_{L^\infty}$,
we get from \eqref{equivA.5} that
\leqn{equivA.12}{
\norm{\phi^\ep_0 u }^p_{L^p}\leq \norm{\phi^\ep_0}^p_{L^\infty}
\norm{u}^p_{L^p(B_{4/\ep})} \leq C \norm{\sigma^{-\delta-n/p}_{\ep}u}^p_{L^p(B_{4/\ep})}
}
for some constant $C>0$ independent of $\ep \geq 0$.
Next,
\alin{equivA.13}{
2^{-p\delta(j-1)}\norm{S_j(\phi^\ep_j u)}^p_{L^p}
& \leq 2^{-p\delta(j-1)}2^{n(1-j)}\norm{\phi^\ep_j u}_{L^p(A_{2^{j-1}/\ep})}
&& \text{by \eqref{scalopB.3} and \eqref{unityC},}  \\
& \leq 2^{-p\delta(j-1)}2^{n(1-j)}\norm{\phi_1}_{L^\infty}
\norm{\phi^\ep_j u}_{L^p(\cup_{k=-1}^{1}a_{2^{j-k}/\ep})} &&
\text{since
$\norm{\phi_1}_{L^\infty}=\norm{\phi^\ep_j}_{L^\infty}$.} } So
there exists a constant $C>0$ independent of $\ep \geq 0$ such
that \leqn{equivA.14}{ 2^{-p\delta{j-1}}\norm{S_j(\phi^\ep_j
u)}^p_{L^p} \leq \left\{\begin{array}{ll}
C\bigl(\norm{u}^p_{L^p(B_{4/\ep})} + \norm{u}^p_{L^p(a_{8/\ep})}
\bigr) &
\text{if $j=1$} \\
C2^{-p\delta(j-1)} 2^{n(1-j)} \sum_{k=0}^2 \norm{u}^p_{L^p(a_{2^{j-1+k}/\ep})}
& \text{if $j\geq 2$}
\end{array} \right. \, .
} Therefore \lalign{equivA.15}{ \norm{\phi^\ep_0& u}^p_{L^p}
+\sum_{j=1}^{\infty} 2^{-p\delta(j-1)}\norm{S_{j}(\phi^\ep_j u)}^p_{L^p}\notag \\
&\leq C\bigl( \norm{\sigma_\ep^{-\delta-n/p} u}^p_{L^p(B_{4/\ep})}
+\sum_{j=1}^{\infty} 2^{n(j-1)}\norm{S_{j}
(\sigma_\ep^{-\delta-n/p}u)}^p_{L^p(a_{4/\ep})}\bigr)
&& \text{by \eqref{equivA.7}, \eqref{equivA.12}, and \eqref{equivA.14}}
\notag \\
& \leq C\norm{u}^p_{L^p_{\delta,\ep}} \notag && \text{ by \eqref{equivA.3}}
}
where $C>0$ is a constant independent of $\ep \geq 0$. The proof then
follows from this inequality and \eqref{equivA.11}.
\end{proof}
The above lemma shows that the norm \eqn{eqnormA}{
\nnorm{u}_{L^p_{\delta,\ep}}^p := \norm{\phi^\ep_0 u}^p_{L^p} +
\sum_{j=1}^{\infty} 2^{-p\delta(j-1)} \norm{S_j(\phi^\ep_j
u)}^p_{L^p} } is equivalent for $1\leq p < \infty$, independent of
$\ep \geq 0$, to the weighted norm $\norm{u}_{L^p_{\delta,\ep}}$.
For $p=\infty$, the appropriate norm is \eqn{eqnormB}{
\nnorm{u}_{L^\infty_{\delta,\ep}} := \sup\{\norm{\phi^\ep_0
u}_{L^\infty}, 2^{-\delta(j-1)}\norm{\phi_{j}^\ep u}_{L^\infty}
\;\; (j\geq 1) \} } and it is easy to see that there exists a
constant $C>0$ independent of $\ep \geq 0$  such that
\eqn{eqnormC}{ \frac{1}{C} \norm{u}_{L^\infty_{\delta,\ep}} \leq
\nnorm{u}_{L^\infty_{\delta,\ep}} \leq C
\norm{u}_{L^\infty_{\delta,\ep}} }

The same arguments used in proving the previous lemma can be used
to establish the following generalization.
\begin{lem} \label{equivB} \mnote{[equivB]}
For $1\leq p < \infty$, let \leqn{equivB.1}{
\nnorm{u}_{W^{k,p}_{\delta,\ep}}^p := \norm{\phi^\ep_0
u}^p_{W^{k,p}} + \sum_{j=1}^{\infty} 2^{-p\delta(j-1)}
\norm{S_j(\phi^\ep_j u)}^p_{W^{k,p}} \, , } and for $p=\infty$ let
\leqn{equivB.2}{\nnorm{u}_{W^{k,\infty}_{\delta,\ep}} :=
\sup\{\norm{\phi^\ep_0
u}_{W^{k,\infty}},2^{-\delta(j-1)}\norm{S_j(\phi^\ep_j
u)}_{W^{k,\infty}} \; (j\geq 1) \}\, . }Then there exists a
constant $C>0$ independent of $\ep \geq 0$ such that
\eqn{equivB.3}{\frac{1}{C} \norm{u}^{p}_{W^{k,p}_{\delta,\ep} }
\leq \nnorm{u}^p_{W^{k,p}_{\delta,\ep}} \leq C
\norm{u}^{p}_{W^{k,p}_{\delta,\ep}}\, . }
\end{lem}
For the remainder of this section, we will use the two equivalent
norms $\norm{\cdot}_{W^{k,p}_{\delta,\ep}}$ and
$\nnorm{\cdot}_{W^{k,p}_{\delta,\ep}}$ interchangeably and refer
to both using the notation $\norm{\cdot}_{W^{k,p}_{\delta,\ep}}$.
From \eqref{equivB.1}, it follows that there exist
a constant $C>0$ independent of $\ep \geq 0$ such that
\leqn{inclA}{ \norm{u}_{W^{k_2,p}_{\delta_2,\ep}} \leq
C\norm{u}_{W^{k_1,p}_{\delta_1,\ep}} \quad \text{whenever $k_2\leq
k_1$ and $\delta_1 \leq \delta_2$.} } Thus we have the inclusion
$W^{k_1,p}_{\delta_1,\ep}\subset W^{k_2,p}_{\delta_2,\ep}$ for
$k_2\leq k_1$ and $\delta_1 \leq \delta_2$.
The representation \eqref{equivB.1} is particularly useful for
extending estimates from the usual Sobolev spaces $W^{k,p}_\delta$
to the weighted ones $W^{k,p}_{\delta,\ep}$ $(\ep > 0)$ as the
next lemma shows. It also makes clear the philosophy behind
deriving weighted Sobolev inequalities which is to derive global estimates
from scaling and local Sobolev inequalities \cite{Bart86}.

We remark that the norm $\nnorm{\cdot}_{W^{k,p}_{\delta,1}}$, as
an alternate representation for the standard weighted norms
$\norm{\cdot}_{W^{k,p}_{\delta,1}}$, was introduced by Maxwell in
\cite{Max04}. There he used the norm to define the weighted
Sobolev spaces for non-integral $k$ (see also \cite{BrKa}). Here
we will only be interested in integral $k$.
\begin{lem} \label{ciA} \mnote{[ciA]}
Suppose $\ep_0 > 0$ and for all $u\in C^\infty(\Rbb^n,V)$, $u
\mapsto F_1(u)$ is a map that satisfies  \alin{ciA.1}{
\phi^\ep_0 F_1(u) & = \phi^\ep_0 F_1((\phi^\ep_0 + \phi^\ep_1)u)\, , \\
\phi^\ep_j F_1(u) & = \phi^\ep_j F_1\Bigl(\sum_{k=-1}^{1}\phi^\ep_{j+k}u\Bigr) \quad (j\geq 1)\, , \\
S_{j} F_1(u) &= 2^{-(j-1)\lambda} F_{1}(S_j u) \quad (j\geq 1) \,
, } where the $F_{\alpha}$  $(\alpha= 2,3,4,5)$ are linear
operators on $V$.

\begin{itemize}
\item[\textbf{(i)}] If there is an estimate of the form
\eqn{ciA.3a}{ \norm{F_1(u)} _{W^{k_1,p_1}} \leq C_1
\norm{F_2(u)}_{W^{k_2,p_2}} } where $p_1\geq p_2$, then
\eqn{ciA.4a}{ \norm{F_1(u)}_{W^{k_1,p_1}_{\delta_1,\ep}} \leq
C\norm{F_2(u)}_{W^{k_2,p_2}_{\delta_2,\ep}} } for some constant
$C>0$ independent of $\ep \in [0,\ep_0]$ provided
$\delta_1+\lambda \geq \delta_2$. \item[\textbf{(ii)}] If there
exists an estimate of the form \eqn{ciA.3}{
\norm{F_1(u)}_{W^{k_1,p_1}} \leq C_1\norm{F_2(u)}_{W^{k_2,p_2}}
\norm{F_3(u)}_{W^{k_3,p_3}} +
C_2\norm{F_2(u)}_{W^{k_4,p_4}}\norm{F_1(u)}_{W^{k_5,p_5}} } where
$\displaystyle{\frac{1}{p_1}= \frac{1}{p_2}+\frac{1}{p_3} =
\frac{1}{p_4} + \frac{1}{p_5}}$ $(1\leq p_1\leq p_\alpha \leq
\infty \quad \alpha=2,3,4,5)$, then \eqn{ciA.4}{
\norm{F_1(u)}_{W^{k_1,p_1}_{\delta_1,\ep}} \leq
C\bigl(C_1\norm{F_2(u)}_{W^{k_2,p_2}_{\delta_2,\ep}}
\norm{F_3(u)}_{W^{k_3,p_3}_{\delta_3,\ep}} +
C_2\norm{F_4(u)}_{W^{k_4,p_4}_{\delta_4,\ep}}\norm{F_5(u)}_{W^{k_5,p_5}_{\delta_5,\ep}}
\bigr)} for some constant $C>0$ independent of $\ep \in [0,\ep_0]$
provided $\delta_1 +\lambda \geq \max\{ \delta_2+\delta_3 ,
\delta_4+\delta_5\}$.
\end{itemize}
\end{lem}
\begin{proof}
We only proof part (ii) for $1\leq p_\alpha < \infty$. Part (i)
can be proved in a similar manner using the inequality 
\leqn{useinq}{(\sum_j
a^p_j)^{1/p} \leq (\sum_j a^q_j)^{1/q} \quad \text{for $a_j\geq 0$ and $
0<q\leq
p$} } instead of H\"{o}lder's and Minkowski's inequalities. See also
the proof of theorem 1.2 in \cite{Bart86}.

Recall H\"{o}lder's and Minkowski's inequalities which state that
for $1\leq p \leq q\leq r < \infty$, $1/p=1/q+1/r$ and any two
sequences $a_j,b_j \geq 0$ that the following holds \leqn{hinq}{
\Bigl(\sum_j a_j^{p} b_j^{p}\Bigr)^{1/p} \leq \Bigl(\sum_j
a_j^{q}\Bigr)^{1/q}\Bigl(\sum_j b_j^{r}\Bigr)^{1/r} }
and 
\leqn{minq}{
\Bigl(\sum_j(a_j+b_j)^p\Bigr)^{1/p} \leq \Bigl(\sum_j
a_j^p\Bigr)^{1/p} + \Bigl(\sum_j a_j^p \Bigr)^{1/p} \, . } Next,
suppose $j\geq 2$. Then \alin{ciA.5}{ \norm{S_j(\phi^\ep_j
F_1(u)&)}^{p_1}_{W^{k_1,p_1}} = \Bnorm{\phi^\ep_1 S_j
F_1\Bigl(\sum_{k=-1}^1\phi^\ep_{j+k}u)\Bigr)}^{p_1}_{W^{k_1,p_1}}
\leq C 2^{-(1-j)p_1\lambda} \Bnorm{
F_1\Bigl(\sum_{k=-1}^1S_{j}\phi^\ep_{j+k}u\Bigr)}^{p_1}_{W^{k_1,p_1}} \\
&\leq C
2^{-(j-1)p_1\lambda}\Bigl(C_1
\Bnorm{F_2\Bigl(\sum_{k=-1}^1S_{j}\phi^\ep_{j+k}u\Bigr)}_{W^{k_2,p_2}}
\Bnorm{F_3\Bigl(\sum_{k=-1}^1S_{j}\phi^\ep_{j+k}u\Bigr)}_{W^{k_3,p_3}}\\
&
\quad+C_2\Bnorm{F_4\Bigl(\sum_{k=-1}^1S_{j}\phi^\ep_{j+k}u\Bigr)}_{W^{k_4,p_4}}
\Bnorm{F_5\Bigl(\sum_{k=-1}^1S_{j}\phi^\ep_{j+k}u\Bigr)}_{W^{k_5,p_5}}
\Bigr)^{p_1}} where $C>0$  is a constant independent of $\ep\geq
0$. Note that in deriving this, we have used the fact that
$\norm{\phi_1^\ep}_{W^{k_1,\infty}}$ is bounded for $\ep \in
[0,\ep_0]$.
From the above inequality, we see that \alin{ciA.6}{
&2^{-\delta_1 p_1(j-1)}\norm{S_j(\phi^\ep_j
F_1(u))}^{p_1}_{W^{k_1,p_1}}  \leq \\ \;\; & C \Bigl(C_1
2^{-\delta_2(j-1)}\Bigl(\sum_{k=-1}^1\norm{F_2(S_{j+k}(\phi^\ep_{j+k}u))}_{W^{k_2,p_2}}\Bigr)
2^{-\delta_3(j-1)}\Bigl(\sum_{k=-1}^1\norm{F_3(S_{j+k}(\phi^\ep_{j+k}u))}_{W^{k_3,p_3}}\Bigr)\\
&+ C_2
2^{-\delta_4(j-1)}\Bigl(\sum_{k=-1}^1\norm{F_4(S_{j+k}(\phi^\ep_{j+k}u))}_{W^{k_4,p_4}}\Bigr)
2^{-\delta_5(j-1)}\Bigl(\sum_{k=-1}^1\norm{F_5(S_{j+k}(\phi^\ep_{j+k}u))}_{W^{k_5,p_5}}\Bigr)
\Bigr)^{p_1}} where we have used $\delta_1+\lambda$ $\geq$ $\max
\{ \delta_2+\delta_3,\delta_4+\delta_5\}$. The above inequality
along with \eqref{hinq} and \eqref{minq} imply
\alin{ciA.7}{ &\Bigl(\sum_{j=1}^\infty 2^{-\delta_1
p_1(j-1)}\norm{S_j(\phi^\ep_j
F_1(u))}^{p_1}_{W^{k_1,p_1}}\Bigr)^{1/p_1} \leq
\\ \;\; & C \Bigl(C_1 \Bigl(\sum_{j=1}^\infty2^{-\delta_2
p_2(j-1)}
\norm{F_2(S_{j}(\phi^\ep_{j}u))}_{W^{k_2,p_2}}^{p_2}\Bigr)^{1/p_2}
\Bigl(\sum_{j=1}^\infty 2^{-\delta_3 p_3(j-1)}
\norm{F_3(S_{j}(\phi^\ep_{j}u))}_{W^{k_3,p_3}}^{p_3}\Bigr)^{1/p_3}\\
&+ C_2 \Bigl(\sum_{j=1}^\infty 2^{-\delta_4 p_4(j-1)}
\norm{F_4(S_{j}(\phi^\ep_{j}u))}_{W^{k_4,p_4}}^{p_4}\Bigr)^{1/p_4}
\Bigl(\sum_{j=1}^\infty 2^{-\delta_5 p_5(j-1)}
\norm{F_5(S_{j}(\phi^\ep_{j}u))}_{W^{k_5,p_5}}^{p_5}\Bigr)^{1/p_5}\Bigr)
} and hence \lalign{ciA.8}{ \Bigl(\sum_{j=1}^\infty 2^{-\delta_1
p_1(j-1)}\norm{S_j(\phi^\ep_j F_1(u))}^{p_1}_{W^{k_1,p_1}}\Bigr)^{1/p_1}
&\leq
C\bigl(C_1\norm{F_2(u)}_{W^{p_2,k_2}_{\delta_2,\ep}}
\norm{F_3(u)}_{W^{p_3,k_3}_{\delta_3,\ep}} \notag\\ \quad &
+
 C_2\norm{F_4(u)}_{W^{p_4,k_4}_{\delta_4,\ep}}
\norm{F_5(u)}_{W^{p_5,k_5}_{\delta_5,\ep}}\bigr)\label{ciA.8.1}\,
.} Similar arguments show that \lalign{ciA.9}{ \Bigl(\norm{\phi^\ep
F_1(u)}_{W^{k_1,p_1}_{\delta_1,\ep} }^{p_1} +
\norm{S_1(\phi^\ep_1F_1(u))}_{W^{k_1,p_1}_{\delta_1,\ep} }^{p_1}
\Bigr)^{1/p_1} &\leq
C\bigl(C_1\norm{F_2(u)}_{W^{p_2,k_2}_{\delta_2,\ep}}
\norm{F_3(u)}_{W^{p_3,k_3}_{\delta_3,\ep}} \notag\\ \quad &
+
 C_2\norm{F_4(u)}_{W^{p_4,k_4}_{\delta_4,\ep}}
\norm{F_5(u)}_{W^{p_5,k_5}_{\delta_5,\ep}}\bigr)\label{ciA.9.1}\,
}for some constant $C>0$ independent of $\ep \in [0,\ep_0]$. The
proof now follows from the two inequalities \eqref{ciA.8.1} and
\eqref{ciA.9.1}.
\end{proof}
The next lemma is a variation of the previous one and can be
proved in the same fashion.
\begin{lem} \label{ciB} \mnote{[ciB]}
Suppose $\ep_0>0$ and for all $u\in C^\infty(\Rbb^n,V)$, $u
\mapsto F_1(u)$ is a map that satisfies \alin{ciB.1}{
\phi^\ep_0 F_1(u) & = \phi^\ep_0 F_1((\phi^\ep_0 + \phi^\ep_1)u)\, , \\
\phi^\ep_j F_1(u) & = \phi^\ep_j F_1\Bigl(\sum_{k=-1}^{1}\phi^\ep_{j+k}u\Bigr) \quad (j\geq 1)\, , \\
S_{j} F_1(u) &= 2^{-(j-1)\lambda} F_{1}(S_j u) \quad (j\geq 1) \,
, } where \eqn{ciB.2}{ F_{2} = \Der P_2 \, ,\quad F_{3} = P_3\, ,
\quad F_{4} = \Der P_4\, , \quad \text{and}\quad F_{5} = P_5\, ,}
and $P_{\alpha}$ $(\alpha= 2,3,4,5)$ are linear operators on $V$.

\begin{itemize}
\item[\textbf{(i)}] If there exists an estimate of the form
\eqn{ciB.3a}{\norm{F_1(u)}_{W^{k_1,p_1}} \leq C_1\norm{
F_2(u)}_{W^{k_2,p_2}}} where $p_1\geq p_2$, then there exists a
constant $C>0$ independent of $\ep \in [0,\ep_0]$ such that
\eqn{ciB.4a}{\norm{F_1(u)}_{W^{k_1,p_1}_{\delta_1,\ep}} \leq
C\bigl(\norm{F_2(u)}_{W^{k_2,p_2}_{\delta_2-1,\ep}}+\ep \norm{P_2
u}_{W^{k_2,p_2}_{\delta_2,\ep}}\bigr) } provided $\delta_1+\lambda
\geq \delta_2$. \item[\textbf{(ii)}] If there exists an estimate
of the form \eqn{ciB.3}{ \norm{F_1(u)}_{W^{k_1,p_1}} \leq
C_1\norm{ F_2(u)}_{W^{k_2,p_2}}\norm{F_3(u)}_{W^{k_3,p_3}} +
C_2\norm{F_2(u)}_{W^{k_4,p_4}}\norm{F_1(u)}_{W^{k_5,p_5}} } where
$\displaystyle{\frac{1}{p_1}= \frac{1}{p_2}+\frac{1}{p_3} =
\frac{1}{p_4} + \frac{1}{p_5}}$ $(1\leq p_1\leq p_\alpha \leq
\infty \quad \alpha=2,3,4,5)$, then \alin{ciB.4}{
\norm{F_1(u)}_{W^{k_1,p_1}_{\delta_1,\ep}} \leq
C\bigl(C_1&\bigl(\norm{F_2(u)}_{W^{k_2,p_2}_{\delta_2-1,\ep}}+\ep
\norm{P_2 u}_{W^{k_2,p_2}_{\delta_2,\ep}} \bigr)
\norm{F_3(u)}_{W^{k_3,p_3}_{\delta_3,\ep}}  \\  &+
C_2\bigl(\norm{F_4(u)}_{W^{k_4,p_4}_{\delta_4-1,\ep}}+\ep\norm{P_4
u}_{W^{k_4,p_4}_{\delta_4,\ep}} \bigr)
\norm{F_5(u)}_{W^{k_5,p_5}_{\delta_5,\ep}} \bigr)} for some
constant $C>0$ independent of $\ep\in [0,\ep_0]$ provided
$\delta_1 +\lambda$ $\geq$ $\max\{\delta_2+\delta_3,
\delta_4+\delta_5\}$.
\end{itemize}
\end{lem}
\begin{rem}
\emph{By using the generalized H\"{o}lder's inequality, part (ii)
of lemmas \ref{ciA} and \ref{ciB} can be extended in the obvious
fashion if there exists estimates of the form \eqn{nest}{
\norm{F_1(u)}_{W^{k_1,p_1}} \leq C
\norm{F_2(u)}_{W^{k_2,p_2}}\norm{F_{3}(u)}_{W^{k_3,p_3}}\cdots\norm{F_{N}(u)}_{W^{k_N,p_N}}}
where $\frac{1}{p_1}$ $=$ $\sum_{i=2}^{N} \frac{1}{p_{i}}$ $(1\leq
p_1 \leq p_i \leq \infty)$, $F_1$ is as in lemma \ref{ciA}, and
$F_{i}$ $(i\geq 2)$ are of the form $F_{i} = P_i$ or $F_{i} = D
P_i$ with $P_i$ a linear operator on $V$.}
\end{rem}
We will now use these two lemmas to extend various inequalities
from the standard Sobolev spaces to the weighted ones. All of
these inequalities have been derived before by various authors,
see for example \cite{Bart86,BrKa,CBC,ChrDel03,Max04,Oli05c}. The
new aspect here is that we show that the constants in the
inequalities are independent of $\ep \geq 0$ and hence we find
inequalities that interpolate between the weighted ($\ep > 0$) and
the standard ones ($\ep = 0$). We begin with a weighted H\"{o}lder
inequality.
\begin{lem}\label{ciC}
Suppose $\ep_0 > 0$, $\delta_1 = \delta_1+\delta_2$ and
$\displaystyle{\frac{1}{p_1}= \frac{1}{p_2}+\frac{1}{p_3}}$. Then
there is a constant $C>0$ independent of $\ep\in [0,\ep_0]$ such that
\eqn{ciC.1}{\norm{uv}_{L^{p_1}_{\delta_1,\ep}} \leq C
\norm{u}_{L^{p_2}_{\delta_2,\ep}}\norm{v}_{L^{p_3}_{\delta_3,\ep}}
} for all $u\in L^{p_2}_{\delta_2,\ep}$ and $v\in
L^{p_3}_{\delta_3,\ep}$.
\end{lem}
\begin{proof}
Follows directly from H\"{o}lder's inequality and lemma \ref{ciA}.
\end{proof}
Next, we consider weighted versions of the Sobolev inequalities.
\begin{lem} \label{ciD}
$\;$\\
\begin{itemize}
\item[\textbf{(i)}] For $\ep_0 > 0$ and $k>n/p$ there exists a
constant $C>0$ independent of $\ep \in [0,\ep_0]$ such that
\eqn{ciD.1}{ \norm{u}_{L^\infty_{\delta,\ep}} \leq C
\norm{u}_{W^{k,p}_{\delta,\ep}} } for all $u\in
W^{k,p}_{\delta,\ep}$. Moreover $u\in C^{0}_{\delta,\ep}$ and for
$\ep > 0$, $u(x) = \text{o}(|x|^\delta)$ as $|x|\rightarrow
\infty$. \item[\textbf{(ii)}] For $\ep_0>0$ and $1\leq p < n$
there exists a constant $C>0$ independent of $\ep \in [0,\ep_0] $
such that \eqn{ciD.2}{\norm{u}_{L^{np/(n-p)}_{\delta,\ep}} \leq C
\bigl(\norm{\Der u}_{L^p_{\delta-1,\ep}} +\ep
\norm{u}_{L^p_{\delta,\ep}} \bigr)} for all $u\in
W^{1,p}_{\delta,\ep}$.
\end{itemize}
\end{lem}
\begin{proof}
\noindent \textbf{(i)} The estimate $\norm{u}_{L^\infty_{\delta,\ep}} \leq C
\norm{u}_{W^{k,p}_{\delta,\ep}}$ for some constant $C>0$ independent
of $\ep\geq 0$ follows from the usual Sobolev inequality
$\norm{u}_{L^\infty} \leq C
\norm{u}_{W^{k,p}}$ $(k>n/p)$ and lemma \ref{ciA}. Since
$\norm{\cdot}_{W^{k,p}_{\delta,\ep}}$ for $\ep > 0$ is equivalent to
$\norm{\cdot}_{W^{k,p}_{\delta,1}}$, the statement
 $u(x) = \text{o}(|x|^\delta)$ as $|x|\rightarrow
\infty$ for $\ep >0$ follows from theorem 1.2 in \cite{Bart86}.
\\

\noindent \textbf{(ii)} Follows from lemma \ref{ciB} and the
Sobolev inequality $\norm{u}_{L^{np/(n-p)}} \leq C \norm{\Der
u}_{L^p} $ which holds for all $u\in W^{1,p}$ where $1\leq p < n$.
\end{proof}
In addition to the Sobolev inequalities, we will also require
weighted versions of the multiplication and Moser inequalities.
We first consider the multiplication inequalities.

\begin{lem} \label{ciG} \mnote{ciG}
Suppose $\ep_0>0$, $1\leq p <\infty$, $k_1,k_2\geq k_3$, $k_3 <
k_1+k_2 -n/p$, $\delta_1+\delta_2 \leq \delta_3$, and $V_1\times
V_2 \rightarrow V_3\; :$ $(u,v) \mapsto uv$ is a
multiplication. Then there exists a constant $C>0$ independent of
$\ep \in [0,\ep_0]$ such that \eqn{ciG1}{ \norm{u
v}_{W^{k_3,p}_{\delta_3,\ep}} \leq C
\norm{u}_{W^{k_1,p}_{\delta_1,\ep}}\norm{v}_{W^{k_2,p}_{\delta_2,\ep}}
} for all $u \in W^{k_1,p}_{\delta_1,\ep}$ and $v\in
W^{k_2,p}_{\delta_2,\ep}$.
\end{lem}
\begin{proof} This proof does not follow directly from lemma \ref{ciA},
but can be proved in a simlar fashion. To see this first
recall the Sobolev mlutiplication inequality
\leqn{mulinq}{\norm{uv}_{W^{k_3,p}}
\leq C \norm{u}_{W^{k_1,p}}\norm{v}_{W^{k_2,p}}
}
which holds
for  $1\leq p <\infty$, $k_1,k_2\geq k_3$, and $k_3 <
k_1+k_2 -n/p$.
So
\alin{ciG.1}{
 &\norm{uv}_{W^{k_3,p}_{\delta_3}} = \bigl(\norm{\phi_0^\ep uv}^p_{W^{k_3,p}}
+ \sum_{j=1}^\infty 2^{-p\delta_3(j-1)}\norm{S_j(\phi_j^\ep uv)}_{W^{k_3,p}}^p \bigr)^{1/p} \\
& \leq C\bigl(\norm{\phi_0^\ep u(\phi^\ep_0+\phi^\ep_1)v}^p_{W^{k_1,p}}
+ \sum_{j=1}^\infty 2^{-p\delta_3(j-1)}\norm{S_j(\phi_j^\ep u)S_j(\sum_{k=-1}^{1}\phi^{\ep}_{j+k}v)}_{W^{k_1,p}}^p \bigr)^{1/p} 
\\
& \leq C\bigl(\norm{\phi_0^\ep u (\phi^\ep_0+\phi^\ep_1)v }_{W^{k_3,p}}^{p/2}
+ \sum_{j=1}^\infty 2^{-(p/2)\delta_3(j-1)}\norm{S_j(\phi_j^\ep u)S_j(\sum_{k=-1}^{1}\phi^{\ep}_{j+k}v)}_{W^{k_1,p}}^{p/2} 
\bigr)^{2/p} \\
&\leq C\bigl(\norm{\phi^\ep_0 u}_{W^{k_1,p}}^{p/2}\norm{\phi^\ep_0 v}_{W^{k_2,p}}^{p/2} 
+ \sum_{j=1}^\infty 2^{-(p/2)\delta_1(j-1)} \norm{S_j(\phi_j^\ep u)}_{W^{k_1,p}}^{p/2}
2^{-(p/2)\delta_2(j-1)} \norm{S_j(\phi^{\ep}_{j}v)}_{W^{k_2,p}}\bigr)^{2/p}
\\
&\leq C\bigl(\norm{\phi_0^\ep u}^p_{W^{k_1,p}}
+ \sum_{j=1}^\infty 2^{-p\delta_1(j-1)}\norm{S_j(\phi_j^\ep u)}_{W^{k_1,p}}^p \bigr)^{1/p} 
\bigl(\norm{\phi_0^\ep v}^p_{W^{k_2,p}}
+ \sum_{j=1}^\infty 2^{-p\delta_2(j-1)}\norm{S_j(\phi_j^\ep v)}_{W^{k_2,p}}^p \bigr)^{1/p} 
\\
&\leq C\norm{u}_{W^{k_1,p}_{\delta_1}} \norm{v}_{W^{k_2,p}_{\delta_2}}  }
where in deriving the third, fourth, and fifth lines we used 
\eqref{useinq},
\eqref{mulinq}, and \eqref{hinq}, respectively. 
\end{proof}

\begin{lem} \label{ciE} \mnote{ciE}
$\;$
\begin{itemize}
\item[\textbf{(i)}] If $\ep_0>0$  and $\delta_1$ $\geq$ $\max\{\delta_2 +
\delta_3,\delta_4 +\delta_5\}$, then there exists a constant $C>0$
independent of $\ep \in [0,\ep_0]$ such that
\eqn{ciE.1}{\norm{uv}_{H^k_{\delta_1,\ep}} \leq
C\bigl(\norm{u}_{H^k_{\delta_2,\ep}}
\norm{v}_{L^\infty_{\delta_3,\ep}} + \norm{v}_{H^k_{\delta_4,\ep}}
\norm{u}_{L^\infty_{\delta_5,\ep}}\bigr)} for all $u\in
H^k_{\delta_2,\ep}\cap L^\infty_{\delta_5,\ep}$ and $v\in
H^k_{\delta_4,\ep}\cap L^\infty_{\delta_3,\ep}$.
\item[\textbf{(ii)}] If $\ep_0>0$ and $\delta_1$ $\geq$ $\max\{\delta_2 +
\delta_3,\delta_4 +\delta_5\}$, then there exists a constant $C>0$
independent of $\ep \in [0,\ep_0]$ such that
\eqn{ciE.2}{\norm{[\Der^{\Ic},u]v}_{L^2_{\delta_1-|I|,\ep}} \leq
C\bigl(\bigl(\norm{\Der
u}_{H^{k-1}_{\delta_2-1,\ep}}+\ep\norm{u}_{L^2_{\delta_2,\ep}}\bigr)\norm{v}_{L^\infty_{\delta_3,\ep}} +
\bigl(\norm{\Der u}_{L^\infty_{\delta_4-1,\ep}}+\ep\norm{u}_{L^\infty_{\delta_4,\ep}}\bigr)
\norm{v}_{H^{k-1}_{\delta_5,\ep}}\bigr)} for all $|\Ic|\leq k$,
$u\in H^{k}_{\delta_2,\ep}\cap W^{1,\infty}_{\delta_4,\ep}$ and
$v\in  H^{k-1}_{\delta_5,\ep}\cap L^\infty_{\delta_3,\ep}$.
\item[\textbf{(iii)}] Suppose $\ep_0>0$, $F \in
C^{\ell}(V,\Rbb^m)$ is a map that satisfies $\Der F\in
C^{k-1}_b(V,\Rbb^m)$, and $1\leq |\Ic|\leq k$. Then there exists a
$C>0$ independent of $\ep \in [0,\ep_0]$ such that
\eqn{ciE.3}{\norm{\Der^\Ic F(u)}_{L^2_{\delta-|\Ic|,\ep}} \leq C
\norm{\Der F}_{C^{k-1}_b} \norm{u}^{k-1}_{L^\infty}\bigl(\norm{\Der
u}_{H^{k-1}_{\delta-1,\ep}}+\ep\norm{u}_{L^2_{\delta,\ep}}\bigr) } for all $u \in H^k_{\delta,\ep}\cap
L^\infty$. \item[\textbf{(iv)}] Suppose $\ep_0>0$ and  $F\in
C^{k}_{b}(V,\Rbb^m)$. Then there exists a $C>0$ independent of
$\ep \in [0,\ep_0]$ such that \eqn{ciE.4}{
\norm{F(u)}_{H^k_{\delta,\ep}} \leq
C\norm{F}_{C^k_b}(1+\norm{u}^{k-1}_{L^\infty})\norm{u}_{H^k_{\delta,\ep}}
} for all $u\in H^{k}_{\delta,\ep}\cap L^\infty$.
\end{itemize}
\end{lem}
\begin{proof} Inequalities $(i)-(iv)$ follow directly from
\eqref{inclA}, lemmas \ref{ciA} and \ref{ciB}, and the following standard
Sobolev inequalities:
\begin{itemize}
\item[\textbf{(i)}] $\norm{uv}_{H^k} \leq C\bigl(\norm{u}_{H^k}
\norm{v}_{L^\infty} + \norm{v}_{H^k} \norm{u}_{L^\infty}\bigr) $
for all $u\in H^k\cap L^\infty$ and $v\in H^k\cap L^\infty$.
\item[\textbf{(ii)}] $\norm{[\Der^{\Ic},u]v}_{L^2} \leq
C\bigl(\norm{\Der u}_{H^{k-1}}\norm{v}_{L^\infty} + \norm{\Der
u}_{L^\infty}
 \norm{v}_{H^{k-1}}\bigr)$ for all $|\Ic|\leq k$,
$u\in H^{k}\cap W^{1,\infty}$ and
$v\in  H^{k-1}\cap L^\infty$.
\item[\textbf{(iii)}] Suppose $F \in C^{\ell}(V,\Rbb^m)$
is a map that satisfies $\Der F\in C^{k-1}_b(V,\Rbb^m)$ and $1\leq
|\Ic|\leq k$. Then
$\norm{\del^\Ic
F(u)}_{L^2} \leq C \norm{\Der F}_{C^{k-1}_b}
\norm{u}^{k-1}_{L^\infty}\norm{\Der
u}_{H^{k-1}}$
for all $u \in H^k\cap L^\infty$.
\item[\textbf{(iv)}] Suppose $F\in C^{k}_{b}(V,\Rbb^m)$.
Then
$\norm{F(u)}_{H^k}
\leq C\norm{F}_{C^k_b}(1+\norm{u}^{k-1}_{L^\infty})\norm{u}_{H^k}$
for all $u\in H^{k}\cap L^\infty$.
\end{itemize}
Note that we have used $\norm{\cdot}_{L^\infty_{0,\ep}}
= \norm{\cdot}_{L^\infty}$.
\end{proof}
In addition to the Moser inequalities, we also need to know when
the map $u \mapsto F(u)$ is locally Lipschitz on $H^{k}_{\delta}$.
\begin{lem}\label{Lip} \mnote{Lip} Suppose $\ep_0>0$, $F \in
C^{\ell}_{b}(V,\Rbb)$, $F(0)=0$, $\delta \leq 0$, and $k \leq \ell
$, and $k>n/2$. Then for each $R > 0$ there exists a $C > 0$
independent of $\ep \in [0,\ep_0]$ such that \eqn{Lip1}{
\norm{F(u_1)-F(u_2)}_{H^{k}_{\delta,\ep}} \leq
C\norm{u_1-u_2}_{H^{k}_{\delta,\ep}} \quad \text{for all
$u_1,u_2\in B_{R}(H^{k}_{\delta,\ep})$.} }
\end{lem} \begin{proof}
See the proof of lemma B.6 in \cite{Oli05c}.
\end{proof}

We conclude this section with a lemma comparing the norms
$\norm{\cdot}_{L^{p}_{\delta}}$ and
$\norm{\cdot}_{L^{p}_{\delta,\ep}}$.
\begin{lem} \label{ciF} \mnote{[ciF]}
$\;$\\
\begin{itemize}
\item[(i)] If $\delta \leq
-n/p$, $1\leq p \leq \infty$, and  $0\leq \ep \leq 1$, then \eqn{ciF.2}{
\ep^{-\delta-n/p}\norm{u}_{L^p_{\delta}}
\leq \norm{u}_{L^p_{\delta,\ep}}
\leq   \norm{u}_{L^p_{\delta}} }
for
all $u\in L^p_{\delta}$. \item[(ii)] If $-n/p < \delta$,
$1\leq p <\infty$,  and $0< \ep \leq 1$,
then
\eqn{ciF.3}{
\norm{u}_{L^p_{\delta}}
\leq \norm{u}_{L^p_{\delta,\ep}}
\leq  \ep^{-\delta-n/p} \norm{u}_{L^p_{\delta}}
} for all $u\in L^p_{\delta}$.
\end{itemize}
\end{lem}
\begin{proof}
\noindent
\noindent \textbf{(i)} By assumption $0\leq \ep \leq 1$, and
so we have $\ep \sigma_1(x) \leq \sigma_{\ep}(x) \leq \sigma_{1}(x)$
for all $x\in \Rbb^n$. By assumption
$-\delta-n/p > 0$ and so we
get $\ep^{-\delta-n/p}\sigma_1^{-\delta-n/p}
\leq \sigma_\ep^{-\delta-np} \leq \sigma_1^{-\delta-np}$.
Therefore, directly from the definition of the weighted
norm, we find
$\ep^{-\delta-n/p}\norm{u}_{L^p_{\delta}}
\leq \norm{u}_{L^p_{\delta,\ep}}
\leq   \norm{u}_{L^p_{\delta}}$.
Part \textbf{(ii)} is proved in a similar fashion.
\end{proof}

\sect{hbloc}{Quasilinear symmetric hyperbolic systems}

In this section we establish a local existence and uniqueness
theorem for a particular form of quasilinear symmetric hyperbolic
system on the weighted Sobolev spaces $H^k_\delta$.
In \cite{Oli05c}, we proved a local existence and uniqueness theorem
for quasilinear parabolic systems on the $H^k_\delta$ spaces by
adapting the approach of Taylor \cite{TayIII} (see theorem 7.2, pg
330, and proposition 7.7, pg 334) which is based on using
mollifiers to construct a sequence of approximate solutions and
then showing that the sequence converges to a true solution. Here,
we will again follow the same approach for quasilinear symmetric
hyperbolic systems and adapt the local existence and uniqueness
theorems of Taylor (see proposition 2.1, pg 370) to work on the
weighted Sobolev spaces. We will only provide a brief sketch of
the proof since the proof is very similar to the one in
\cite{Oli05c} and the details can easily be filled in by the
reader. Related existence results have
been derived independently in \cite{BrKa} using a different
method.

The hyperbolic equations that we will consider are of the form
\lalign{hb}{ &b^0(u,v)\partial_{t}v = b^j(u,v)\partial_i v +
f(u,v)v + h
, \label{hb.1}\\
&v|_{t=0}  = v_{0} \label{hb.2}}
where
\begin{itemize} \item[(i)] the map $u=u(t,x)$ is $\Rbb^r$-valued while
the maps
$v=v(t,x)$ and $h=h(t,x)$ are $\Rbb^{m}$-valued,
\item[(ii)] $b^0, b^j, f \in C^{k}_b(\Rbb^r\times\Rbb^m, \Mbb_{m\times m})$
 $(j=1,\ldots,n)$,
\item[(iii)] $b^0$ and $b^j$ $(j=1,\ldots,n)$ are symmetric, and
\item[(iv)] there exists a constant $\omega > 0$ such that
\leqn{bb}{ b^0(\xi_1,\xi_2) \geq \omega \id_{\!m\times m}
\quad \text{for all $(\xi_1, \xi_2) \in \Rbb^r\times\Rbb^{m}$.}}
\end{itemize}

\subsect{galerk}{Galerkin method}

Let $j \in \Co(\Rbb^{n})$ be any function that satisfies $j\geq 0$,
$j(x)=0$ for $|x|\geq 1$,
and $\int_{\Rbb^{n}}j(x)\,d^n x =1$.
Following the standard prescription, we construct from $j$ the
mollifier $j_{\eta}(x) := \eta^{-n}j(x/\eta)$
 $(\eta > 0)$ and the smoothing operator \eqn{smooth}{
J_{\eta}(u)(x) := j_{\epsilon}* u(x) =
\int_{\Rbb^{n}}j_{\eta}(x-y)u(y)\, d^{n}y \, .}

Following Taylor (\cite{TayIII},Ch. 16, sect. 1 \& 2), we first
solve the approximating equation \lalign{approx}{ &
b^0(u,\Je\ve)\partial_t\ve = \Je b^j(u,\Je\ve)\partial_i \Je\ve +
\Je f(u,\Je\ve)\Je \ve + \Je h
\label{approx.1} \\
& \ve|_{t=0} = v_{0}, \label{approx.2} }
and latter show that the solutions $v_\ep$ converge to
a solution of \eqref{hb.1}-\eqref{hb.2} as $\et \rightarrow 0$.

\begin{prop} \label{parA} \mnote{[parA]} Suppose $T_1, T_2 > 0$, $\et > 0$, $\delta\leq \gamma \leq 0$, $k>n/2$,
$v_0\in H^k_{\delta}$, $u\in C^{0}([-T_1,T_2],H^k_\gamma)$, and
$h\in C^{0}([-T_1,T_2],H^{k}_{\delta})$ for some $T>0$. Then there
exists a $T_{*}>0$ $(T_*<T_1,T_2)$ and a unique $v_{\eta}\in
C^{1}((-T_*,T_{*}),H^{k}_{\delta})$ that solves the initial value
problem \eqref{approx.1}-\eqref{approx.2}. Moreover if
 $\sup_{0\leq  t < T_{*}}\norm{v_{\et}(t)}_{H^k_{\delta}} < \infty$
then there exists a $T^{*} \in (T_{*},T_2]$ such that $\ve$
extends to a unique solution on $(-T_*,T^{*})$.\end{prop}
\begin{proof} Fix $\et > 0$ and define
\eqn{parA.1}{F(t,v) := (b^0(u,\Je v))^{-1} (\Je b^j(u,\Je
v)\partial_i \Je v  + \Je f(u,\Je v)\Je v + \Je h) \, . } Then the
approximating equations \eqref{approx.1}-\eqref{approx.2} can be
written as the first order differential equation $\dot{v} = F(v)
\; ; \; v(0)= v_{0}$ on $H^{k}_{\delta}$. If we can show that $F$
is continuous and is Lipshitz in a neighborhood of $v_0$ in
$H^k_{\delta}$, then the proof follows immediately from standard
existence, uniqueness, and continuation theorems for ODEs on
Banach spaces.

To prove that $F$ is locally Lipshitz, we first prove the following lemma.
\begin{lem} \label{parA1} \mnote{[parA1]}
Suppose $\delta\leq  \gamma \leq 0$, $\ell > n/2$ and that $f \in
C^\ell_b(\Rbb^m\times \Rbb^m,\Mbb_{m\times m})$. Then for each
$u\in H^\ell_\gamma$ and $R>0$ there exists a constant $C>0$ such
that \eqn{parA1.1}{ \norm{f(u,v_1)v_1 -
f(u,v_2)v_2}_{H^\ell_{\delta}} \leq
C\norm{v_1-v_2}_{H^\ell_{\delta}}\, } for all $v_1,v_2\in
B_{R}(H^\ell_{\delta})$.
\end{lem}
\begin{proof}
Let f(0,0) = c and $g(x,y) = f(x,y) - c$ so that $g(0,0) =0$. Then
\eqn{parA1.2}{ f(u,v_1)v_1 - f(u,v_2)v_2 = c(v_1-v_2) +
(g(u,v_1)-g(u,v_2))v_1 + g(u,v_2)(v_1-v_2) \, . } Since $\gamma
\leq 0$ and $\ell > n/2$, we get from
lemma \ref{ciG} that \eqn{parA1.3}{ \norm{f(u,v_1)v_1 -
f(u,v_2)v_2 }_{H^\ell_{\delta}} \leq C\bigl( 1 + \norm{
g(u,v_2)}_{H^\ell_{\gamma}})\norm{v_1-v_2}_{H^\ell_{\delta}} +
\norm{v_1}_{H^\ell_{\delta}}\norm{g(u,v_1)-g(u,v_2)}_{H^\ell_{\gamma}}
\, . } By lemma \ref{Lip}, lemmas \ref{ciD} and \ref{ciE}, and
\eqref{inclA}, we get from the above inequality that
\eqn{parA1.4}{ \norm{f(u,v_1)v_1 - f(u,v_2)v_2 }_{H^\ell_{\delta}}
\leq C(\norm{u}_{H^\ell_\gamma},\norm{v_1}_{H^\ell_{\delta}},
\norm{v_2}_{H^\ell_{\delta}})\norm{v_1-v_2}_{H^\ell_{\delta}} }
where $P(y_1,y_2,y_3)$ is a polynomial. This proves the lemma.
\end{proof}
Using lemma A.7 of \cite{Oli05c}, it is not difficult to prove the
following variation of the above lemma.
\begin{lem} \label{parA2} \mnote{[parA2]}
Suppose $\delta \leq \gamma \leq 0$, $\et > 0$, $\ell > n/2$ and
that $f \in C^\ell_b(\Rbb^m\times \Rbb^m,\Mbb_{m\times m})$. Then
for each $u\in H^\ell_\gamma$ and $R>0$ there exist a constant
$C>0$ such that \eqn{parA2.1}{ \norm{f(u,\Je v_1)D\Je v_1 -
f(u,\Je v_2)D\Je v_2}_{H^\ell_{\delta}} \leq
C\norm{v_1-v_2}_{H^\ell_{\delta}}\, } for all $v_1,v_2\in
B_{R}(H^\ell_{\delta})$.
\end{lem}

The proof now follows easily from the above lemmas, lemma A.7 of
\cite{Oli05c}, and the estimates of appendix \ref{winq}, which
show that for any $R>0$ the map $F : ([-T_1,T_2]\times
B_R(H^k_{\delta-1}) \rightarrow H^k_{\delta-1}$ is continuous and
moreover there exists a constant $C>0$ such that
$\norm{F(t,v_1)-F(t,v_2)}_{H^k_{\delta}} \leq
C\norm{v_1-v_2}_{H^k_{\delta}}$ for all $v_1,v_2\in
B_R(H^k_{\delta})$.
\end{proof}

\subsect{energy}{Energy estimates}

Fix $k>n/2+1$. By proposition \ref{parA}, we
have a sequence of solutions $\ve \in
C^{1}([-T(\et),T(\et)],H^{k}_{\delta})$ $(0<T(\et)\leq T_1,T_2)$
to the approximating equation \eqref{approx.1}-\eqref{approx.2}.
The goal is to derive bounds on $\ve$ in the $H^k_\delta$ spaces
independent of $\et$. To do this, we use energy estimates which we
now describe.
\begin{lem} \label{engA} \mnote{[engA]}
Suppose $a^0\in C^1([0,\tau],W^{1,\infty})$,
$a^j \in C^0([0,\tau],W^{1,\infty})$, $f\in
C^0([0,\tau],L^2_\lambda)$ and that $w\in
C^1([0,\tau],L^2_\lambda)$ satisfies the equation
\eqn{engA.1}{ a^0\del_t w = \Je a^j\del_j \Je w + g \, . } Then
there exists a constant $C>0$ independent of $\eta > 0$ such that
\eqn{engA.2}{ \frac{d\,}{dt} \ip{w}{a^0 w}_{L^2_\lambda} \leq
C\bigl[\bigl(1+ \norm{\emph{\Div} a}_{L^\infty} +
\norm{\vec{a}}_{L^\infty}\bigr) \norm{w}_{L^2_\lambda}^2 +
\norm{g}_{L^2_\lambda}\norm{w}_{L^2_\lambda}\bigr] } where
$\emph{\Div} a = \del_t a^0 + \del_j a^j$ and $\vec{a} =
(a^1,\ldots,a^n)$.
\end{lem}
\begin{proof}
First, we have \eqn{engA.3}{ \frac{d\,}{dt} \ip{w}{a^0 w}_{L^2_\lambda} =
2\ip{w}{a^0\del_tw}_{L^2_\lambda} + \ip{w}{\del_t a^0
w}_{L^2_\lambda} = 2\ip{w}{\Je a^j\del_j  \Je w}_{L^2_\lambda} +
2\ip{w}{g}_{L^2_\lambda} + \ip{w}{\del_t a^0 w}_{L^2_\lambda} \, .
} Letting $\Jde$ denote the adjoint of $\Je$ with respect to the
inner-product \eqref{L2ip}, we can write the above expression as
\leqn{engA.4}{ \frac{d\,}{dt} \ip{w}{a^0 w}_{L^2_\lambda} =
2\ip{\Jde w}{a^j\del_j  \Je w}_{L^2_\lambda} +
2\ip{w}{g}_{L^2_\lambda} + \ip{w}{\del_t a^0 w}_{L^2_\lambda} \, .
} Integration by parts shows that \leqn{engA.5}{ \ip{\Jde
w}{a^j\del_j  \Je w}_{L^2_\lambda} = - \ip{\del_j\Jde w}{a^j\Je
w}_{L^2_\lambda} -\ip{\Jde w}{(\del_j a^j + a^j
\rho^{-1}\del_j\rho)\Je w}_{L^2_{\lambda}} } where $\rho =
\sigma_1^{-2\lambda -n}$. Since
$\norm{\rho^{-1}\partial_j\rho}_{L^\infty}< \infty$, together
lemmas B.7
and B.8 of \cite{Oli05c} and \eqref{engA.5} imply that
\leqn{engA.6}{ \ip{\Jde w}{a^j\del_j  \Je w}_{L^2_\lambda} \leq -
\ip{\del_j\Je w}{a^j\Je w}_{L^2_\lambda} + C(1+\norm{\del_i
a^i}_{L^\infty}+ \norm{\vec{a}}_{L^\infty})
\norm{w}^2_{L^2_{\lambda}} \, . }
Again integrating by parts and using
lemma B.8 of  \cite{Oli05c}, we find that
\leqn{engA.7}{
 -\ip{\del_j\Je w}{a^j\Je w}_{L^2_\lambda} \leq C(1+\norm{\del_i a^i}_{L^\infty}+
\norm{\vec{a}}_{L^\infty})\norm{w}^2_{L^2_{\lambda}}  \, . } The
proof now follows from the Cauchy-Schwartz inequality and
equations \eqref{engA.4}, \eqref{engA.6}, and \eqref{engA.7}.
\end{proof}

Let $\ve^\alpha = \Der^\alpha\ve$, $\be^0 = b^0(u,\Je\ve)$, $\be^j
= b^j(u,\Je \ve)$and $\fe = f(u,\Je \ve)\Je \ve$. The evolution
equation \eqref{approx.1} implies that \leqn{eng1}{ \del_t \ve =
(\be^0)^{-1}\Je\be^j\del_j\Je\ve + (\be^0)^{-1}\fe +
(\be^0)^{-1}h\, .} Differentiating this equation yields
\leqn{eng2}{ \be^0\partial_t\ve^\alpha = \Je \be^j\del_j
\Je\ve^\alpha + g^\alpha } where \leqn{eng3}{g^\alpha =
\be^0[\Der^\alpha,(\be^0)^{-1}\Je \be^j]\del_j \Je \ve +
\be^0\Der^\alpha\bigl((\be^0)^{-1}\Je \fe\bigr) + \be^0
\Der^\alpha\bigl( (\be^0)^{-1}\Je h\bigr) \, .}

To  simplify the following estimates, we will assume that
$b^j(0,0)=0$. It is not difficult to treat the case where
$b^j(0,0)\neq 0$.  Recalling that $\delta\leq \gamma \leq 0$ and
$k>n/2+1$, we get from the calculus inequalities of appendix
\ref{winq} and lemma A.7 of \cite{Oli05c} the following estimate
\alin{eng3a}{ \norm{\be^0&[\Der^\alpha,(\be^0)^{-1}\Je
\be^j]\del_j \Je \ve}_{L^2_{\delta-|\alpha|}} \leq
\norm{\be^0}_{L^\infty} \norm{[\Der^\alpha,(\be^0)^{-1}\Je
\be^j]\del_j \Je
\ve}_{L^2_{\delta-|\alpha|}} \\
& \leq C\bigl(\norm{(\be^0)^{-1}\Je \be^j}_{H^k_{0}}\norm{\del_j
\Je \ve}_{L^\infty_{\delta-1}} + \norm{(\be^0)^{-1}\Je
\be^j}_{W^{1,\infty}} \norm{\del_j \Je \ve}_{H^{k-1}_{\delta-1}}
\bigr)\\ & \leq C \bigl[(1+(\norm{u}_{L^\infty} +
\norm{\ve}_{L^\infty})^{k-1})\bigl(\norm{u}_{H^k_0}+\norm{\ve}_{H^k_0}\bigr)\norm{\ve}_{W^{1,\infty}_\delta}
+(1+\norm{u}_{W^{1,\infty}}+\norm{\ve}_{W^{1,\infty}})\norm{\ve}_{H^{k}_{\delta}}
\bigr] } where $C$ is independent of $\eta$. By the Sobolev
inequality (lemma \ref{ciD}) we have
\eqn{eng3b}{\norm{u}_{W^{1,\infty}} \leq C\norm{u}_{H^k_\eta}\, ,
\quad  \norm{v}_{W^{1,\infty}} + \norm{v}_{W^{1,\infty}_{\delta}}
\leq C\norm{\ve}_{H^k_\delta}\, ,} and hence \eqn{eng3c}{
\norm{\be^0[\Der^\alpha,(\be^0)^{-1}\Je \be^j]\del_j \Je
\ve}_{L^2_{\delta-|\alpha|}} \leq
P(\norm{u}_{H^k_\eta},\norm{\ve}_{H^k_\delta} )} for a $\eta$
independent polynomial $P(y_1,y_2)$ . The other terms in
$g^\alpha$ can be estimated in a similar fashion to get
\leqn{eng4.1}{\norm{g^\alpha}_{L^2_{\delta-|\alpha|}} \leq
P(\norm{u}_{H^k_\gamma},\norm{\ve}_{H^k_\delta},\norm{h}_{H^k_\delta})
} where as above $P(y_1,y_2,y_3)$ is an $\eta$ independent
polynomial . It can also be shown using the calculus inequalities
and \eqref{eng1} that \leqn{eng4.2}{ \norm{\Div b}_{L^\infty} \leq
P(\norm{u}_{H^k_\gamma},\norm{\ve}_{H^k_\delta},\norm{h}_{H^k_\delta})\,
. } Finally, we note that \leqn{eng5}{ \norm{\vec{b}}_{L^\infty}
\leq C\, . }

Next, if we define \eqn{enormA}{ \nnorm{\ve}^2_{k,\delta} :=
\sum_{|\alpha|\leq k} \ip{\Der^\alpha \ve}{\be^0\Der^\alpha \ve
}_{L^2_{\delta-|\alpha|}}\, ,} then by \eqref{bb} and \eqref{eng5}
there exists a constant $C>0$ independent of $\et$ such that
\leqn{enormB}{ C^{-1}\norm{\ve}_{H^k_\delta} \leq
\nnorm{\ve}_{k,\delta} \leq C \norm{\ve}_{H^k_\delta} \, . } Since
$\sup_{0\leq t\leq T}\norm{u(t)}_{H^k_\gamma}<\infty$ and
$\sup_{0\leq t\leq T]}\norm{h(t)}_{H^k_\delta}<\infty$, lemma
\eqref{engA} and \eqref{eng4.1}, \eqref{eng4.2}, \eqref{eng5}, and
\eqref{enormB} imply that \leqn{enormBa}{
\frac{d\;}{dt}\nnorm{\ve}^2_{k,\delta} \leq
C(\nnorm{\ve}_{k,\delta} ) \nnorm{\ve}_{k,\delta} } or
equivalently \eqn{enormC}{ \frac{d\;}{dt}\nnorm{\ve}_{k,\delta}
\leq P(\nnorm{\ve}_{k,\delta} ) } for a polynomial $P(y)$ with
positive coefficients that are independent of $\et>0$. By
Gronwall's inequality, \eqref{enormBa}, and proposition \ref{parA},
this implies that there exists constants $T_*,K>0$, both
independent of $\et>0$, such that $T(\et)\geq T_*$ and
\leqn{eng6}{ \sup_{0\leq t \leq T_*}\norm{\ve(t)}_{H^k_\delta}
\leq K \, . } Using the time reversed version of the equation
(i.e. sending $t\mapsto -t$) we also get, shrinking $T_*$ if
necessary, that \leqn{eng7}{ \sup_{-T_*\leq t \leq
0}\norm{\ve(t)}_{H^k_\delta} \leq K \, . } Finally, from
\eqref{eng1}, \eqref{eng6}, \eqref{eng7}, lemma A.7 of
\cite{Oli05c}, lemmas \ref{ciD} and \ref{ciE}, and \eqref{inclA},
we see, increasing $K$ if necessary, that \leqn{neg8}{
\sup_{-T_*\leq t\leq T_*}\norm{\del_t\ve(t)}_{H^{k-1}_\delta} \leq
K . }

\subsect{le}{Local existence and uniqueness}

To get local existence following the approach of Taylor (see theorem 1.2, pg 362
in \cite{TayIII}), we let $\eta \searrow 0$ and use the bounds
\eqref{eng6}-\eqref{neg8} obtained from the energy estimates to extract a weakly convergent
subsequence of $v_\eta$ which has a limit
that solves the initial value problem \eqref{hb.1}-\eqref{hb.2}.
Since the
proof is very similar to that of theorem B.2 in \cite{Oli05c}, we
omit the details.
\begin{prop} \label{leA} \mnote{[leA]}
Suppose $T_1,T_2 >0 $, $\delta\leq \gamma \leq 0$, $k>n/2+1$,
$v_0\in H^k_\delta$, $u\in C^0([-T_1,T_2],H^k_\gamma)$ and $h\in
C^0([-T_1,T_2],H^k_\delta)$. Then there exists a $T_*>0$
$(T_*<T_1,T_2)$ and a $v \in L^\infty((-T_*,T_*),H^k_\delta)$ $\cap$
$\emph{\text{Lip}}((-T_*,T_*),H^{k-1}_\delta)$ that solves the
initial value problem \eqref{hb.1}-\eqref{hb.2}.
\end{prop}
Using the estimates of appendix B of \cite{Oli05c} and of appendix \ref{winq}
and \ref{energy} of this paper, it is not difficult to adapt the proofs
of propositions 1.3-1.5, pgs. 364-365, in \cite{TayIII} to get the following theorem.
\begin{thm} \label{leB} \mnote{[leB]}
The solution $v$ from proposition \ref{leA} is unique in
$L^\infty((-T_1,T_2),H^k_\emph{\text{loc}})$ $\cap$
$\emph{\text{Lip}}((-T_1,T_2),H^{k-1}_\emph{\text{loc}})$ and
satisfies the additional regularity \eqn{leB.1}{ v \in
C^0((-T_*,T_*),H^k_\delta)\cap C^1((-T_*,T_*),H^{k-1}_{\delta}) \,
. } Moreover, if $T_* < T_2$ and $\sup_{0\leq
T<T_*}\norm{v(t)}_{W^{1,\infty}}<\infty$, then there exists a $T^*
\in (T_*,T_2]$ such that the solution can be extended to a
solution of \eqref{hb.1}-\eqref{hb.2} on $(-T_*,T^*)$.
\end{thm}
For linear systems, the energy estimate (see lemma \ref{nlimA}
with $\ep =1$)
ensures, via the
continuation principle of the above theorem,  that the solutions
can be continued as long as the functions $u(t)$ and $h(t)$ are defined.
\begin{prop} \label{leC} \mnote{[leC]}
Suppose $T_1,T_2 >0 $, $\delta\leq \gamma \leq 0$, $k>n/2+1$,
$v_0\in H^k_\delta$, $u\in C^0([-T_1,T_2],H^k_\gamma)$ and $h\in
C^0([-T_1,T_2],H^k_\delta)$. Then the initial value problem
\lalign{hblin}{ &b^0(u)\partial_{t}v = b^j(u)\partial_i v +
f(u)v + h
, \label{hblin.1}\\
&v|_{t=0}  = v_{0} \label{hblin.2}} has a solution \eqn{leC.1}{ v
\in C^0([-T_1,T_2],H^k_\delta)\cap
C^1([-T_1,T_2],H^{k-1}_{\delta})) \, . } that is unique in
$L^\infty((-T_1,T_2),H^k_\emph{\text{loc}})$ $\cap$
$\emph{\text{Lip}}((-T_1,T_2),H^{k-1}_\emph{\text{loc}})$.
\end{prop}
Let $[n/2]$ denote the largest integer with $[n/2]\leq n/2$ and
$k_0 = [n/2] +2$. Then differentiating the solution from
theorem \ref{leB}, with respect to $t$, and using
proposition \ref{leC} yields the following result.
\begin{cor} \label{leD} \mnote{[leD]}
Suppose $k = k_0 + s$, $u \in  \bigcap_{\ell=0}^{s} C^\ell([-T_1,T_2],H^{k-\ell}_{\gamma}) $
and $h \in  \bigcap_{\ell=0}^{s} C^\ell([-T_1,T_2],H^{k-\ell}_{\delta}) $. Then the solution from \ref{leA}
satisfies the additional regularity
\eqn{leD.1}{
v \in \bigcap_{\ell=0}^{s+1} C^\ell((-T_*,T_*),H^{k-\ell}_{\delta}) \, .
}
\end{cor}


\end{document}